%% file: main-v2.tex
\documentclass[letter,11pt]{article}
\pdfoutput=1
\usepackage{siunitx}
\usepackage{jheppub-wExecSum}
\usepackage{graphicx}
\usepackage{epstopdf}
\usepackage{amsmath}
\usepackage{amsfonts}
\usepackage{amssymb}
\usepackage{appendix}
\usepackage{comment}
\usepackage{color}
\usepackage{slashed}
\usepackage{subfigure}
\usepackage{braket}
\usepackage{multirow}
\usepackage{epsfig}
\usepackage{mathrsfs}
\usepackage{latexsym}
\usepackage{url}
\usepackage{etoolbox}
\usepackage{bbm}
\usepackage{ulem}
\usepackage{mathtools}
\usepackage{footnote}
\usepackage{hyperref}
\usepackage{sectsty}
\usepackage{enumitem}
\usepackage{xpatch}
\usepackage{scalerel}
\usepackage[usenames,dvipsnames,table]{xcolor}
\usepackage{lineno}

\definecolor{nicered}{rgb}{0.7,0.1,0.1}
\definecolor{nicegreen}{rgb}{0.1,0.5,0.1}
\definecolor{header_color}{HTML}{15588c}

\newcommand{\beq}{\begin{equation}}
\newcommand{\eeq}{\end{equation}}
\newcommand\bout{\bgroup\markoverwith{\textcolor{blue}{\rule[0.5ex]{4pt}{0.8pt}}}\ULon}

\renewcommand{\figurename}{\color{header_color}{Figure}}
\hypersetup{colorlinks,citecolor= nicegreen,linkcolor= nicered}
\DeclareMathAlphabet{\mathbbold}{U}{bbold}{m}{n}    
\sectionfont{\color{header_color}}
\subsectionfont{\color{header_color}}
\subsubsectionfont{\color{header_color}}
\makeatletter
\xpatchcmd{\@lbibitem}
 {\item[\hfil}
 {\item[\hfil\color{header_color}}
 {}{}
\makeatother

 \DeclareUnicodeCharacter{0301}{\'{e}}

\begin{document}

\title{High-Energy and Ultra-High-Energy Neutrinos:\\
A Snowmass White Paper}

\renewcommand*{\thefootnote}{\alph{footnote}}

\collaboration{
Editors: 
Markus Ackermann\footnote{markus.ackermann@desy.de}, 
Mauricio Bustamante\footnote{mbustamante@nbi.ku.dk}, 
Lu Lu\footnote{lu.lu@icecube.wisc.edu},
Nepomuk Otte\footnote{otte@gatech.edu},
Mary Hall Reno\footnote{mary-hall-reno@uiowa.edu},
Stephanie Wissel\footnote{wissel@psu.edu}
\vspace{1cm}}

\input{authors}
\input{affiliations}

\input{endorsers}
 \input{affiliations-more}
\date\today

\abstract{Astrophysical neutrinos are excellent probes of astroparticle physics and high-energy physics. With energies far beyond solar, supernovae, atmospheric, and accelerator neutrinos, high-energy and ultra-high-energy neutrinos probe fundamental physics from the TeV scale to the EeV scale and beyond. They are sensitive to physics both within and beyond the Standard Model through their production mechanisms and in their propagation over cosmological distances. They carry unique information about their extreme non-thermal sources by giving insight into regions that are opaque to electromagnetic radiation. This white paper describes the opportunities astrophysical neutrino observations offer for astrophysics and high-energy physics, today and in coming years.
}

\setcounter{tocdepth}{2}
\maketitle
\flushbottom


\section{Introduction and overview}

\label{sec:introduction}

Cosmic neutrinos are unique probes of extreme environments surrounding the most energetic sources in the Universe and a unique test beam for weak interactions at energies inaccessible through accelerators.  Within the last decade, the discovery of high-energy (HE, TeV to 100 PeV) astrophysical neutrinos by IceCube \cite{Aartsen:2013jdh} has opened a new window to learn more about cosmic accelerators and neutrino interactions at the highest energies.
With next-generation experiments pushing sensitivity and energy reach, we anticipate that the wealth of information will expand dramatically.  Ultra-high-energy (UHE, $>$100~PeV) neutrinos, long-sought but not yet detected, provide the only means of directly investigating processes that occur at energy scales of EeV ($\equiv 10^{18}$~eV) and above in the distant Universe.  Discovering them would open new regimes of exploration in high-energy physics, astrophysics, and cosmology. In this white paper, we describe the significant physics opportunities offered by cosmic neutrinos and map out the experimental landscape in the coming decades. 

\begin{figure}[t!]
  \centering
  \includegraphics[width=\textwidth]{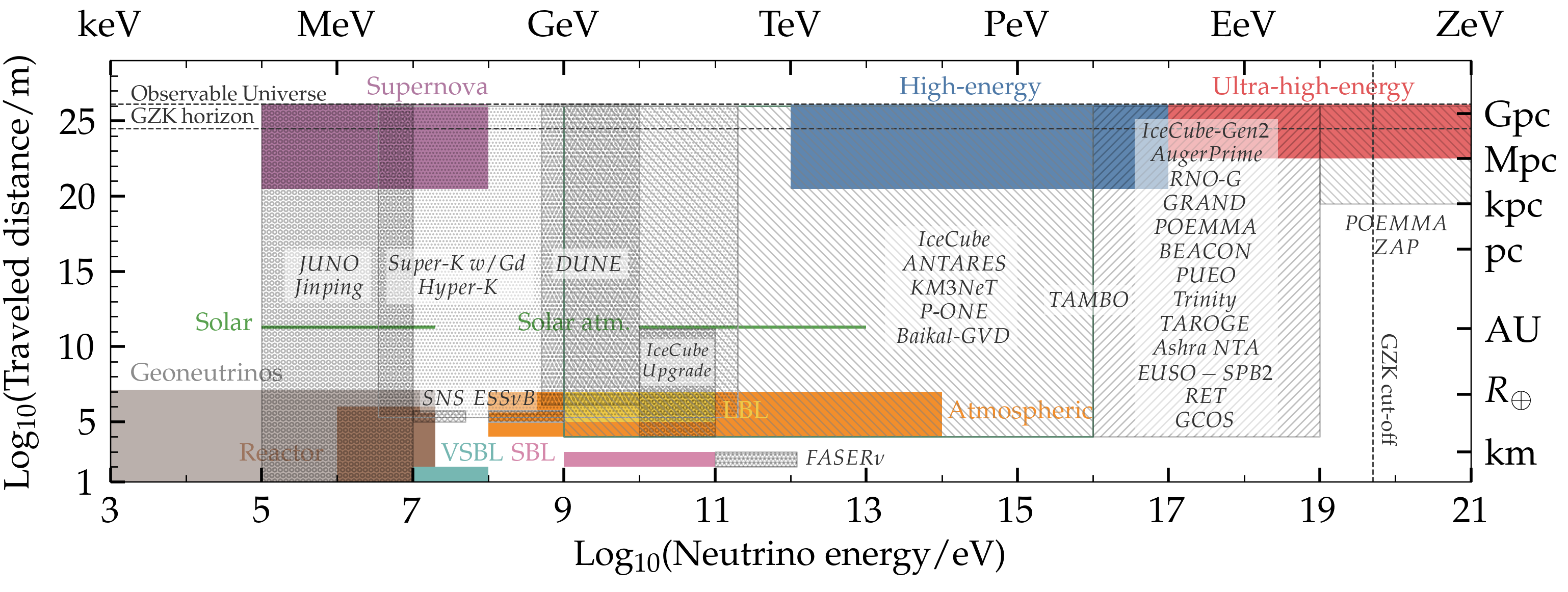}
  \caption{\label{fig:scales}Distribution of neutrino sources in energy and distance traveled to the detector, and present and future experiments aimed at detecting them.  We focus on high-energy and ultra-high energy neutrinos. Updated from Ref.~\cite{Ackermann:2019cxh}. \label{fig:1}}
\end{figure}

Observations of neutrinos from different sources, across different energies and traveled distances, have led to the fundamental-physics conclusions that neutrinos have mass and mix among flavors.
These Nobel-prize winning experimental tests include measurements of neutrinos in the sub-GeV-to-10-PeV energy range from cosmic-ray interactions in the atmosphere
\cite{Super-Kamiokande:1998kpq} and of neutrinos from the Sun
\cite{Super-Kamiokande:2001ljr,Super-Kamiokande:2001bfk,SNO:2001kpb,SNO:2002tuh,Super-Kamiokande:2002ujc,SNO:2005oxr}. Indeed, neutrinos access important questions in the complementary fields of high-energy physics and astrophysics. The wide range of neutrino energies and traveled distances allow us to explore neutrino properties, their interactions, and fundamental symmetries across a wide breadth of parameter space, as shown in Fig.~\ref{fig:1}.  And because they are neutral and weakly interacting, they carry information about the physical conditions at their points of origin; at the highest energies, even from powerful cosmic accelerators at the edge of the observable Universe.

Recently, the discovery of a diffuse flux of HE astrophysical neutrinos, in the TeV--PeV range~\cite{Aartsen:2013jdh,Albert:2017nsd} opened a new view to the Universe. They have made possible the direct measurement of weak interactions in a new energy regime, including the neutrino-nucleon cross section \cite{IceCube:2017roe,Bustamante:2017xuy,IceCube:2018pgc}, inelasticity distribution~\cite{IceCube:2018pgc}, and the first Glashow resonance ($\bar\nu_e e$) candidate~\cite{Lu:ICRC2019,IceCube:2021rpz}. Ongoing and future observations will refine the measurements of the astrophysical neutrino observables (energy spectrum, flavor composition, distribution of arrival directions and arrival times) and extend them beyond 10~PeV.  For particle physics, this means gaining sensitivity to smaller predicted effects and extending the energy scale of fundamental physics that can be tested.  For astrophysics, this means probing the most energetic non-thermal sources of the Universe indirectly through the diffuse flux, and directly through the discovery of point sources.  Further,  neutrinos from transient astrophysical events, detected in spatial or temporal coincidence with cosmic rays and electromagnetic radiation
\cite{IceCube:2018cha,Stein:2020xhk,Giommi:2020hbx,ANTARES:2017bia}, will improve our understanding of the extreme physical processes in these environments.  UHE neutrinos with energies exceeding 100~PeV, first predicted more than fifty years ago~\cite{Beresinsky:1969qj} but still undiscovered, are the next frontier in probing fundamental physics and astrophysics at the ultimate neutrino energies. 

The preceding decade has ushered in a new era of astroparticle physics, including high-energy neutrino detection.  Figure~\ref{fig:1} shows that the potential outlined above will be achieved by a rich experimental program of detectors in the next 10--20~years that are presently in different stages of planning, design, and construction.  We anticipate that the next decade will result in the construction of multiple high-energy neutrino detectors spanning complementary regions of the sky, with differing sensitivity to different energy ranges between TeV and EeV, and complementary flavor-identification capabilities.  While the preceding decade was one of neutrino discovery at high energies, the coming years will be of higher-precision studies at high energies and, plausibly, of discovery at ultra-high energies. These studies are further enhanced by observations with all four messengers -- cosmic rays, neutrinos, photons, and gravitational waves. For reviews, see complementary Snowmass Whitepapers on each of these messengers~\cite{Coleman:2022abf, Abraham:2022jse, Engel:2022bgx, Berti:2022wzk} and the broad scope of multi-messenger physics~\cite{Engel:2022yig}.

\subsection{HE and UHE cosmic neutrinos in particle physics}

There is a vast landscape of physics to explore at the highest energies, and high-energy cosmic neutrinos are uniquely well-equipped for the task~\cite{Ackermann:2019cxh}.  Their potential as probes of fundamental physics~ \cite{Anchordoqui:2005is,Anchordoqui:2013dnh,Ahlers:2018mkf,Ackermann:2019cxh,Arguelles:2019rbn} was identified early, but they were only discovered recently, in 2013, when the IceCube Neutrino Observatory observed a diffuse flux of TeV--PeV cosmic neutrinos \cite{IceCube:2013cdw, Aartsen:2013jdh, IceCube:2014stg, IceCube:2015qii, Aartsen:2016xlq}. Since then, there has been a gradual shift of focus from proposing prospective tests of high-energy neutrino physics to performing real, data-driven tests, of increasing sophistication and based on progressively more and better experimental data.  
This, paired with a rich present and future experimental program, provides a valuable opportunity to make significant progress.

To seize this opportunity, the neutrino community needs a comprehensive exploration plan that maximizes the potential to probe neutrino properties and the discovery of new physics, identifies target experimental sensitivities, and exploits synergies between TeV--PeV neutrino experiments and their upcoming counterparts at lower and higher energies.  

Figure \ref{fig:scales} shows why high-energy (TeV--PeV) and ultra-high energy ($\geq$ EeV) neutrinos are incisive probes of new physics.  
Because they have the highest neutrino energies known---TeV to EeV---they can probe physics at energy scales that are inaccessible to us in the laboratory.  Because they travel unscathed for the longest distances---up to a few Gpc, the size of the observable Universe---even tiny effects can accumulate and become observable. 

Section~\ref{subsec:lessons_particle} outlines the current results for UHE neutrino interactions, lepton flavor mixing and on neutrino production and propagation, including standard model and beyond the standard model (BSM) approaches. In Sec. \ref{subsec:future_particle}, we discuss the goals for the future and summarize efforts underway.

\begin{figure}[t!]
  \centering
  \includegraphics[width=0.5\textwidth]{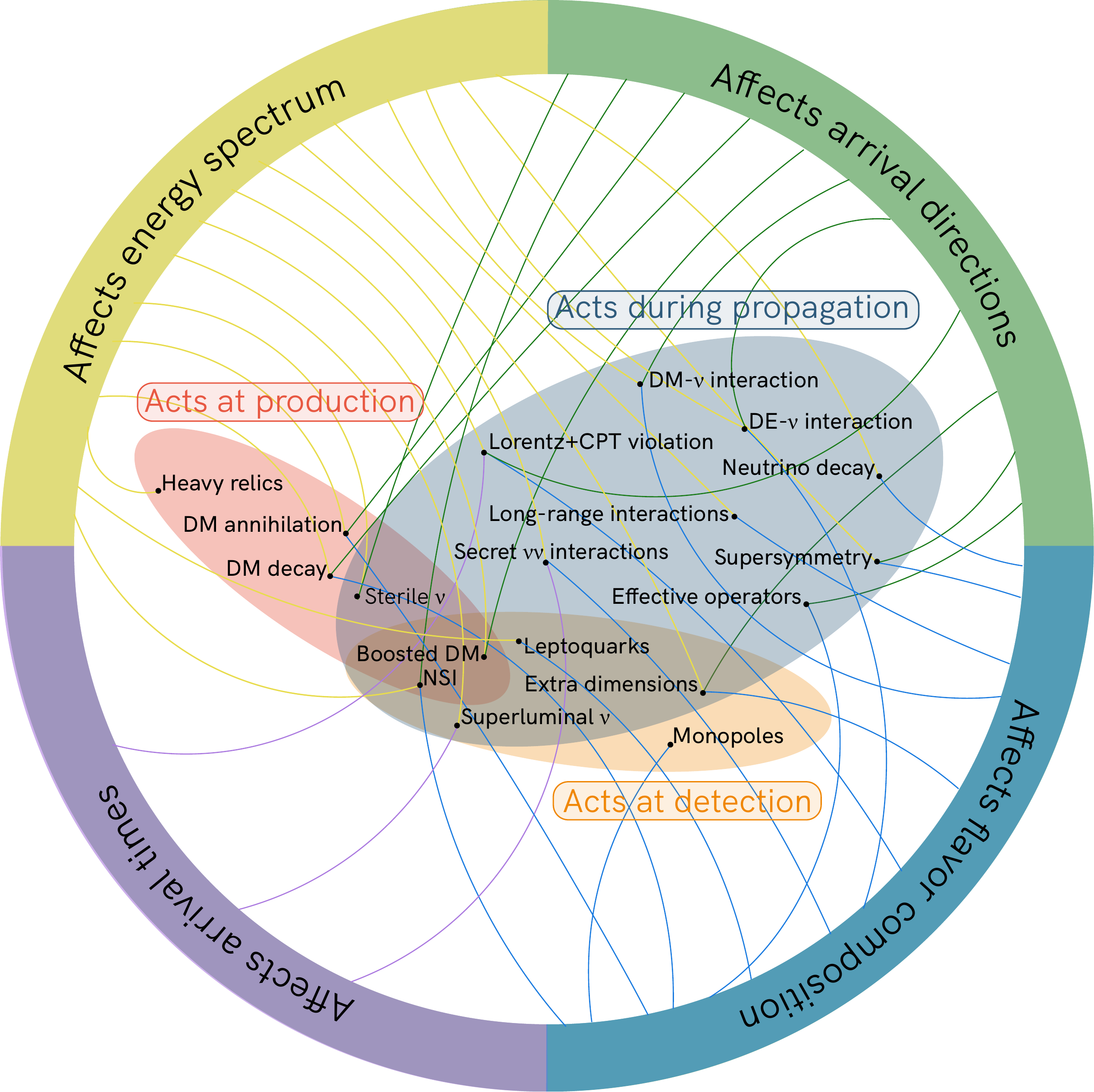}
  \caption{\label{fig:models}\small Models of new neutrino physics and other new physics classified according to the stage at which they act---at production, during propagation, at detection---and what feature they affect---energy spectrum, arrival directions, flavor composition, arrival times.  Figure reproduced from~Ref.\cite{Arguelles:2019rbn}.}
\end{figure}


\subsection{HE and UHE cosmic neutrinos in astrophysics}

That neutrinos can emerge from and point back to their sources make them exceptional probes of the extreme environments that produce ultra-high-energy cosmic rays (UHECRs). The luminosity densities of diffuse sub-TeV gamma rays, high-energy neutrinos and UHECRs  are comparable, signaling the possible connection between cosmic acceleration and the $pp$ and $p\gamma$ interactions that would produce high-energy neutrinos and photons \cite{Murase:2013rfa,Katz:2013ooa,Murase:2018utn}. Measurements of the diffuse high-energy neutrino spectrum and flavor composition, i.e., the relative contribution of $\nu_e$, $\nu_\mu$, and $\nu_\tau$ in the incoming flux, and a nascent multi-messenger program that includes neutrino telescopes will help identify characteristics of source populations.

The origin and acceleration mechanism of UHECRs, with energies in excess of EeV, remains a fundamental outstanding question in astroparticle physics~\cite{AlvesBatista:2019tlv}. A cut-off is observed at energies around 50~EeV~\cite{HiRes:2007lra, PierreAuger:2008rol, Castellina:2019huz, TA:ICRC2019, PierreAuger:2020kuy}. This cut-off coincides with the predictions by Greisen, Zatsepin, and Kuzmin\ \cite{Greisen:1966jv, Zatsepin:1966jv} for UHE protons interacting with the cosmic microwave background. Alternatively, the cut-off could be a consequence of the maximal energy attainable by cosmic accelerators. In both cases there should be a flux of ``cosmogenic neutrinos'' produced through interactions of UHECRs with cosmic background photons\ \cite{Beresinsky:1969qj, Stecker:1978ah, Hill:1983xs, Yoshida:1993pt, Engel:2001hd} which could reach energies well above $\sim$~1~EeV. 
Measurements or constraints on the neutrino energy spectrum would provide much-needed insight into high-energy particle acceleration, the evolution of sources over cosmological length scales, and the mass composition of UHECRs.
            
UHECRs can also interact with gas or radiation inside the sources themselves to produce UHE neutrinos. Because their interactions are weak, these astrophysical neutrinos can travel long distances across the Universe, undisturbed, and point back to their sources, thus tracing a population of high-energy non-thermal sources over Gpc length scales.
Despite many measurements of UHECRs, the sources of UHECRs are still unidentified, though they are believed to be extreme astrophysical environments where charged particles can be accelerated up to $10^{21}$~eV.  To reveal them, complementary information from the neutrinos born in UHECR interactions may be essential.  Further, because high-energy neutrinos should be copiously produced in astrophysical environments, they are also rich probes of high-energy astrophysics~\cite{Ackermann:2019ows}.

The current status and future prospects for diffuse flux and neutrino point source measurements, and of multi-messenger modeling, are described in Sec.~\ref{subsec:lessons_astro} and Sec.~\ref{subsec:future_astro}. The quest for measurements of the cosmogenic neutrino flux and potential BSM sources of neutrinos are also outlined in Secs.~\ref{subsec:lessons_astro} and \ref{subsec:future_astro}.


\subsection{Detector requirements to achieve the science goals} 
\label{sec:requirements}

\begin{figure*}[t!]
  \centering
  \includegraphics[width=\textwidth]{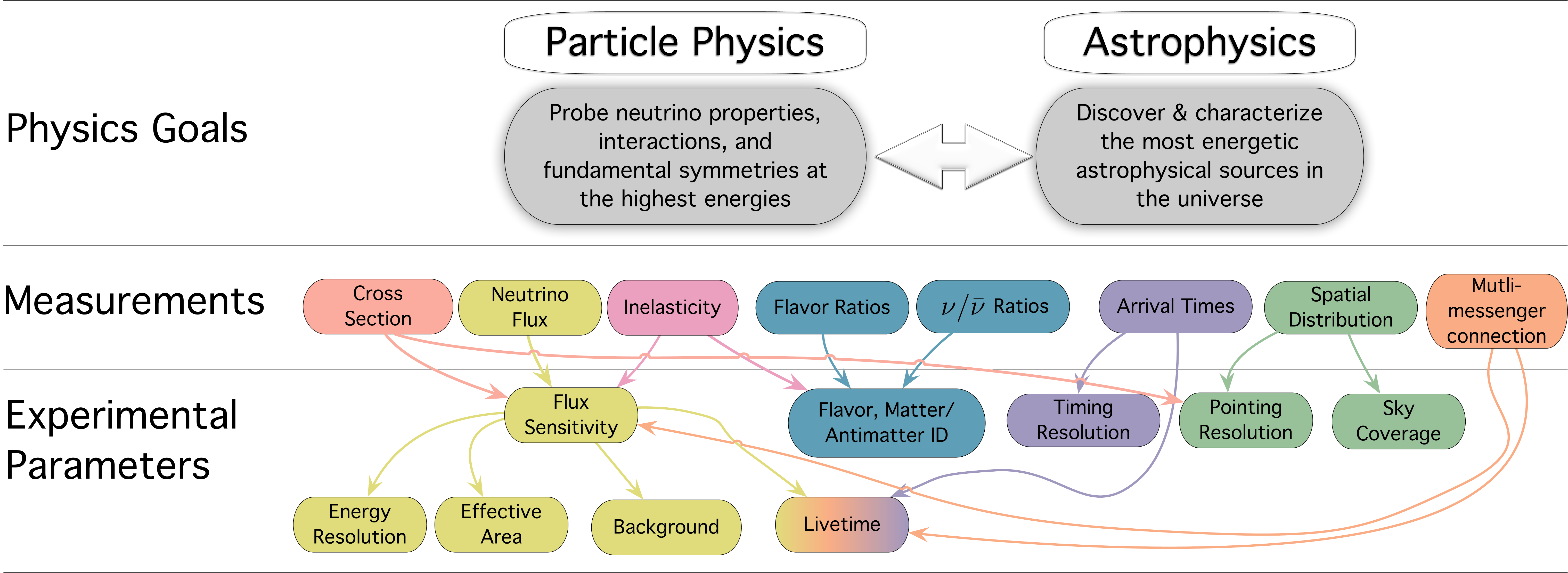}
  \caption{\label{fig:flowchart}\small Cosmic neutrinos probe fundamental particle physics and astrophysics topics in complementary ways. These physics goals dictate requirements on both physical parameters and experimental design.  
  }
\end{figure*}

To probe fundamental physics and astrophysics with neutrinos, we can make several measurements with neutrinos. There are four important observables measured by neutrino telescopes that may be used, individually or jointly, to search for new physics and probe astrophysics: the energy spectrum, the distribution of arrival directions, the flavor composition, and the arrival times.  Each observable has a standard expectation: respectively, a power law in energy\ \cite{IceCube:2015gsk}, an isotropic diffuse flux\ \cite{IceCube:2017trr, Denton:2017csz}, a flavor composition at Earth of $\nu_e:\nu_\mu:\nu_\tau \approx 1:1:1$\ \cite{Aartsen:2015ivb, IceCube:2015gsk, Palomares-Ruiz:2015mka, Bustamante:2015waa, Bustamante:2019sdb}, and coincident arrival of neutrinos and other messengers from transient astrophysical sources \cite{Huang:2019etr, Jacob:2006gn, Addazi:2021xuf}.
Indeed, observations of the neutrino characteristics can be made in concert with similar observables with the other messengers --- photons, cosmic rays, and gravitational waves. New physics can be imprinted on these observables in several ways as shown in Fig.~\ref{fig:models}, such that combined information from several experimental parameters can enhance discovery. The flow from physics goals to experimental parameters is shown schematically in Fig.~\ref{fig:flowchart}.

Fundamentally, large exposures are required to achieve the required sensitivity needed to observe  low neutrino fluxes, requiring an order of magnitude improvement in flux sensitivity in the HE range and two orders of magnitude improvement in the UHE range. As the detectors scale to larger volumes, it is critical to maintain low backgrounds to maximize the improvement in flux sensitivity. Energy resolutions of 0.1 in $\log_{10}(E/{\rm GeV})$ at the HE scale, and in half-decade energy binning, are sufficient to resolve bumps and dips that may indicate new physics~\cite{Hooper:2007jr, Ioka:2014kca, Ng:2014pca, Dhuria:2017ihq, Bustamante:2020mep, Creque-Sarbinowski:2020qhz, Esteban:2021tub} while also distinguishing among models of astrophysical neutrino production~\cite{IceCube:2015gsk, IceCube:2020wum}.

Flavor and $\nu/\bar{\nu}$ ratios provide complementary probes of new neutrino physics and neutrino production mechanisms. Because neutrinos are predominately expected from the decays of muons and charged pions, the nominal expectation is that only electron and muon neutrinos are generated at the sources, and that $\nu$ and $\bar{\nu}$ are produced in comparable numbers.   After leaving the sources, oscillations over cosmological distances are expected to distribute the flux nearly evenly among all flavors by the time the neutrinos reach Earth.
In reality, however, different neutrino production channels become accessible at different energies and, as result, the flavor and $\nu/\bar{\nu}$ ratios should vary with energy; see, e.g., Refs.~\cite{Barenboim:2003jm,  Anchordoqui:2003vc, Kashti:2005qa, Kachelriess:2006ksy, Lipari:2007su, Hummer:2010ai, Hummer:2011ms, Barger:2014iua, Bustamante:2015waa, Biehl:2016psj, Morejon:2019pfu}. As a result, the expected flavor ratios at Earth might deviate from an equi-flavor composition, and might do so as a function of energy.  Thus, we can use the flavor ratios measured at Earth~\cite{Aartsen:2015ivb, IceCube:2015gsk, IceCube:2018pgc, IceCube:2020abv}, combined with information about the values of the neutrino mixing parameters~\cite{Esteban:2020cvm}, to infer the flavor ratios at the sources~\cite{Palomares-Ruiz:2015mka, Bustamante:2019sdb, Song:2020nfh}.
However, large deviations are possible if there are BSM effects in the oscillations.  They can be significantly altered by, for example, Lorentz invariance violation, neutrino decay, and neutrino interactions during propagation over long propagation distances~\cite{Barenboim:2003jm, Keranen:2003xd, Beacom:2003nh, Pakvasa:2007dc, Bustamante:2010bf, Bustamante:2010nq, Mehta:2011qb, Arguelles:2015dca, Bustamante:2015waa, Shoemaker:2015qul, Gonzalez-Garcia:2016gpq, Rasmussen:2017ert, Rasmussen:2017ert, Ahlers:2018yom, Ahlers:2018yom, Ahlers:2020miq, Song:2020nfh}. Large event statistics and complementary flavor-specific detection techniques are needed to identify flavor-specific signals and to measure the flavor composition statistically in a sample of collected events. In the TeV--PeV range, water Cherenkov neutrino telescopes are sensitive to all flavors, though with different efficiency.  We advocate for the exploration of new techniques to improve flavor separation, like muon and neutron echoes~\cite{Li:2016kra}.  In the EeV range, some instruments will be sensitive only to certain flavors, while others will be sensitive to all flavors. We advocate for a comprehensive approach that may allow us to combine flavor information from multiple experiments.

Sub-degree pointing resolution is needed to resolve the neutrino sky~\cite{Murase:2016gly,Ahlers:2014ioa,Fang:2016hop,Bartos:2016wud,Bartos:2021tok} while also reducing the systematic uncertainties on cross section~\cite{Connolly:2011vc, Bustamante:2017xuy, IceCube:2017roe, Denton:2020jft, IceCube:2021jhz, Huang:2021mki, Valera:2022ylt, Esteban:2022uuw} and inelasticity measurements~\cite{IceCube:2018pgc}.  Resolving the neutrino sky will be important to search for BSM physics that causes anisotropies. These can be due to a variety of effects  like BSM matter interactions, dark matter clumping~\cite{Bai:2013nga, ANTARES:2019svn,Iovine:2019rmd,ANTARES:2015vis, Reno:2021cdh,Zentner:2009is,IceCube:2016aga}, or Lorentz invariance violation~\cite{IceCube:2010fyu,Addazi:2021xuf}.

The flux sensitivity, energy resolution, and pointing resolution are all key parameters for measuring both the neutrino-nucleon cross section~\cite{Connolly:2011vc, Bustamante:2017xuy, IceCube:2017roe, Denton:2020jft, IceCube:2021jhz, Huang:2021mki, Valera:2022ylt, Esteban:2022uuw} and inelasticity~\cite{IceCube:2018pgc}, complementary probes of deep inelastic scattering. Key to both is a large number of events~\cite{Hussain:2006wg}. In the TeV-PeV energy range, we can expect improved precision as instruments gain in sensitivity, but at the EeV energy scale, next-generation detectors will need tens of events to measure the cross section to within an order of magnitude.

Measuring the arrival times of neutrinos is important for both time-domain transient and multi-messenger astrophysics, but also to search for evidence for new physics that would cause photons, neutrinos, and gravitational waves to arrive at Earth at different times~\cite{Huang:2019etr, Jacob:2006gn, Addazi:2021xuf}. Being able to capture the transient behavior of sources requires the ability for instruments to send and respond to real-time alerts. Continuous operation is ideal for detecting transient events and improving overall flux sensitivity.

To make advances, we plan to build on existing, mature experiments while also exploring new technologies. There are several operating and planned experiments, many of which are complementary in terms of their energy range, flavor sensitivity, and sky coverage. We describe them below and in Sec.~\ref{sec:experiments}.


\begin{figure*}[t!]
  \centering
  \includegraphics[width=0.9\textwidth]{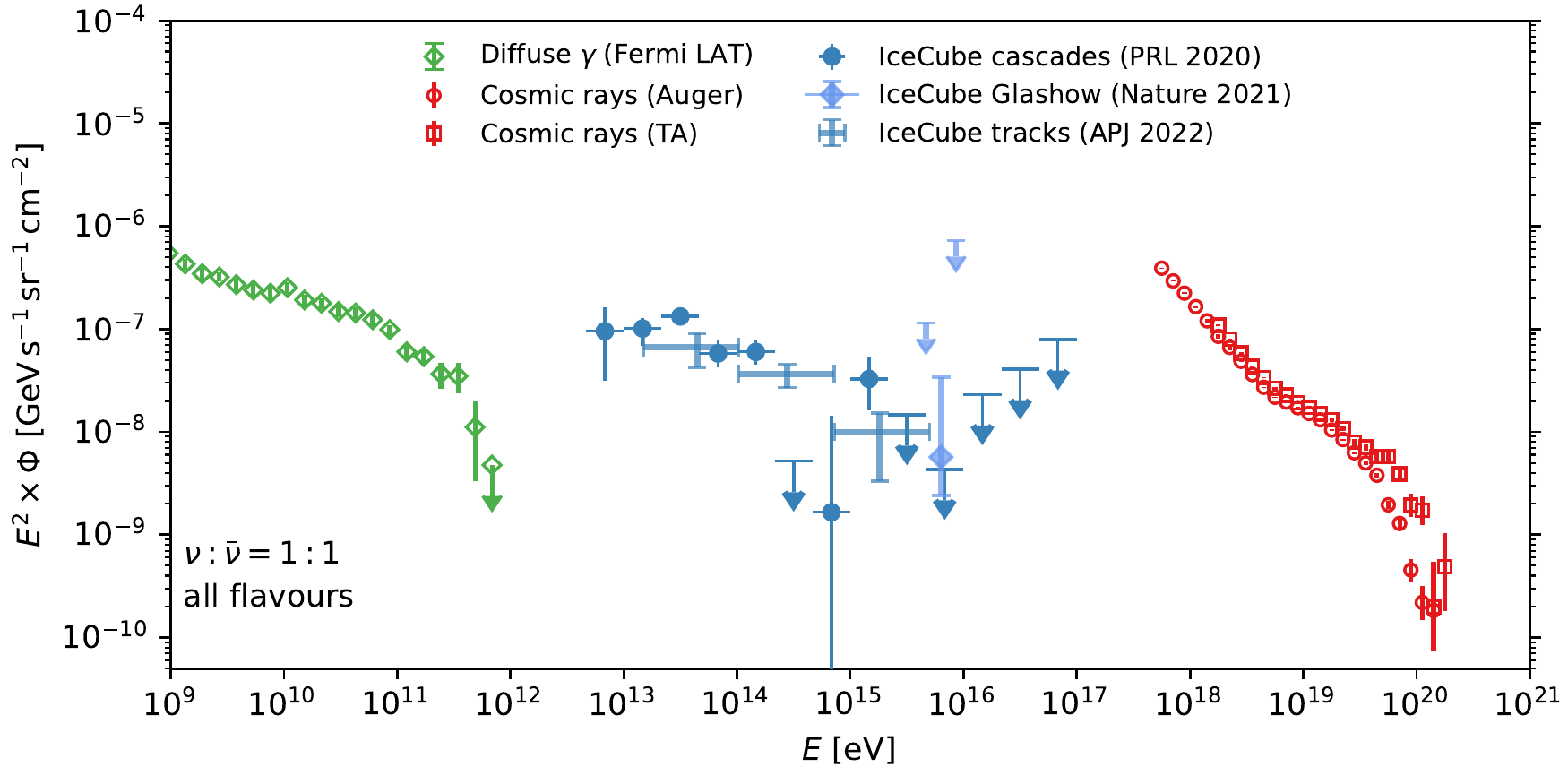}
  \caption{\label{fig:messengers_overview}\small Selected present-day measurements of high-energy gamma rays by Fermi-LAT~\cite{Fermi-LAT:2014ryh}, high-energy neutrinos by IceCube~\cite{IceCube:2020acn,IceCube:2021rpz,IceCube:2021uhz}, and ultra-high-energy cosmic rays by the Pierre Auger Observatory~\cite{PierreAuger:2021hun} and the Telescope Array (TA)~\cite{Ivanov:2020rqn}. The three neutrino measurements are independent and utilise different event topology: contained cascades (solid points), partially-contained cascades (light blue diamonds) and through-going tracks (blue crosses). Error bars and upper limits represent 68\% confidence intervals. For the track sample only bins in the sensitive energy range are shown while the fit was done from 100\,GeV to 100\,PeV.}
\end{figure*}

\subsection{Present and future experimental landscape}

Figure~\ref{fig:messengers_overview} shows the present-day landscape of measurements of high-energy cosmic messengers: gamma rays, neutrinos, and ultra-high-energy cosmic rays.  A complete picture of the high-energy Universe is necessarily multi-messenger in nature.  Below, we focus on neutrinos, but point out scenarios where gamma rays and cosmic rays offer complementary information.

IceCube, presently the largest neutrino telescope, 
is an in-ice Cherenkov detector in Antarctica. It instruments 1~km$^3$ of deep underground ice with thousands of photomultipliers that collect the light emitted by particle showers initiated by high-energy neutrino interactions.  From the amount of light collected and its spatial and temporal profiles, IceCube infers the energy, flavor, and arrival direction of the neutrinos.  Because the bulk of their arrival directions is broadly consistent with an isotropic distribution, the diffuse neutrino flux that IceCube sees is likely of predominant extragalactic origin, though the sources are unknown, save for two promising source associations \cite{IceCube:2018cha, Stein:2020xhk, Giommi:2020hbx}. In particle physics, IceCube has measured the TeV--PeV neutrino-nucleon cross section  \cite{IceCube:2017roe, Bustamante:2017xuy, IceCube:2020rnc} and inelasticity distribution  \cite{IceCube:2018pgc} for the first time, probed charm production in neutrino interactions  \cite{IceCube:2018pgc}, and seen hints of the first high-energy $\nu_\tau$  \cite{Stachurska:2019wfb}, and the Glashow resonance (indicating $\bar{\nu}_{e}$)~\cite{Lu:ICRC2019,IceCube:2021rpz}.  ANTARES, a Cherenkov detector in the Mediterranean Sea operating until recently, nears the sensitivity to the IceCube diffuse neutrino flux~\cite{Albert:2017nsd, Fermani:2020oxx}.  Three new telescopes under construction, KM3NeT\ \cite{KM3Net:2016zxf, Fermani:2020oxx}, P-ONE\ \cite{P-ONE:2020ljt}, and Baikal-GVD\ \cite{Avrorin:2019dli}, will improve our sensitivity to TeV--PeV neutrinos to the Southern Sky. The next generation of IceCube, IceCube-Gen2~\cite{IceCube-Gen2:2020qha}, will improve our sensitivity across a broad energy range from the TeV scale to the EeV scale.

In the next 10--20 years, new detectors may improve our sensitivity to neutrino energies above the energy range where that of IceCube becomes too small to detect a significant flux. 
There are several planned experiments targeting the PeV energy range to determine the high-energy spectrum of the astrophysical flux observed by IceCube. Observation of a spectral cut-off or the continuation of the power-law spectrum would help reveal the sources of these neutrinos while also extending our observations of neutrinos into a new energy scale. These experiments take different approaches including radar (RET-N~\cite{Prohira:2019glh}), particle showers in a valley (TAMBO~\cite{Romero-Wolf:2020pzh}), and Earth-skimming tau neutrinos from a mountain (Trinity\ \cite{Otte:2019knb}, Ashra NTA\ \cite{Sasaki:2014mwa}, CTA\ \cite{Fiorillo:2020xst}), balloons (EUSO-SPB~\cite{Adams:2017fjh}), and satellites (POEMMA \cite{Anchordoqui:2019omw}).

EeV neutrinos have long been predicted as coming from the interaction of ultra-high energy cosmic rays with cosmic photon backgrounds\ \cite{Greisen:1966jv, Zatsepin:1966jv, Beresinsky:1969qj} or photons inside astrophysical sources~\cite{Fang:2013vla, Padovani:2015mba, Fang:2017zjf, Muzio:2019leu, Rodrigues:2020pli, Muzio:2021zud}, but they have not been discovered yet \cite{Lu:2017amt, Zas:2017xdj, IceCube:2018fhm, Persichilli:2019wyh, Gorham:2019guw}.  The flux of these neutrinos is expected to be low \cite{ Fang:2013vla, Padovani:2015mba, Fang:2017zjf, Romero-Wolf:2017xqe, AlvesBatista:2018zui, Heinze:2019jou, Muzio:2019leu, Rodrigues:2020pli, Muzio:2021zud}.  At even higher energies, ZeV neutrinos might come from cosmic strings \cite{Berezinsky:2011cp, Anchordoqui:2018qom}.
Next-generation, multi-purpose EeV--ZeV detectors are being planned or under construction with a variety of detection strategies: in-water and in-ice Cherenkov (IceCube-Gen2\ \cite{IceCube-Gen2:2020qha}), in-air Cherenkov, fluorescence, and particle showers (EUSO-SPB2\ \cite{Adams:2017fjh}, POEMMA\ \cite{Anchordoqui:2019omw}, AugerPrime\ \cite{Aab:2016vlz}, GCOS\ \cite{Horandel:2021prj}), and radio (GRAND\ \cite{GRAND:2018iaj}, RNO-G\ \cite{RNO-G:2020rmc}, PUEO\ \cite{Deaconu:2019rdx}, BEACON\ \cite{Wissel:2020sec}, TAROGE\ \cite{Nam:2020hng}, AugerPrime\ \cite{Aab:2016vlz}, GCOS\ \cite{Horandel:2021prj}). 

Current and planned detectors are detailed in Sec.~\ref{sec:experiments}.


\section{Current status and lessons learned}
\label{sec:lessons_learned}


\subsection{In particle physics}
\label{subsec:lessons_particle}

\noindent
{\bf\color{header_color} HE neutrino interactions}: With the beam of neutrinos at TeV to PeV energies, we are already exploring tests of neutrino interactions in a new energy regime through their cross sections~\cite{IceCube:2017roe, Bustamante:2017xuy, IceCube:2020rnc}, inelasticity distributions~\cite{IceCube:2018pgc}, and the Glashow resonance~\cite{Glashow:1960zz, IceCube:2021rpz}. As statistics grow and new experiments come online, more subtle BSM effects can be explored.

\noindent
{\bf\color{header_color} UHE neutrino interactions:} The interactions of UHE neutrinos with nucleons have center-of-mass energies of $\sqrt s\sim30$~TeV (vs.~$\sim$1~TeV using TeV--PeV neutrinos\ \cite{IceCube:2017roe, Bustamante:2017xuy}), providing an excellent opportunity to probe models of the nucleon and nuclear structure\ \cite{Connolly:2011vc, Cooper-Sarkar:2011jtt, Bertone:2018dse, Anchordoqui:2019ufu}, and new physics in neutrino-nucleon interactions\ \cite{Berezinsky:1974kz, Cornet:2001gy, Kusenko:2001gj, AlvarezMuniz:2001mk, Anchordoqui:2001cg, Kowalski:2002gb, AlvarezMuniz:2002ga, Hooper:2002yq, Friess:2002cc, Anchordoqui:2002vb, Anchordoqui:2005pn, Anchordoqui:2006ta, Hussain:2006wg, Borriello:2007cs, Hussain:2007ba, Barger:2013pla, Marfatia:2015hva, Ellis:2016dgb, IceCube:2017roe, Bustamante:2017xuy, Bertone:2018dse, Klein:2019nbu, Anchordoqui:2019ufu, Mack:2019bps, Denton:2020jft}.
\smallskip

\noindent
{\bf\color{header_color} Flavor transitions at the highest energies:}  Both astrophysical and cosmogenic neutrinos are expected to arrive at Earth with nearly equal fluxes of each of the three flavors, $\nu_e:\nu_{\mu}:\nu_{\tau} = 1:1:1$, as a result of the neutrino production processes at the sources and the values of the oscillation parameters measured in terrestrial experiments.
As some experiments have flavor sensitivity (particularly for $\nu_\tau$), UHE neutrinos provide a key test for flavor changing processes up to the highest energies\ \cite{Beacom:2003nh, Pakvasa:2007dc, Bustamante:2010bf, Bustamante:2010nq, Mehta:2011qb, Bustamante:2015waa, Arguelles:2015dca, Shoemaker:2015qul, Gonzalez-Garcia:2016gpq, Rasmussen:2017ert, Ahlers:2018yom}.
\smallskip

\noindent
{\bf\color{header_color} Tests of neutrino properties during propagation:}  Some of the fundamental properties of neutrinos may have an energy dependence that manifests only at the highest energies.  
Notable examples that have received attention include the breaking of known symmetries or the appearance of new ones\ \cite{AmelinoCamelia:1997gz, Hooper:2005jp, GonzalezGarcia:2005xw, Anchordoqui:2005gj, Bazo:2009en, Bustamante:2010nq, Kostelecky:2011gq, Diaz:2013wia, Stecker:2014oxa, Stecker:2014xja, Tomar:2015fha, Ellis:2018ogq, Laha:2018hsh}, neutrino self-interactions\ \cite{Lykken:2007kp, Ioka:2014kca, Ng:2014pca, Blum:2014ewa, Shoemaker:2015qul, Altmannshofer:2016brv, Barenboim:2019tux, Murase:2019xqi, Bustamante:2020mep}, neutrino-dark matter interaction\ \cite{Feldstein:2013kka, Esmaili:2013gha, Higaki:2014dwa, Rott:2014kfa, Dudas:2014bca, Ema:2013nda, Zavala:2014dla, Fong:2014bsa, Murase:2015gea, Anchordoqui:2015lqa, Boucenna:2015tra, Cohen:2016uyg, Dev:2016qbd, Arguelles:2017atb, Hiroshima:2017hmy, Chianese:2017nwe, Sui:2018bbh, Murase:2019xqi, Arguelles:2019ouk}, and neutrino-dark energy interaction\ \cite{Anchordoqui:2007iw, Klop:2017dim}. 
\smallskip

Below we focus on a selection of tests with high-energy neutrinos that have the potential to yield important new physical insight.


\subsubsection{Cross sections}\label{subsec:crosssections} 

\begin{figure}[t]
  \centering
  \includegraphics[width=\textwidth]{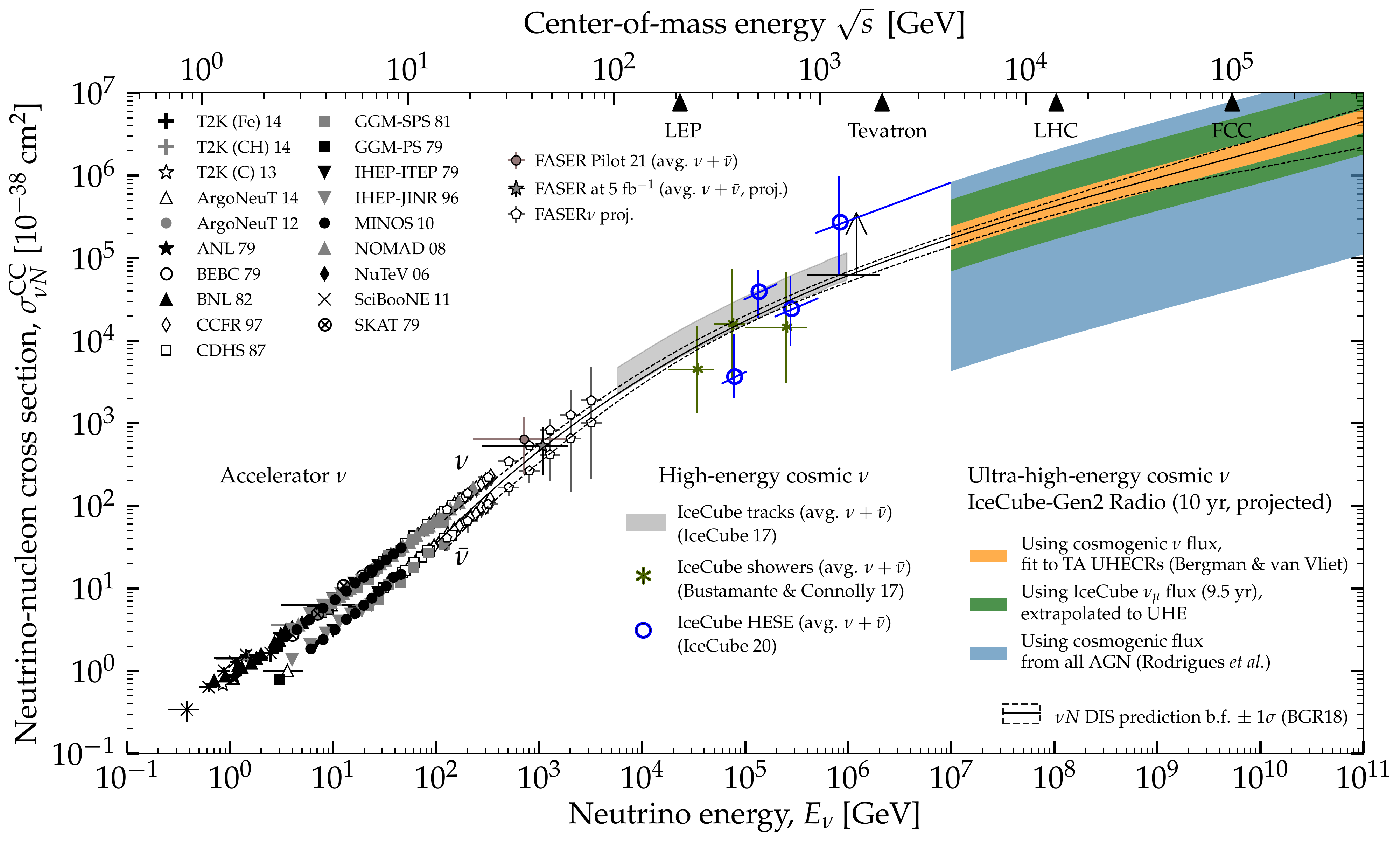}
  \caption{Neutrino-nucleon cross section measurements, compared to deep-inelastic-scattering (DIS) cross section predictions from Ref.~\cite{Bertone:2018dse} (BGR18).  In the TeV range, FASER and FASER$\nu$ have started measurements~\cite{Arakawa:2022rmp}. Measurements in the TeV--PeV range are based on IceCube showers~\cite{Bustamante:2017xuy, IceCube:2020rnc} and tracks~\cite{IceCube:2017roe}.  Projected measurements at energies above 100~PeV~\cite{Valera:2022ylt} are based on 10~years of operation of the radio component of IceCube-Gen2, assuming a resolution in energy of 10\% and a resolution in zenith angle of 2$^\circ$.  Since the flux at these energies remains undiscovered, projections for the measurement of the cross section are for different flux predictions.  Figure adapted from Ref.~\cite{Valera:2022ylt}.}
  \label{fig:cross_section_uhe}
\end{figure}

The cross section for neutrino interactions with nucleons is a unique probe of SM and BSM physics.  In SM physics, measuring the cross section probes the parton distribution functions (PDFs) indirectly.  In BSM physics, measuring the cross section may identify dramatic deviations predicted by various models~\cite{Kusenko:2001gj, Anchordoqui:2001cg, Anchordoqui:2018qom, Klein:2019nbu}. Figure~\ref{fig:cross_section_uhe} shows the neutrino-nucleon interaction cross section, $\sigma_{\nu N}$, measured from GeV to PeV energies, and its projected measurements at hundreds of PeV, compared to a recent SM prediction~\cite{Bertone:2018dse}. In recent years, significant work has gone into predictions for the cross section both within and outside of the SM, motivated by improvements in the determination of PDFs from new collider data and enhanced analysis techniques, and anticipating the advent of next-generation neutrino experiments.

At energies above a few GeV, neutrinos primarily interact with matter via deep inelastic scattering (DIS), where a neutrino exchanges a $W$ (charged-current) or $Z$ (neutral-current) boson with a parton of a nucleon, i.e., a quark or a gluon~\cite{CTEQ:1993hwr, Conrad:1997ne, Giunti:2007ry, Formaggio:2012cpf}. The differential cross section for this process can be expressed in terms of DIS structure functions, which describe the underlying QCD dynamics of the nuclear medium. The structure functions depend on Bjorken-$x$---the fraction of nucleon momentum carried by the interacting parton---and on $Q^2$---the four-momentum transferred to the mediating boson.  Structure functions are computed by convoluting PDFs with coefficient functions using perturbation theory. In the last two decades, different groups have studied high-energy neutrino cross sections \cite{Cooper-Sarkar:2007zsa,Gluck:2010rw,Goncalves:2010ay,Connolly:2011vc,Cooper-Sarkar:2011jtt,Albacete:2015zra,Arguelles:2015wba,Bertone:2018dse} and they have identified several effects that play a significant role in this calculation.

Charm- and top-quark production are important in high-energy neutrino-nucleon interactions. Therefore, heavy-quark mass effects must be calculated to provide an adequate description. At leading order, the slow-rescaling and the modification of light cone momentum fraction in the PDFs must be included~\cite{Aivazis:1993kh}. At higher orders, different formalisms have been developed to account for this effect \cite{Aivazis:1993pi, Thorne:1997ga, Cacciari:1998it}. Nevertheless, the contribution of charm or top production can be significantly different depending on the approach~\cite{Jeong:2010za, Garcia:2020jwr}.

PeV and EeV neutrinos probe the small-$x$ and high-$Q^2$ region of DIS, a kinematic region with very limited data available from colliders and fixed-target experiments to perform PDF fits~\cite{Hou:2019efy}. State-of-the-art PDFs have significantly lowered the minimum value of $x$ used to $\mathcal{O}(10^{-8})$~\cite{Harland-Lang:2014zoa, NNPDF:2017mvq, Sato:2019yez, Hou:2019efy}, but extrapolation must be done if lower values are probed.  Uncertainties in the extrapolation of PDFs translate into uncertainties in the predicted cross section, especially at the highest neutrino energies.  In addition, the stability of the perturbative expansion is spoiled in the very-small-$x$ region, so resummation corrections must be included in the DGLAP formalism~\cite{Bonvini:2016wki,Bonvini:2017ogt,Ball:2017otu,Bertone:2018dse}. An efficient way to account for resummation and saturation effects is the color dipole approach \cite{Gluck:2010rw, Goncalves:2010ay, Albacete:2015zra, Arguelles:2015wba}. Depending on the formalism and input PDFs, the predictions can differ by a factor of 3 in the ultra-high-energy regime; see, e.g., Fig.~\ref{fig:cross_section_uhe} and Fig.~3 in Ref.~\cite{Bustamante:2017xuy}.

High-energy neutrinos are attenuated by interacting with matter during their passage through Earth; the neutrino-nucleon cross section is extracted from measuring the attenuation.
The flux is attenuated more strongly the higher the neutrino energy and the longer the path traversed by the neutrino inside the Earth. This feature can be exploited to probe the neutrino cross section at high
energies. The first measurement of the neutrino cross section from 6.3~TeV to 980~TeV was done by IceCube using up-going muon neutrinos~\cite{IceCube:2017roe}. Later, two independent measurements were carried out using starting events with energies between 60~TeV and 10 PeV~\cite{Bustamante:2017xuy, IceCube:2020rnc}. The results are compatible with the SM predictions described previously. However, it is difficult to draw conclusions beyond that due to the large uncertainties of these measurements.

The neutrino cross section measured in the TeV--PeV region using the astrophysical flux of neutrinos discovered by IceCube has improved our understanding of the cross section since the last Snowmass study a decade ago~\cite{Klein:2013xoa}. We can expect that in the coming decade, improved precision with IceCube, KM3NeT, and Baikal-GVD will reduce the systematic uncertainties, but not necessarily improve the energy reach. The cross section can be measured either by assuming a known flux and inferring the cross section from the observed number of events~\cite{IceCube:2021jhz}, or by comparing the absorption in the Earth as a function of the observed trajectories that neutrinos travel through~\cite{Connolly:2011vc, Bustamante:2017xuy, IceCube:2017roe, IceCube:2020rnc, Denton:2020jft, IceCube:2021jhz, Huang:2021mki, Valera:2022ylt, Esteban:2022uuw}. The latter requires significant absorption of neutrinos through the Earth and therefore has an energy threshold of 5--10 TeV.

In the PeV range, $W$-boson production becomes relevant from two processes: electron anti-neutrino scattering on atomic electrons, and neutrino-nucleus interactions in which the hadronic coupling is via a virtual photon. The former produces the distinct Glashow resonance~\cite{Glashow:1960zz}, which peaks at 6.3~PeV. The latter can reach up to 5-10\% of the DIS cross section in the PeV range \cite{Seckel:1997kk, Alikhanov:2015kla, Gauld:2019pgt, Zhou:2019vxt, Zhou:2019frk}, and it becomes increasingly more important for heavy nuclei as it scales with $Z^2$.

IceCube recently reported the detection of a particle shower compatible with it being due to a Glashow resonance~\cite{IceCube:2021rpz}, with a deposited energy of $6.05\pm0.72$\,PeV. The inferred neutrino energy is $\sim$6.3~PeV, after correcting for the invisible energy from particles that do not radiate visible photons. 
Signatures of low-energy muons were also observed in the event, with energies consistent with expectations of a $W^-$ decay, tagging it as a hadronic cascade.  
Future analyses with more data could reach a $5\sigma$ detection. On the other hand, $W$-boson production ~\cite{Seckel:1997kk, Alikhanov:2015kla, Zhou:2019vxt, Zhou:2019frk} deserves more attention, especially in the near future as much more data is collected. In fact, recent studies have shown that these interactions can play a significant role in the detection of tau neutrinos from cosmic origin~\cite{Soto:2021vdc}. 

Going beyond standard DIS interactions, it is well known that the quark and gluon PDFs of nucleons bound in nuclei are modified compared to free nucleons. Therefore, high-energy neutrino interactions in ice, water, or rock are modified due to the presence of nuclear effects. The most relevant effect for the calculation of the neutrino DIS is that of shadowing~\cite{Frankfurt:2011cs}, namely, the depletion of nuclear structure functions as compared to their free-nucleon counterparts. In the last years, several collaborations have produced PDF sets in the nuclear sector \cite{Kovarik:2015cma, Eskola:2016oht, AbdulKhalek:2019mzd, Walt:2019slu}. First calculations of the neutrino cross section using nuclear PDFs find 5--15\% suppression in the PeV range, but the uncertainty of the nuclear corrections are still large \cite{Bertone:2018dse, Klein:2020nuk, Garcia:2020jwr}. Future Electron-Ion Collider (EIC) measurements of nuclear structure functions will reduce the uncertainties in the neutrino cross section due to nuclear corrections \cite{Khalek:2021ulf}.


\subsubsection{Inelasticity}

Inelasticity is the measure of the fraction of neutrino energy transferred to a hadronic target in deep inelastic scattering. The cross section and inelasticity are complementary probes of new physics since a new interaction should have an inelasticity distribution distinct from that expected from conventional charged current deep inelastic scattering, which is modeled well in the Standard Model~\cite{Anchordoqui:2006wc}. Recent results from IceCube made the first measurement of the $\nu_\mu$ inelasticity distribution above 1 TeV by tracking both muons from a charged-current interaction and the shower deposited in the detector~\cite{IceCube:2018pgc}.  This inelasticity data was also used to constrain the atmospheric $\overline\nu:\nu$ ratio. With more data and good atmospheric neutrino rejection, it may be possible to apply this calculation also to astrophysical neutrinos.  


\subsubsection{Neutrino decay}

\begin{figure}[t]
\centering
\includegraphics[width=0.9\textwidth]{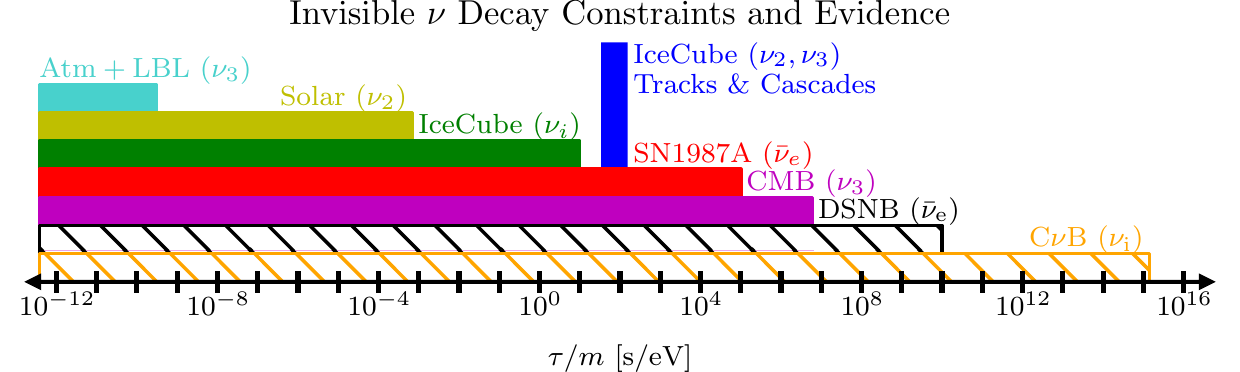}
\caption{Constraints on invisible neutrino decay from a range of experiments.
The blue region represents a hint for neutrino decay \cite{Denton:2018aml,Abdullahi:2020rge} in IceCube data.
Dashed regions represent anticipated sensitivities from future measurements.
CMB constraint depends sensitively on the details of the decay; see Ref.~\cite{Chen:2022idm}. Adapted from Ref.~\cite{Abdullahi:2020rge}.}
\label{fig:neutrino_decay}
\end{figure}

Because neutrinos have mass, they decay.  In the Standard Model, they can decay radiatively, e.g., $\nu_j\to\nu_i+\gamma$ \cite{Petcov:1976ff, Marciano:1977wx}, where $\nu_j$ and $\nu_i$ are different neutrino mass eigenstates.  However, the associated lifetimes are longer than the age of the Universe.
Yet, if neutrinos couple to a new light or massless mediator, then the decay rate could be enhanced.  This can be tested in a variety of environments.
In general, terrestrial and solar constraints are not very strong.
The strongest existing constraints come from the cosmic microwave background~\cite{Hannestad:2005ex, Escudero:2019gfk, Chacko:2020hmh, Escudero:2020ped}, although the bounds may be significantly relaxed~\cite{Barenboim:2020vrr,Chen:2022idm} and there may even be hints of neutrino decay~\cite{Escudero:2019gfk}.  In addition, these constraints are somewhat more model-dependent than the others.
A measurement of the diffuse supernova neutrino background would provide the next most sensitive probe of neutrino decay~\cite{Ando:2003ie, Fogli:2004gy}, followed by a Galactic supernova, although lifetime bounds from SN1987A~\cite{Kamiokande-II:1987idp} can be evaded depending on the flavor structure. The existing constraints from terrestrial and astrophysical experiments as well as projected sensitivities are shown in Fig.~\ref{fig:neutrino_decay}.

After that, the next most relevant constraint comes from high-energy astrophysical neutrinos observed at IceCube \cite{Beacom:2002vi, Beacom:2003zg, Baerwald:2012kc, Bustamante:2016ciw, Rasmussen:2017ert, Denton:2018aml, Abdullahi:2020rge}.
Both the fact that neutrinos have been detected as well as the detailed spectral and flavor information have been used to probe neutrino decay.
A weak hint for neutrino decay was identified by comparing the spectra of different flavors of neutrinos which makes certain predictions, in particular for the tau neutrino flux \cite{Denton:2018aml, Abdullahi:2020rge}.


\subsubsection{Dark matter} \label{sec:dm_current}

\begin{figure}[t]
 \centering
 \includegraphics[width=0.9\textwidth]{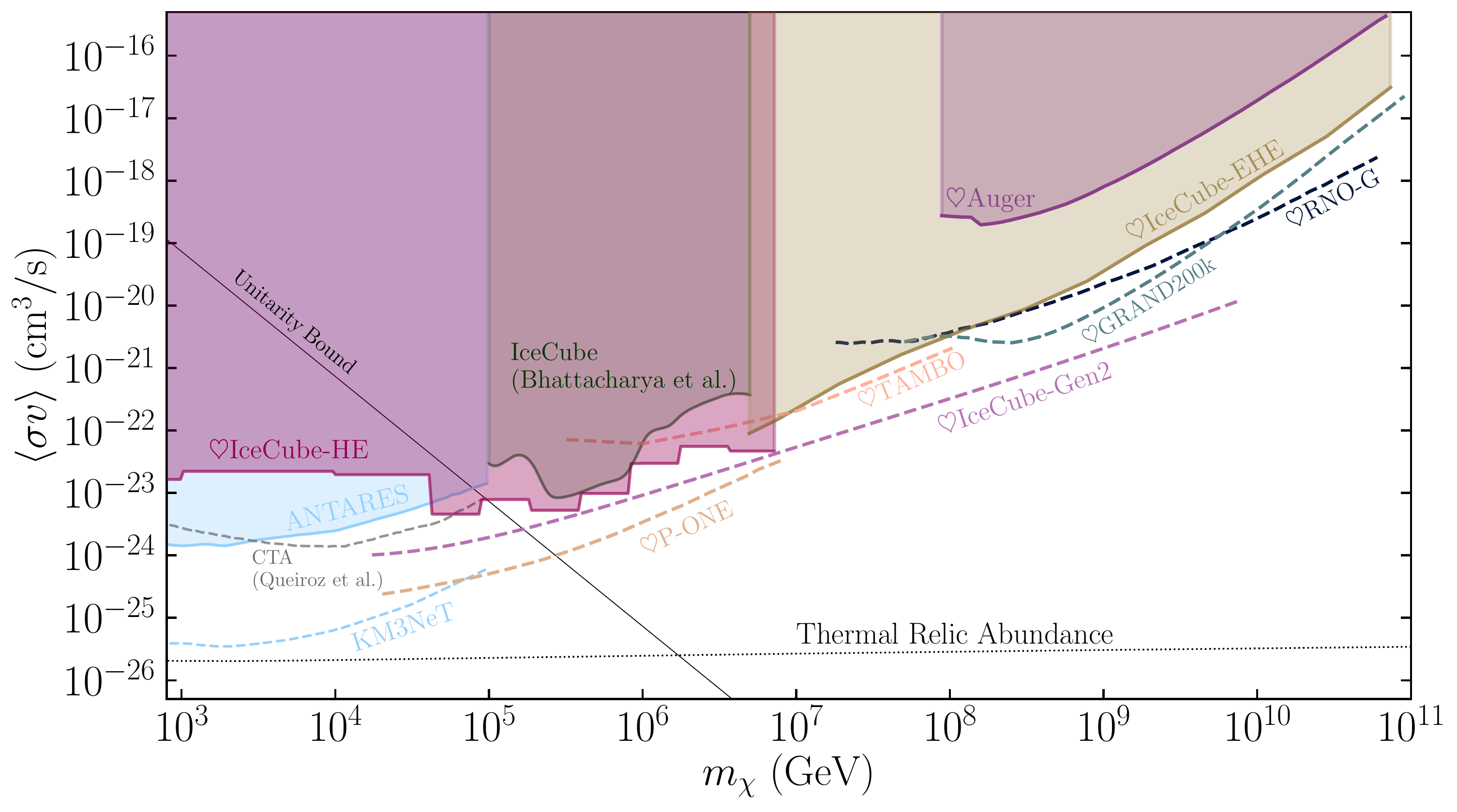}
 \caption{Constraints on the velocity-averaged annihilation cross section, $\langle \sigma v \rangle$, of supra-TeV dark-matter particles, with mass $m_\chi$, into neutrinos.  Solid and dashed lines represent, respectively, 90\%~C.L.~existing limits and sensitivities.  Figure reproduced from Ref~\cite{Arguelles:2019ouk}; heart symbols show new constraints introduced therein.}
 \label{fig:nuDM_annihilation}
\end{figure}

Weakly interacting massive particles (WIMPs) are the primary candidates for particle dark matter (DM).
The WIMP hypothesis yields a thermally average cross-section rate, $\langle \sigma v \rangle \simeq 3 \times 10^{-26} \rm \, cm^3~s^{-1}$, which can explain the observed relic abundance after the freeze-out and is independent of the annihilation products.
Meanwhile, thermal production of WIMPs in the early Universe implies possible ongoing annihilation of DM to Standard Model (SM) particles. This possibility has facilitated the indirect search for dark matter. There is a distinct possibility that neutrinos might be the principal portal to the dark sector. Such possibility is motivated by the {\it scotogenic} models where the neutrinos mass is achieved via interaction with DM, see, e.g.,  Ref.~\cite{Boehm:2006mi}. Furthermore, the upper limit on the DM annihilation cross-section to neutrinos serves as an upper bound on DM annihilation to SM particles, as the latter is larger~\cite{Beacom:2006tt,Yuksel:2007ac}.

In principle, DM could annihilate to all SM particles. Annihilation to most SM particles results in the production of gamma rays and neutrinos. While the most constraining limits on DM annihilation cross-section for sub-PeV DM are obtained from the absence of signal in multi-wavelength observations, especially from the Milky Way and its satellite galaxies, neutrino observations are shown to be providing stronger probes of very heavy DM with masses $> \rm PeV$ \cite{Murase:2012xs}. Moreover, there is a distinct possibility for DM directly annihilating to neutrinos, which makes indirect search via neutrinos important. 

Remarkably, for energies $\gtrsim 0.1 \rm \, MeV$, there exists an interrupted coverage of the neutrino flux from the Universe. An extraordinary amount of data has been collected in this range for measuring neutrino fluxes which can be utilized to search for DM annihilation to neutrinos which have been largely used to impose constraints on DM annihilation to neutrinos. While for the low-mass regime ($m_{\rm DM}< \rm TeV)$, the limits are already approaching the thermal relic density values (see \cite{Arguelles:2019ouk} for details), the current and upcoming neutrino telescopes aiming at very high energies are opening a new avenue for the indirect search of DM. 

Figure \ref{fig:nuDM_annihilation} shows the upper limits on DM annihilation cross section for DM particles heavier than 1 TeV. Currently, the upper limit on DM annihilation cross section is obtained from the observations of ANTARES and IceCube neutrino observatories. For DM in the mass range of 1--100 TeV, ANTARES dominates the upper limit landscape, thanks to its optimal location to observe the Galactic center. For heavier masses, the best constraints are given by IceCube observation of high-energy cosmic neutrinos, as well as IceCube's search for extremely-high-energy (EHE) neutrinos. The ability of neutrino telescopes to probe DM annihilation will be significantly improved with the upcoming and planned neutrino experiments.   


\subsubsection{Secret neutrino interactions}
\label{subsec:secret_neutrino_interactons}

\begin{figure}[t]
 \centering
 \includegraphics[width=.45\linewidth]{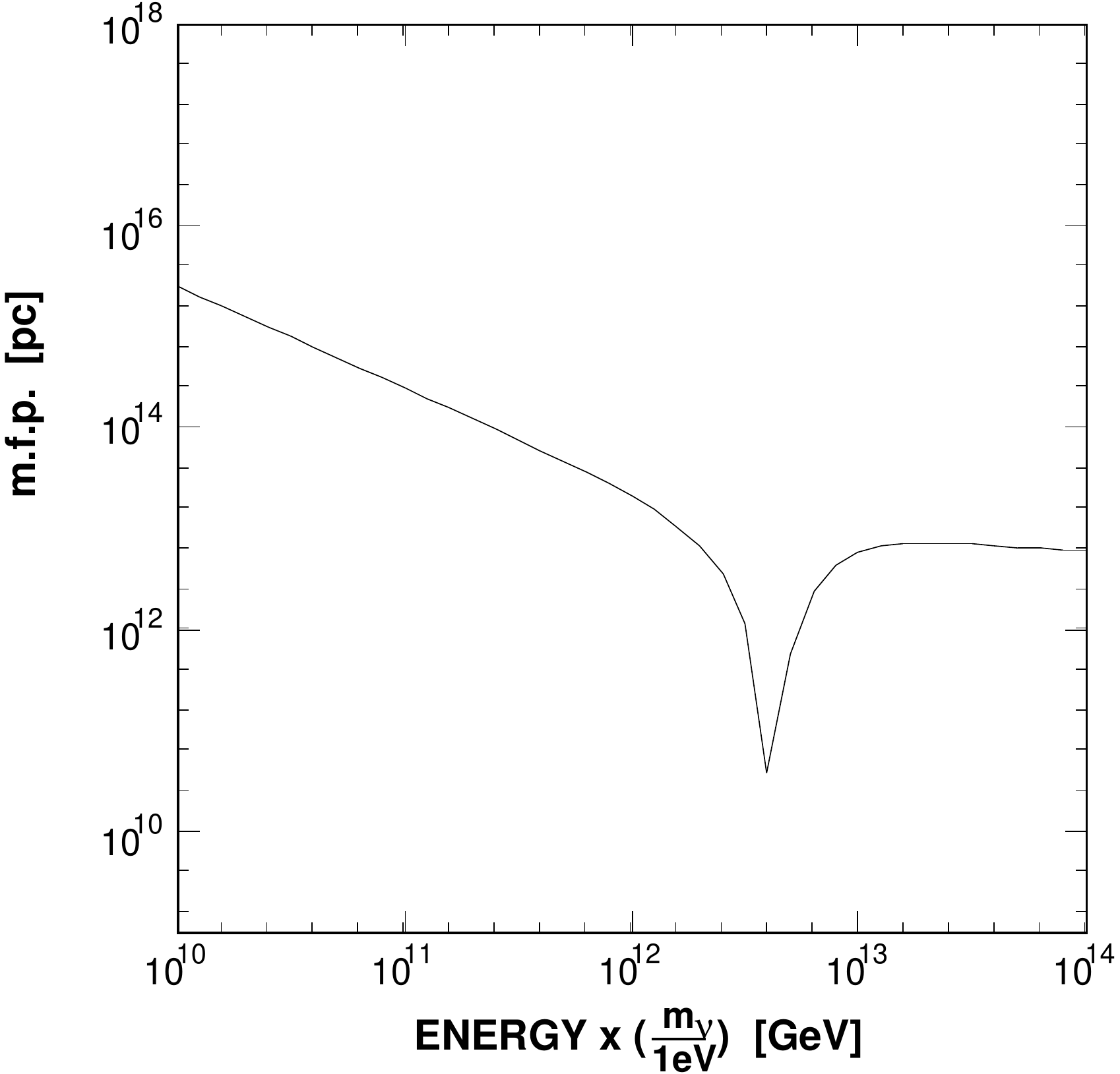}
 \caption{\label{Fig:mfp_CnuB} Mean free path of UHE neutrinos in C$\nu$B field. Only the SM neutrino-neutrino interactions are involved. The sharp resonance structure appears due to the interactions via s-channel of $Z^0$ boson. The possible BSM interactions involving neutrinophilic mediator are expected to form a similar structure at $E_\nu\gtrsim 10^6$~GeV range, which would distort spectral shape of neutrinos. Figure reproduced from Ref.~\cite{Yoshida:1996ie}.}  
\end{figure}

While the SM allows for interactions among neutrinos, these interactions are all highly suppressed by the electroweak scale -- any interaction cross section involving only neutrinos scales as the square of the Fermi constant, $G_F^2$. It is still unknown whether there are additional BSM interactions solely among neutrinos that are stronger than this, with some effective interaction scale $G_{\rm eff.} \gg G_F$. In a UV-complete BSM model, this implies the existence of some new electrically-neutral mediator that couples to neutrinos, significantly lighter than the $Z^0$ boson.

New interactions of this type can significantly modify the character of the HE and UHE neutrino flux arriving at the Earth by the scattering of the HE/UHE neutrinos off of nearly-at-rest cosmic neutrino background (C$\nu$B) neutrinos. The main feature of this effect is the absorption of neutrinos with a characteristic energy $E \approx m_\phi^2/(2m_\nu)$, where $m_\phi$ is the mass of the new neutrinophilic mediator and $m_\nu$ is the mass of one of the light neutrino eigenstates. The spectral distortion features may appear at PeV-EeV region, depending on the absolute mass of neutrinos. 
This model is independently motivated by the neutrino mass generation mechanism~\cite{Blum:2014ewa}, muon $g-2$ anomaly~\cite{Araki:2015mya}, small-scale problems in dark matter substructures~\cite{vandenAarssen:2012vpm,Tulin:2017ara}, and apparent Hubble tension~\cite{Cyr-Racine:2013jua,Kreisch:2019yzn,Blinov:2019gcj,Carpio:2021jhu}. 
A standard model example of a sharp resonance structure in the mean free path of neutrinos scattering in the C$\nu$B field via $Z^0$ boson production is shown in Fig. \ref{Fig:mfp_CnuB}.

If we assume $m_\nu = 0.1$ eV and $m_\phi = 10$ MeV, this predicts absorption of HE, 500 TeV neutrinos. This effect has been explored in, for instance, Refs.~\cite{Hooper:2007jr, Lykken:2007kp, Ioka:2014kca, Ng:2014pca, Ibe:2014pja, Blum:2014ewa, DiFranzo:2015qea, Cherry:2016jol, Kelly:2018tyg, Barenboim:2019tux, Murase:2019xqi, Bustamante:2020mep, Creque-Sarbinowski:2020qhz}. Beyond the absorption of a characteristic energy of HE/UHE neutrinos, additional effects can be detected, including additional neutrino flux below this energy from a cascade effect~\cite{Ioka:2014kca,Ng:2014pca} and a delayed arrival of post-scattering neutrinos relative to the ones that do not scatter~\cite{Murase:2019xqi}. 

Constraints on such interactions using current IceCube data have been derived in Refs.~\cite{Bustamante:2020mep,Esteban:2021tub}. Depending on the flavor structure of the interaction between this new mediator and neutrinos of flavor $\alpha$ and $\beta$, these searches can place the most stringent constraints on the interactions for mediator masses between roughly $4$ and $40$ MeV, even stronger than some laboratory and cosmology constraints~\cite{Berryman:2018ogk,Kelly:2019wow,Blinov:2019gcj}. In general, the measurement of astrophysical HE/UHE neutrinos constrains $G_{\rm eff} \lesssim 10^{10} G_F$ given current data.


\subsubsection{Prompt neutrinos} 

Prompt neutrinos come from decays of heavy-flavor hadrons, predominantly from charm mesons. In nature, prompt neutrinos can be produced through cosmic-ray interactions with nuclei in the Earth's atmosphere.
Due to very short lifetimes of heavy-flavor hadrons, the energy dependence of the flux of prompt atmospheric neutrinos scales roughly according to the cosmic-ray flux, while the energy dependence of conventional atmospheric neutrinos that come from pion and kaon decays falls with an additional power of $1/E_\nu$.
At energies of $E_\nu \gtrsim$ 100 TeV--1 PeV, the prompt atmospheric neutrino flux dominates over the conventional neutrino flux and becomes the main component of the flux of atmospheric neutrinos. 
Meanwhile, high-energy neutrino telescopes such as IceCube and KM3NeT search for astrophysical neutrinos at TeV--10 PeV energies \cite{IceCube:2013cdw,IceCube:2014stg,IceCube:2015qii,IceCube:2020acn,KM3NeT:2018wnd}. 
Probing the prompt atmospheric neutrinos is not only interesting for detection itself, but also important in that they are the primary background to astrophysical neutrinos. 

\begin{figure}[t]
 \centering
 \includegraphics[width=.55\linewidth]{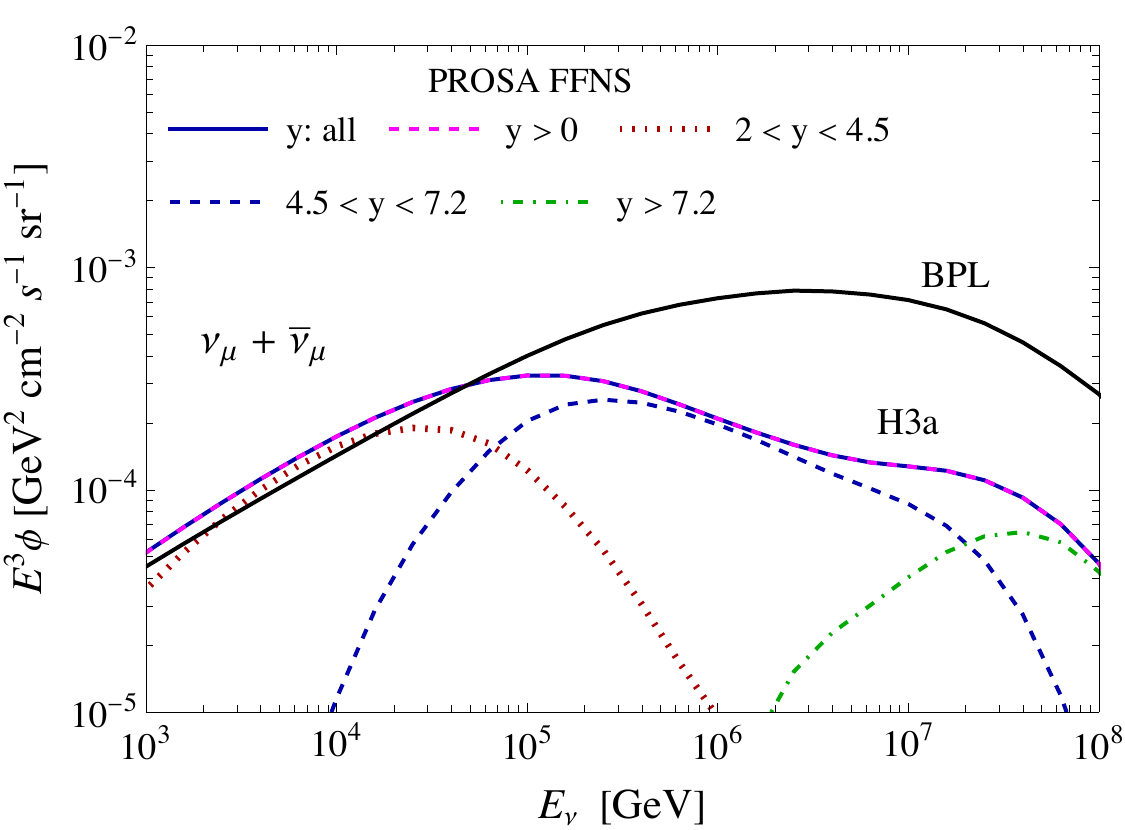}
 \caption{\label{Fig:AtmNu-H3a}Prompt atmospheric neutrino fluxes for $\nu_\mu+\bar\nu_\mu$ assuming a broken power law (BPL) and the H3a~\cite{Gaisser:2011klf} cosmic ray flux. The dashed lines show the contributions to the flux from different collider center-of-mass rapidity ranges for charm production, described in more detail in Ref.~\cite{Jeong:2021vqp}.}  
\end{figure}

Prompt atmospheric neutrinos have not yet been experimentally detected, and the theoretical prediction of their flux has large uncertainties that depend on various factors such as the cosmic-ray spectrum and mass composition, the model for heavy-flavor production, and the parton distribution functions~\cite{Bhattacharya:2015jpa, Gauld:2015kvh, Bhattacharya:2016jce, Zenaiev:2019ktw}.  
One of most significant contributions to large uncertainties in the prompt neutrino flux originates from the limited knowledge of relevant kinematic regions of heavy-flavor production cross sections in hadron collisions. In evaluating the prompt neutrino flux, models for heavy-flavor production for the charm meson production cross section are compared with the data collected by the LHCb Collaboration~\cite{LHCb:2015swx}, which provides data in the most forward region so far, for charm meson rapidities of $2.0<y<4.5$. 
To illustrate the charm meson rapidities relevant to the prompt atmospheric neutrino flux as a function of neutrino energy,
Fig.~\ref{Fig:AtmNu-H3a} shows the prompt atmospheric neutrino fluxes for $\nu_\mu + \bar{\nu}_\mu$ from the charm produced at different collider rapidity ranges in $pp$ collisions \cite{Jeong:2021vqp}. 
These results are evaluated with one of the modern cosmic-ray spectra analyzed taking into account cosmic-ray composition, H3a~\cite{Gaisser:2011klf}. 
The result with the traditional broken power law (BPL) spectrum is presented for comparison.
In the figure, one can see that for the energies of $E_\nu \gtrsim 10^5 - 10^6$ GeV, where the prompt neutrinos are most important, the flux depends on  charm produced in the equivalent collider rapidity range of $y \gtrsim 4.5$.

As noted, current experimental data for charm production are available for $y < 4.5$ from LHCb measurements. Recently, two experiments at the LHC, FASER$\nu$~\cite{FASER:2020gpr} and SND@LHC~\cite{SHiP:2020sos}, were approved. They are in preparation to probe more forward regions, $y \gtrsim 7.2$, during Run 3 of the LHC. Already in 2018, a pilot detector for FASER$\nu$ collected data from 12.2 fb$^{-1}$ of $pp$ collisions at the LHC, from which candidate neutrino interaction events were observed~\cite{FASER:2021mtu}. Combined, the Run 3 measurements by FASER$\nu$ and SND@LHC will allow for tests of theoretical approaches to forward charm production, for example as reviewed in Refs.~\cite{Anchordoqui:2021ghd,Feng:2022inv}, to refine theoretical predictions of the prompt atmospheric neutrino flux.


\subsubsection{Fundamental symmetries}

\begin{figure}[t!]
 \begin{minipage}{0.50\textwidth} 
    \centering
    \includegraphics[width=1.0\textwidth]{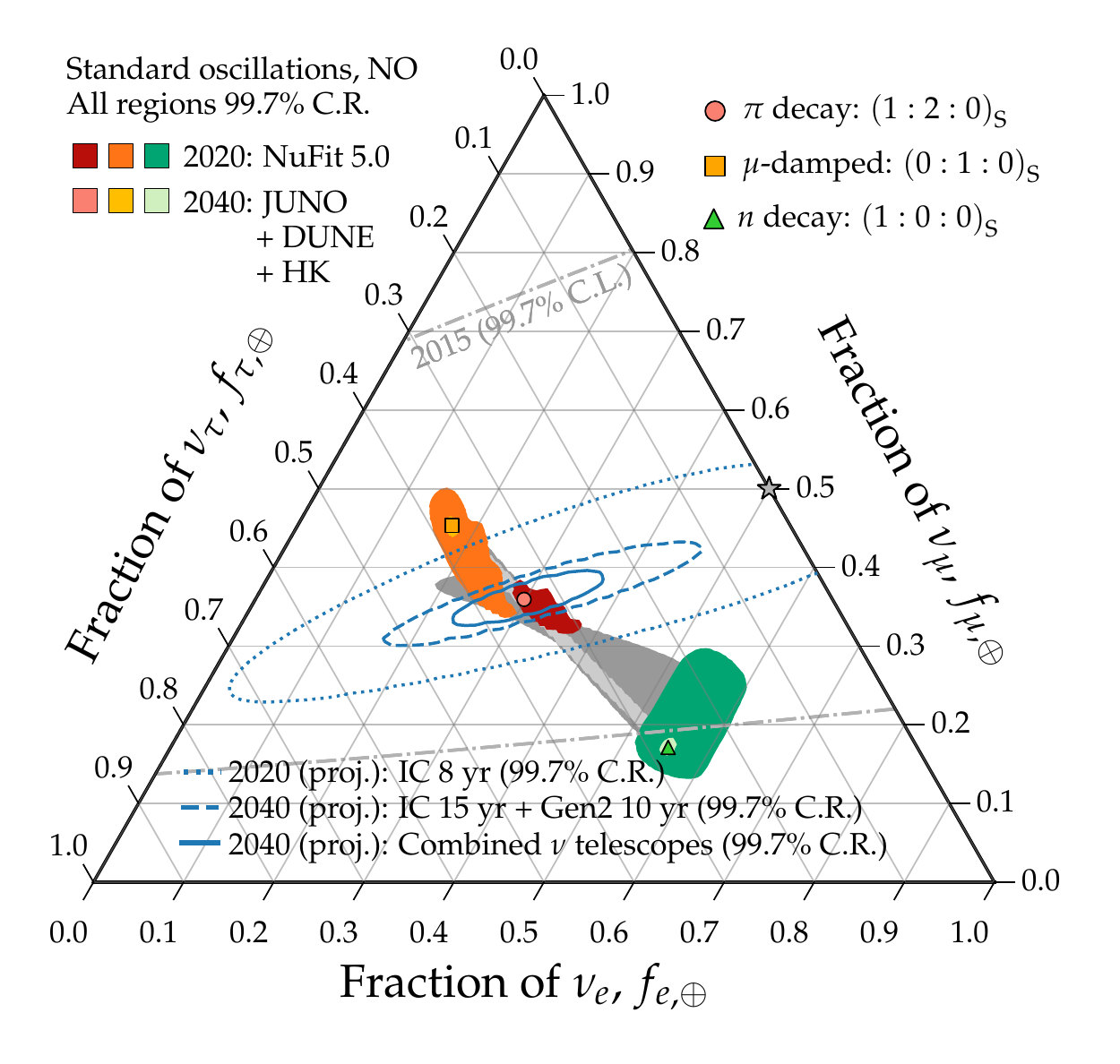}
    \caption{Expected measurements of the flavor composition measured at Earth in both 2020 and 2040. Lines show the 99.7\% credibility regions of using the astrophysical neutrino flux assuming a composition at Earth of $0.3\nu_e:0.36\nu_\mu:0.34\nu_\tau$. These can be compared to predictions from different assumed source compositions ($\pi$ decay, $\mu$-damped $\pi $ decay, and $n$ decay dominated scenarios), unitarity, and standard oscillations. These predictions  will improve over time as oscillation parameters are further refined with new experiments. Figure reproduced from Ref.~\cite{Song:2020nfh}.}
    \label{fig:flavor_ratios_std}
  \end{minipage}
  \hspace*{0.2cm}
  \vspace*{-0.5cm}
  \begin{minipage}{0.48\textwidth}  
   \centering
   \includegraphics[width=0.9\textwidth]{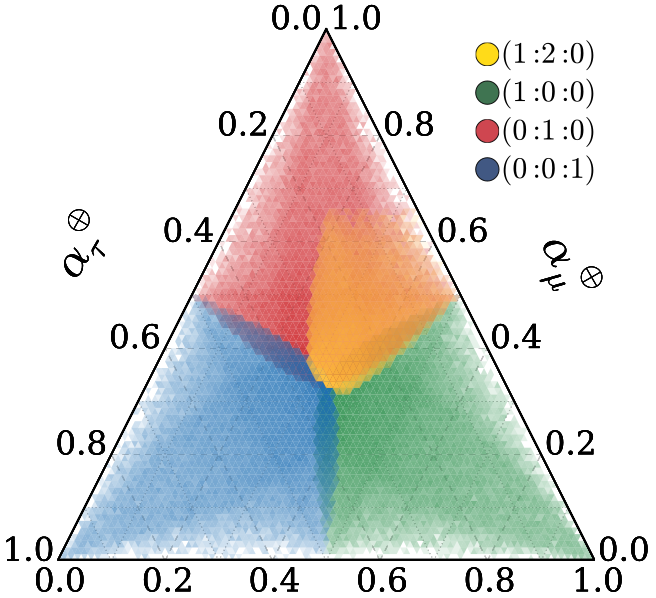}
  \vspace*{0.9cm}
   \caption{Expected flavor ratios observed at Earth under a generic class of new physics assumptions, which can include Lorentz invariance and CPT violation, violation of the equivalence principle, cosmic torsion, and non-standard interactions, among others. The colors represent different assumed flavor compositions at the sources. Figure reproduced from Ref.~\cite{Arguelles:2015dca}. \medskip\medskip\medskip}
   \label{fig:flavor_ratios_liv}
  \end{minipage}
\end{figure}

The Standard Model rests on two fundamental symmetries: the Lorentz and CPT symmetries.  According to the CPT theorem, any Lorentz-invariant local quantum field theory with a Hermitian Hamiltonian, like the Standard Model, must respect CPT symmetry.  So far, all attempts to find experimental evidence for Lorentz-invariance breaking have failed.

However, because the Standard Model is regarded to be an effective field theory, the Lorentz symmetry upon which it rests might not be truly fundamental, but might be broken at high energies.  The full theory of which the Standard Model is an effective one is presently known, but effective-theory extensions of the Standard Model exist that, while not ultraviolet-complete, contain additional Lorentz-violating interactions.  The Standard Model Extension (SME) is the best-developed effective field theory of this kind.  In it, the coupling strength of the Lorentz-invariance-violating (LIV) interactions, i.e., their Wilson coefficients, have a priori undetermined values.  The SME is regularly used to systematically test Lorentz invariance, by constraining the values of the couplings.  For neutrinos, the presence of LIV modifies flavor-transition probabilities, possibly significantly.  For high-energy neutrinos, the presence of LIV may manifest as deviations in flavor composition of the neutrino flux that reaches Earth, i.e., of the relative amount of $\nu_e$, $\nu_\mu$, and $\nu_\tau$ in the flux.

In the standard case, i.e., if Lorentz symmetry holds, the Hamiltonian that drives flavor transitions in vacuum, is
\begin{equation}
 H_{\rm std}
 =
 \frac{1}{2E} U \rm{diag}(0, \Delta m_{21}^2, \Delta m_{31}^2) U^\dagger \;,
\end{equation}
where $E_\nu$ is the neutrino energy, $\Delta m_{21}^2 \equiv m_2^2 - m_1^2$ and $\Delta m_{31}^2 \equiv m_3^2 - m_1^2$ are the mass-squared differences between neutrino mass eigenstates, and $U$ is the Pontecorvo-Maki-Nakagawa-Sakata (PMNS) mixing matrix that connects the neutrino mass and flavor bases.  The PMNS matrix is conventionally parametrized via four mixing parameters: three mixing angles, $\theta_{12}$, $\theta_{23}$, and $\theta_{13}$, and one CP-violation phase, $\delta_{\rm CP}$, whose values are known from oscillation experiments.  For high-energy cosmic neutrinos, the flavor-transition probability derived from the above Hamiltonian oscillates rapidly.  Because of limited precision in the knowledge of the distance traveled from its source, and in the measurement of neutrino energy, in practice neutrino telescopes are sensitive only to the average flavor-transition probability.  For the transition $\nu_\alpha \to \nu_\beta$ ($\alpha, \beta = e, \mu, \tau$), this is
\begin{equation}
 P_{\alpha\beta}^{\rm std}
 =
 \sum_{i=1}^3
 \lvert U_{\alpha i} \rvert^2
 \lvert U_{\beta i} \rvert^2 \;,
\end{equation}
where $U_{\alpha i}$ and $U_{\beta i}$ arec components of the PMNS matrix.  Thus, the standard average flavor-transition probability is energy-independent, and depends only on the standard mixing parameters.

If neutrinos are produced with a flavor composition of $f_S \equiv (f_{e, {\rm S}}, f_{\mu, {\rm S}}, f_{\tau, {\rm S}})$, then due to oscillations, the flavor composition at Earth will be $f_{\alpha, \oplus} = \sum_{\beta=e,\mu,\tau} P_{\beta \alpha}^{{\rm std}} f_{\beta, {\rm S}}$, for $\alpha = e, \mu, \tau$.  Figure~\ref{fig:flavor_ratios_std} shows the expected flavor ratios at Earth for three benchmark neutrino production scenarios, and accounting for the uncertainties in the neutrino mixing parameters.

The LIV operators in the SME introduce additional terms to the Hamiltonian, i.e.,
\begin{equation}
 \label{equ:hamiltonian_liv}
 H_{\rm LIV}
 =
 \sum_n
 \left( \frac{E_\nu}{\Lambda_n} \right)^n
 \tilde{U}_n
 {\rm diag}(O_{n,1}, O_{n,2}, O_{n,3})
 \tilde{U}_n^\dagger \;,
\end{equation}
where $O_{n,i}$ are the eigenstates of the $n$-dimensional Lorentz-violating SME operator, $\tilde{U}_n$ is a PMNS-like mixing matrix, but parametrized by new mixing parameters, $\xi_i$, and $\Lambda_n$ is the energy scale characteristic of the $n$-dimensional operator.  For instance, $n = -1$ corresponds to neutrino decay; $n = 0$, to CPT-odd LIV; and $n \geq 1$ to CPT-even LIV.  Equation~(\ref{equ:hamiltonian_liv}) reveals the  advantage of using high-energy neutrinos to look for LIV: because $H_{\rm LIV} \propto E_\nu^n$, with $n \geq 0$, its relative importance compared to $H_{\rm std}$ grows with energy.  Therefore, even if the values of the LIV parameters are tiny, the LIV effects may become apparent at high energies.  Below, we show how this would manifest.

The total Hamiltonian is $ H_{\rm tot} = H_{\rm std} + H_{\rm LIV}$.  Unlike $H_{\rm std}$ or $H_{\rm LIV}$ by themselves, the total Hamiltonian $H_{\rm tot}$ is no longer diagonalized by the PMNS matrix $U$ or by a group of $U_n$ matrices, but rather by a new matrix, $V$, that depends on both the standard and LIV mixing parameters.  Because of $H_{\rm LIV}$, the average flavor-transition probability depends not only on the standard mixing parameters, but also on the LIV parameters and the neutrino energy, i.e.,
\begin{equation}
 P_{\alpha\beta}^{\rm LIV}
 =
 \sum_{i=1}^3
 \lvert V_{\alpha i} \rvert^2
 \lvert V_{\beta i} \rvert^2 \;,
\end{equation}
where $V_{\alpha i} \equiv V_{\alpha i}(\theta_{12}, \theta_{23}, \theta_{13}, \delta_{\rm CP}, \xi_i, E_\nu)$ and similarly for $V_{\beta i}$.

Analogously to the standard-oscillation scenario, the flavor composition at Earth with LIV is $f_{\alpha, \oplus} = \sum_{\beta=e,\mu,\tau} P_{\beta \alpha}^{{\rm LIV}} f_{\beta, {\rm S}}$.  Figure~\ref{fig:flavor_ratios_liv} shows the flavor ratios at Earth in the presence of LIV, for the same benchmark production scenarios.  Because the values of the LIV parameters are unknown, when they are allowed to vary large deviations from the standard flavor-composition expectation are possible.  Therefore, detecting a large deviation away from the standard $(f_{e, \oplus}, f_{\mu, \oplus}, f_{\tau, \oplus}) = (1/3, 1/3, 1/3)$ could be interpreted as being due Lorentz-invariance violation.

Another possible test of fundamental symmetries comes from augmenting the Standard Model with a new $U(1)^{\prime}$ gauge symmetry~\cite{Foot:1994vd,He:1991qd,Foot:1990mn}. Such an extension results in a new gauge boson $Z^{\prime}$ responsible for a new interaction between neutrinos and other elementary massive particles. In the ultra-light limit of the $Z^{\prime}$ mass, the resulting interaction is long-range in nature, which would affect the propagation of high-energy astrophysical neutrinos. If the $Z^{\prime}$ exists, neutrinos will in addition interact with matter particles while propagating through the Universe, thus modifying the standard flavor composition $f_{\oplus}$ at Earth~\cite{Bustamante:2018mzu}. 
	
Some well explored BSM extensions are $U(1)^{\prime}$ models with $L_{\alpha}-L_{\beta}$ ($\alpha,\beta = e, \mu, \tau$) symmetries, where $L_\alpha$ is the $\alpha$-flavor lepton number. For $L_{e}-L_{\beta}$ ($\beta = \mu, \tau$), neutrinos would interact with electrons via the $Z^{\prime}$ vector boson subject to the potential $V_{e\beta} = \alpha_{e\beta}\frac{N_e}{R}$, where $\alpha_{e\beta}=g'^2/4\pi$ is the fine structure constant for $L_e-L_\beta$ symmetries. However, for $L_\mu-L_\tau$, since muons and tauons are short-lived and, therefore, rarely encountered by astrophysical neutrinos, the interaction between neutrinos and electrons is only possible via $Z-Z^{\prime}$ mixing, where $Z$ is the SM-boson of the weak interaction. The effective potential is then given by $V_{\mu\tau} = \pm g'(\xi-s_w\chi)\frac{e}{4s_wc_w}\frac{N_n}{4\pi R}$ where $(\xi-s_w\chi)$ is due to the $Z-Z^{\prime}$ mixing~\cite{Heeck:2010pg}. 

In the presence of the new interaction potential, the neutrino propagation Hamiltonian is described by the matrix $H= H_{vac} +V_{mat}+ V_{\alpha\beta}$, where $H_{vac}$ is the vacuum neutrino propagation Hamiltonian in the flavor basis, and $V_{mat}$ is the standard neutrino-matter charged-current interaction term. The new interaction matrix $V_{\alpha\beta}$ is the flavor-diagonal matrix having non-zero  $\alpha \alpha$ and $\beta \beta$ elements as strength of the respective potential with opposite sign.
	
Astrophysical neutrinos are ideal probes because they travel cosmological distances over which any minute extension to the SM can potentially accumulate to a measurable effect.
Any measured deviation from the standard flavor composition could, therefore, be translated into measurements of the parameters of $U(1)^{\prime}$ models. Several of the operating neutrino telescopes like IceCube~\cite{IceCube:2021fyh}, KM3NeT~\cite{KM3Net:2016zxf}, Baikal~\cite{Baikal-GVD:2019fko} with sensitivity to $>$TeV astrophysical neutrinos and their flavors are able to test for $f_{\oplus}$ deviations. 


\subsubsection{ANITA anomalous events}

ANITA has observed a handful of anomalous events; these are events whose observational properties are, at least at first sight, compatible with them being neutrinos. ANITA is a balloon-borne Antarctic UHE particle detector sensitive to the radio emission from neutrino interactions in the ice and air showers that can be generated by both cosmic rays and Earth-skimming neutrinos. The ANITA anomalies come in two types: steeply upcoming air shower and near-horizon air shower events. Air-shower events in ANITA are observed both directly pointed at the payload and reflected off the ice; they can be distinguished with their observed arrival direction and polarity (or phase) inversion due to the reflection off of the ice in the later case. Two events---one each in both the first~\cite{ANITA:2016vrp} and third ANITA flights~\cite{ANITA:2018sgj}---were observed that were consistent with the geomagnetic signal from an air shower in terms of their polarization and spectrum, but their steep arrival direction (roughly $30^{\circ}$ below the horizon) were inconsistent with their observed polarity. In the fourth flight of ANITA, four events near, but below the horizon were observed with an inconsistent polarity indicating that the air shower signal had not been reflected with a significance of 2--3$\sigma$ level when considering possible anthropogenic backgrounds and pointing and polarity reconstruction errors.

Several analyses considered the possibility that the steeply upcoming anomalies are due to a possible Earth-skimming tau neutrino origin and find that that they are in strong tension with predictions from Standard Model cross sections and experimental limits at UHE (including the in-ice neutrino channel within the ANITA)~\cite{PierreAuger:2015ihf, PierreAuger:2021lea,PierreAuger:2021gci,IceCube:2016uab, ANITA:2016vrp, Romero-Wolf:2018zxt, IceCube:2020gbx, ANITA:2021xxh}. This is true for both the diffuse~\cite{Romero-Wolf:2018zxt,IceCube:2020gbx} and point-source neutrino hypotheses~\cite{IceCube:2020gbx, ANITA:2021fuf}. Limits from the Pierre Auger Observatory consider the aperture to both generic upgoing air showers~\cite{PierreAuger:2021gci} and tau lepton induced air showers~\cite{PierreAuger:2021lea}.

The near-horizon ANITA anomalies are more consistent with an Earth-skimming neutrino origin than the steep events in terms of the distributions of their arrival direction and spectral characteristics, but the fluence implied by the detection of four events in one flight is in strong tension with the Pierre Auger Observatory across all energies and is also in tension with ANITA's own in-ice neutrino limits above $10^{19.3}$~eV~\cite{ANITA:2021xxh}.

Several explanations have been proposed to explain the steeply upcoming anomalous events. Mundane effects such as transition radiation of cosmic ray air showers piercing the Antarctic ice sheet~\cite{deVries:2019gzs, Motloch:2016yic} and subsurface reflections due to anomalous ice features~\cite{Shoemaker:2019xlt} may be possible, although the later is disfavored by the observed waveforms of the events~\cite{Smith:2020ecb}. Beyond-the-Standard Model explanations predict the expected signature of new particles exiting the Earth~\cite{Chauhan:2018lnq,Collins:2018jpg,Fox:2018syq,Esmaili:2019pcy,Esteban:2019hcm,Borah:2019ciw,Heurtier:2019git,Heurtier:2019rkz,Hooper:2019ytr}.  As of this white paper, the origin of both sets of
anomalous events remains unclear; follow-up observations of these unusual events are well-motivated \cite{PierreAuger:2021gci}.


\subsection{In astrophysics}\label{subsec:lessons_astro}

{\bf\color{header_color} Particle acceleration in extreme environments:}
The origin of cosmic rays and the associated particle acceleration mechanisms are among the biggest open questions in the astroparticle physics. Possible mechanisms including shock acceleration~\cite{Drury:1983zz}, one-shot/shear acceleration~\cite{Rieger:2019uyp,Mbarek:2021bay}, and magnetic reconnection~\cite{Guo:2020fni}.  In most cases, the accelerated cosmic rays will interact with surrounding matter ($pp$)~\cite{Kelner:2006tc, Kelner:2008ke} and light ($p\gamma$)~\cite{Stecker:1978ah} inside their sources, and produce high-energy neutrinos as a result.  Thus, high-energy neutrinos provide a unique probe of acceleration mechanisms in astrophysical environments. 

Inside astrophysical sources, $pp$ and $p\gamma$ interactions make charged pions that, upon decaying, make high-energy neutrinos: $\pi^+ \to \mu^+ + \nu_\mu$, followed by $\mu^+ \to e^+ + \nu_e + \bar{\nu}_\mu$, and the charge-conjugated processes.  (The beta-decay of neutrons from in $pp$ and $p\gamma$ interactions makes additional $\bar{\nu}_e$, but those neutrinos have an energy approximately a hundred times smaller than neutrinos from pion decay.)  The final-state neutrinos typically carry $\sim3-5$\% of the parent proton energy, given that energy losses of pions and muons are negligibly small.  (The exception is when magnetic fields are intense and synchrotron losses of charged particles are significant; see, e.g., Refs.~\cite{Winter:2013cla, Bustamante:2020bxp}.)  Thus, the detection of PeV neutrinos enables us to study the sources of cosmic rays around $\sim100$~PeV, i.e., around the second knee of the cosmic-ray spectrum, possibly probing the Galactic-to-extragalactic transition in cosmic rays. Higher-energy neutrinos, especially UHE neutrinos in the EeV range, will be crucial to investigate the accelerators of UHECRs. UHE gamma rays would not be directly observed except for nearby UHECR accelerators within 10--100~Mpc, while UHE neutrinos can be used to study distant sources located in the cosmological distance. Gamma-ray bursts (GRBs), tidal disruption events (TDEs), and blazars are among the candidate accelerators of UHECRs.  However, models that may explain the observed neutrinos and UHECRs often require a large baryonic loading, i.e., a large fraction of the available energy imparted to cosmic rays, which may be theoretically challenging.  

Neutrinos are also important as a probe of dense environments that are not visible with photons. Cosmic rays can be accelerated in the vicinity of a black hole~\cite{Ber77,Eichler:1979yy,Stecker:1991vm}, a jet inside a star~\cite{Meszaros:2001ms}, and dense circumstellar material~\cite{Murase:2010cu}. There has been significant progress in numerical simulations of particle acceleration in shocks and accretion flows~\cite{Hoshino:2013pza,Ball:2018icx,Kimura:2018clk}, and in spark gaps~\cite{Levinson:2018arx,Kisaka:2020lfl}. Non-jetted AGN~\cite{Kimura:2014jba,Kalashev:2015cma,Murase:2019vdl,Kimura:2020thg}, TDEs~\cite{Dai:2016gtz,Senno:2016bso,Lunardini:2016xwi}, choked jets~\cite{Murase:2013ffa,Senno:2015tsn,Tamborra:2015fzv,Denton:2018tdj}, and interacting supernovae~\cite{Murase:2017pfe,Petropoulou:2017ymv} are actively discussed as promising targets for high-energy neutrino detection.     


\vspace{0.45em}
\noindent
{\bf\color{header_color} Flavor physics at sources:}
From the full pion decay chain, we nominally expect the flavor composition leaving the neutrino sources to be $f_{e,{\rm S}}:f_{\mu,{\rm S}}:f_{\tau,{\rm S}} = 1:2:0$, where $f_{\alpha, {\rm S}}$ is the fraction of $\nu_\alpha + \bar{\nu}_\alpha$ ($\alpha = e, \mu, \tau$) at production.  However, because neutrinos oscillate, the flavor composition that reaches Earth is different.  In particular, for high-energy neutrinos traveling over cosmological-scale distances, the flavor-transition probabilities are averaged~\cite{Pakvasa:2008nx}.  As a result, the nominal expectation is for the flavor composition at Earth to be the same for all flavors, i.e., $f_{e,\oplus}:f_{\mu,\oplus}:f_{\tau,\oplus} \approx 1:1:1$~\cite{Learned:1994wg,Beacom:2003nh}.  There are relatively small variations around this prediction due to uncertainties in the values of the neutrino mixing parameters~\cite{Bustamante:2015waa, Song:2020nfh}.
However, in highly magnetized sources such as GRBs, pions and muons cool via synchrotron so strongly that the flavor composition at Earth can be modified to $f_{e,\oplus}:f_{\mu,\oplus}:f_{\tau,\oplus} \approx 1:1.8:1.8$ at high energies~\cite{Kashti:2005qa, Bustamante:2015waa, Song:2020nfh}.  Alternately, in high-gradient sources (more than 1.65~keV~cm$^{-1}$), muons from pion decay will be significantly accelerated before they decay, leading to an equal flux of $\nu_e$ and $\nu_\mu$, and a harder neutrino spectrum~\cite{Klein:2012ug}.  In choked jets associated with supernovae and double neutron star mergers, where the neutrino production region is embedded in the progenitor star and merger ejecta respectively, the flavor ratios can further be affected by matter effects~\cite{Mena:2006eq,Razzaque:2009kq,Sahu:2010ap,Xiao:2015gea,Carpio:2020app}.  Measuring the flavor composition of high-energy neutrinos is challenging, since signals from $\nu_e$ and $\nu_\tau$ are easy to confuse; see, however, Ref.~\cite{Li:2016kra}.  Nevertheless, IceCube has performed several measurements of the flavor composition~\cite{IceCube:2015rro, IceCube:2015gsk, IceCube:2018pgc, IceCube:2020abv}; most recently, they have singled out the first $\nu_\tau$ candidates~\cite{IceCube:2020abv}.  We can use these measurements, combined with information on the values of the neutrino mixing parameters~\cite{Esteban:2020cvm}, to infer the flavor composition at the sources and, in turn, constrain their identity~\cite{Mena:2014sja, Palomares-Ruiz:2015mka, Vincent:2016nut, Bustamante:2019sdb, Song:2020nfh}.


\vspace{0.45em}
\noindent
{\bf\color{header_color} Multi-messenger connection and improved source modeling:} Neutrinos are produced by hadronuclear and/or photomeson production processes, in which the neutrino--gamma-ray connection is naturally expected. In particular, the {\it Fermi} gamma-ray data give us stringent constraints on the candidate sources of IceCube neutrinos. For a simple power-law (SPL) spectrum, the spectral index of the neutrino sources is constrained to be $\gamma_{\rm SPL}<2.1$--$2.2$~\cite{Murase:2013rfa}. 
On the other hand, recent shower~\cite{IceCube:2020acn}, HESE~\cite{IceCube:2020wum}, and track data~\cite{IceCube:2021uhz} suggest that the best-fit spectral indices are all softer, and the fact that the all-sky neutrino flux at 10~TeV is larger hints at the existence of hidden neutrino sources that are opaque to GeV--TeV gamma rays~\cite{Murase:2015xka,Capanema:2020rjj,Capanema:2020oet}.    

Because the energy generation rate densities of UHECRs, TeV--PeV neutrinos, and sub-TeV gamma rays are comparable~\cite{Murase:2018utn}, it is natural that, at least in some cases, they are produced in the same network of processes in the same sources. 
This can be done by cosmic-ray reservoirs, where cosmic rays confined in magnetized environments produce neutrinos and gamma rays while escaping cosmic rays contribute to the observed UHECR flux~\cite{Murase:2016gly,Fang:2017zjf}. The candidate source classes include galaxy clusters/groups~\cite{Murase:2008mr,Kotera:2009ms,Fang:2017zjf,Hussain:2021dqp} and starburst galaxies~\cite{Loeb:2006tw,Thompson:2006np,Tamborra:2014xia,Chang:2014hua,Senno:2015tra}. 
In $p\gamma$ scenarios, gamma rays are more likely to be obscured, but the connection between UHECRs and sub-PeV neutrinos is still possible in some source models~\cite{Yoshida:2020div}.    


\vspace{0.45em}
\noindent
{\bf\color{header_color} Cosmogenic neutrinos:} 
While predictions and modeling of the flux of cosmogenic neutrinos~\cite{Romero-Wolf:2017xqe, AlvesBatista:2018zui, Heinze:2019jou} have been steadily improving in recent years due to more information from UHECR experiments~\cite{Castellina:2019huz, TA:ICRC2019, Deligny:2020gzq}, the uncertainty is still about an order of magnitude.
A measurement of the cosmogenic neutrino spectrum will provide valuable inputs to constrain UHECR source properties~\cite{Takami:2007pp, Ahlers:2009rf, Kotera:2010yn, Ahlers:2012rz, Baerwald:2014zga, Aloisio:2015ega, Heinze:2015hhp, Moller:2018isk, vanVliet:2019nse}.


\vspace{0.45em}
\noindent
{\bf\color{header_color} UHE neutrinos from point sources:}
Predictions of the flux of cosmogenic neutrinos are rather uncertain due to not only the composition but also the cosmic-ray maximum energy and redshift evolution of the sources. 
The first detected UHE astrophysical neutrinos may predominantly come from powerful point sources. First, there are source models that predict diffuse neutrino fluxes overwhelming the cosmogenic neutrino flux especially if the UHECR composition is dominated by intermediate or heavy nuclei. 
Second, the UHECR acceleration requires sufficiently powerful sources that may be rare in the Universe. This would make the source identification easier especially at extremely high energies, where atmospheric backgrounds are negligible~\cite{Fang:2016hop, Fiorillo:2022ijt}.
Candidate UHECR accelerators that can also be powerful UHE neutrino emitters include GRBs and hypernovae \cite{Waxman:1999ai,Dermer:2003zv,Murase:2007yt,Razzaque:2013dsa,Thomas:2017dft,Biehl:2017zlw,Zhang:2018agl,Heinze:2020zqb}, magnetar/pulsar-driven supernovae~\cite{Murase:2009pg,Fang:2013vla,Fang:2018hjp,Carpio:2020wzg}, TDEs~\cite{Murase:2008zzc,Biehl:2017hnb,Guepin:2017abw}, and blazars~\cite{Murase:2014foa, Padovani:2015mba,Oikonomou:2019djc,Rodrigues:2020pli,Righi:2020ufi}. 


\subsubsection{Diffuse flux of TeV--100 PeV astrophysical neutrinos}

There are two ways to disentangle the diffuse flux of astrophysical neutrinos from the atmospheric neutrino flux. The first is based on our knowledge of their distinct energy spectrum. The second is based on the expectation that down-going atmospheric neutrinos can be tagged if accompanied by muons produced in the same air shower~\cite{Schonert:2008is,Gaisser:2014bja,Arguelles:2018awr}. Combining both approaches has led to the discovery of an astrophysical flux at the TeV-to-PeV energies with a significance well above $5\sigma$. However, above roughly \SI{10}{\peta \eV}, the picture remains unclear and likely requires next-generation detectors, possibly in combination~\cite{Schumacher:2021Pv}, for a high-significance measurement of the cosmic neutrino flux. Further exploring this energy regime, as well as refining our current knowledge of the diffuse astrophysical flux will help shed light on source production mechanisms. It also opens an avenue to probe fundamental neutrino physics and BSM physics at an energy scale that would be unreachable otherwise.

\begin{figure}[t]
 \centering
 \includegraphics[width=0.6\linewidth]{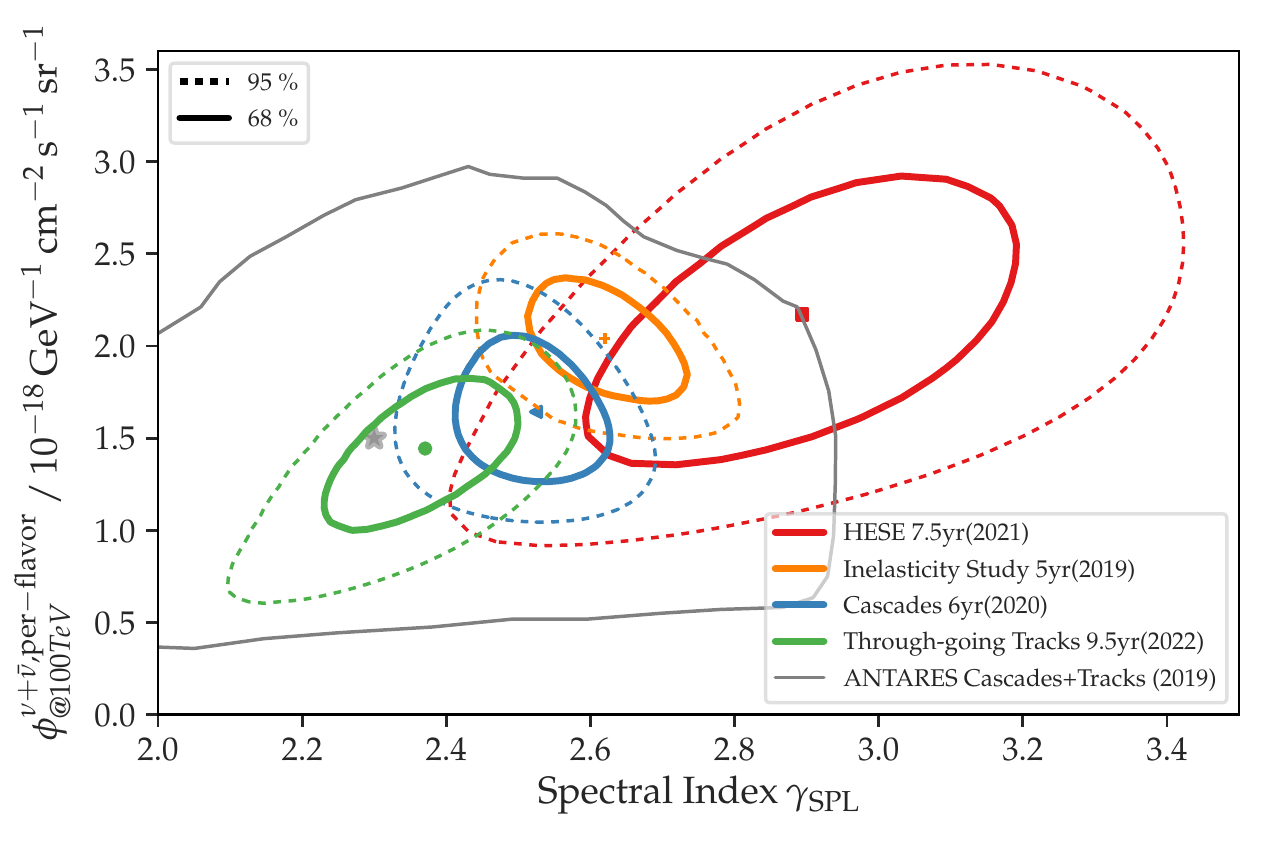}
 \caption{\label{fig:diffuse} 
 Overview of astrophysical diffuse flux measurements to date. The green point and contours shows the most recent best-fit and uncertainties using a sample of predominantly up-going tracks in IceCube with 9.5 years of data taking~\cite{IceCube:2021uhz}. It is overlaid with IceCube results using high-energy starting tracks and cascades (red)~\cite{IceCube:2020wum}, predominantly contained cascades (blue)~\cite{IceCube:2020acn}, and a sample of predominantly starting tracks (orange)~\cite{IceCube:2018pgc}. A recent ANTARES result (gray) using both cascades and tracks with 9 years of data is included for comparison~\cite{Fusco:2020owb}.}  
\end{figure}

The global picture of our current knowledge of the diffuse astrophysical neutrino flux is shown in Fig.~\ref{fig:diffuse} assuming a single-power-law spectrum. 
It summarizes spectral constraints derived from the analysis of various IceCube event samples, as well as ANTARES data~\cite{Fusco:2020owb}. Specifically, the IceCube data sets are a sample of high-energy neutrinos which includes both tracks and cascades with interaction vertices within the instrumented volume~\cite{IceCube:2020wum}, a sample of up-going tracks (mostly muon neutrinos)~\cite{IceCube:2021uhz}, a sample of cascade-like events (mostly electron and tau neutrinos)~\cite{IceCube:2020acn}, and a sample of tracks that start within the instrumented volume~\cite{IceCube:2018pgc}.
An apparent slight tension between the different measurements could be due to differences in flavor composition, energy range, the accounting of atmospheric backgrounds and the spectral model used. The combination of individual samples in a multi-sample analysis~\cite{IceCube:2021jmr}, improved calibration and simulations will lead to improvements of the accuracy of the measurement and a reduction of systematic uncertainties. 

\begin{figure}[t]
 \centering
 \includegraphics[width=0.6\linewidth]{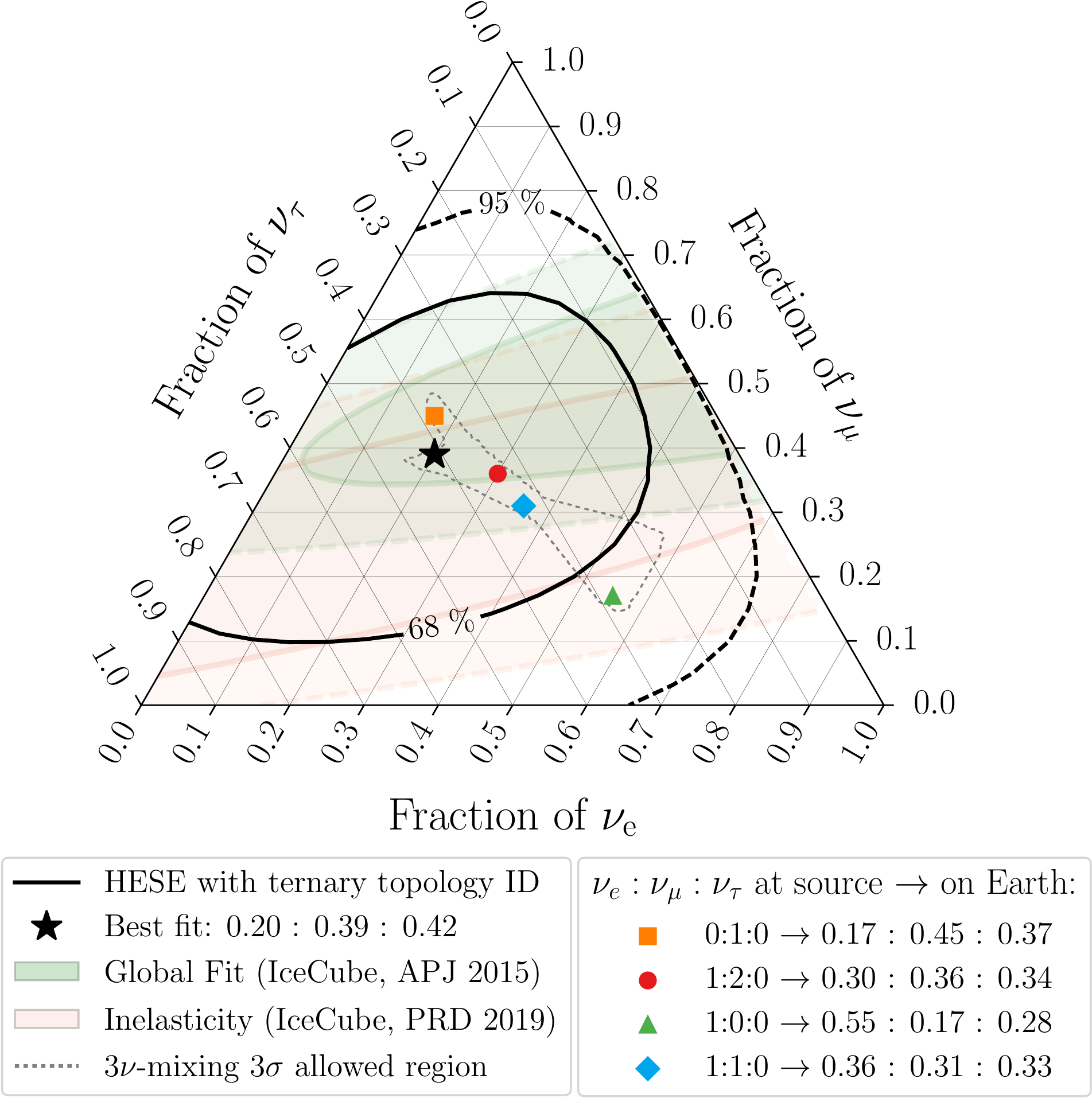}
 \caption{\label{fig:flavorcomp} 
 Flavor constraints on the cosmic neutrino flux from various analyses of IceCube data. The constraints from an IceCube analysis identifying first tau neutrino candidates~\cite{IceCube:2020abv} is shown as black contours. Constraints from earlier measurements, a fit encompassing several IceCube data sets~\cite{IceCube:2015gsk}  and an analysis of the inelasticity distribution of IceCube high-energy events~\cite{IceCube:2018pgc} are shown as shaded regions. They are compared to different scenarios of neutrino production in astrophysical sources and the full range of possible flavor compositions assuming Standard Model flavor mixing (gray dotted region).  Figure reproduced from Ref.~\cite{IceCube:2020abv}.}
\end{figure}

The measurement of the flavor composition of astrophysical neutrinos provides important clues about their origin, production mechanisms and physical properties of their sources. The canonical production of neutrinos in astrophysical sources with flavor ratios $f_{e, {\rm S}}:f_{\mu, {\rm S}}:f_{\tau, {\rm S}} = 1:2:0$ from the decay of charged pions, can be altered by energy losses of pions and muons in the vicinity of the sources, e.g., due to the presence of strong magnetic fields~\cite{Barenboim:2003jm, Kashti:2005qa, Kachelriess:2006ksy, Lipari:2007su, Mehta:2011qb, Winter:2013cla, Bustamante:2020bxp}. Several measurements of the flavor composition have been performed with IceCube~\cite{IceCube:2015rro, IceCube:2015gsk, IceCube:2018pgc, IceCube:2020abv}. Figure~\ref{fig:flavorcomp} summarizes the constraints derived. An important milestone was the recent identification of two $\nu_{\tau}$ candidate events~\cite{IceCube:2020abv}. Current constraints are compatible with several astrophysical production scenarios~\cite{Mena:2014sja, Watanabe:2014qua, Palomares-Ruiz:2015mka, Bustamante:2015waa, Palladino:2015zua, Vincent:2016nut, DAmico:2017dwq, Palladino:2019pid, Bustamante:2019sdb, Song:2020nfh} and standard neutrino oscillations on their propagation to Earth~\cite{Bustamante:2019sdb, Song:2020nfh}, while high-energy neutrino production from the beta-decay of neutrons is strongly disfavored~\cite{Bustamante:2019sdb, Song:2020nfh}. 

The Glashow resonance is an enhancement of the $s$-channel neutrino charged-lepton cross section over the DIS cross section. Due to the preponderance of electrons in matter, it peaks for electron-antineutrinos at \SI{6.3}{\peta \eV} in the electron rest frame. Sixty years after its initial proposal~\cite{Glashow:1960zz}, IceCube detected an astrophysical neutrino interacting at the resonance energy for the first time~\cite{IceCube:2021rpz}. As the resonance occurs on Earth only for electron-antineutrinos, it opens an additional identification channel of both the neutrino flavor and charge. As a result, sources of high-energy astrophysical neutrinos can be expected to produce both neutrinos and antineutrinos. Even with only one event detected thus far, the diffuse flux is now expected to extend above the resonance energy, and constraints on cutoff models can be placed. With additional statistics in the future, we should expect more precise constraints on all fronts including the flux, flavor composition, and production mechanisms; see, e.g., Ref.~\cite{Barger:2014iua, Biehl:2016psj, Bustamante:2020niz}.

\subsubsection{Diffuse flux of ultra-high energy neutrinos: cosmogenic and from sources} 

UHECRs with energies reaching $\sim 10^{11}$~GeV are expected to produce UHE neutrinos of energies well above the PeV scale. UHECR nucleons colliding with the cosmic microwave background (CMB) and extragalactic background light (EBL) produce secondary pions via the photomeson production in extragalactic space. The subsequent decays of pions generated by these interactions, also known as the Greisen-Zatsepin-Kuzmin (GZK) process~\cite{Greisen:1966jv, Zatsepin:1966jv}, generates the a flux of UHE neutrinos~\cite{Beresinsky:1969qj}.
The GZK process is effective when the energies of UHECR nucleons are higher than $\sim 5\times 10^{10}$~GeV, because of the threshold effect on the photopion production process. As approximately $\sim 5$~\% of the parent nucleon energy goes into secondary neutrinos, these cosmogenic neutrinos see their main flux around and below $\sim 10^{9}$~GeV~\cite{Yoshida:1993pt,Takami:2007pp,Kotera:2010yn,Ahlers:2010fw,AlvesBatista:2018zui}. In the case of UHECR nuclei with energies below the threshold for photomeson production, beta-decays of unstable cosmic-ray nuclei produced during the photodisintegration chain can also produce cosmogenic neutrinos~\cite{AlvesBatista:2015jem, AlvesBatista:2019rhs}. In the absence of a significant proton component above $\sim$~50 EeV, this is the dominant mechanism for cosmogenic neutrino production.

Because the CMB field evolves with redshift following the history of Big Bang cosmology, the neutrino production yield highly depends on the cosmic time at which UHECR nucleons collide with CMB photons. Consequently, the cosmogenic neutrino intensity is a good probe for understanding UHECR source evolution with redshift, i.e., the cosmological evolution of UHECR sources~\cite{Yoshida:2012gf}. The present upper limits on the cosmogenic neutrino fluxes obtained by IceCube~\cite{IceCube:2018fhm}, the Pierre Auger Observatory~\cite{PierreAuger:2019ens}, and ANITA~\cite{Gorham:2019guw} have placed constraints on the source evolution parameter space, putting highly evolved astronomical objects such as FSRQs and GRBs into question as UHECR origins~\cite{IceCube:2016uab}.

UHECR sources themselves may yield UHE neutrinos via photonuclear or photohadronic processes, as many of the sources are expected to sit in neighborhood of interstellar gas or photon radiation field~\cite{Fang:2013vla, Padovani:2015mba, Fang:2017zjf, Muzio:2019leu, Rodrigues:2020pli, Muzio:2021zud}. The fact that the UHECR luminosity density is comparable to that of sub-PeV neutrinos measured by IceCube and that of sub-TeV gamma rays measured by {\it Fermi} may also indicate that the three messenger particles share the same origin~\cite{Murase:2018utn}. This astroparticle ``Grand-Unification'' theory has been extensively discussed in the literature. UHECR sources are indeed expected to produce neutrinos and cascaded gamma rays in this scenario, with energies even reaching to $100$~PeV and beyond~\cite{Murase:2013rfa,Murase:2016gly,Fang:2017zjf,Muzio:2021zud}. The generic requirements for being both UHECR accelerators and UHE neutrinos emitters in the unification framework have been extensively discussed for the photohadronic scenario~\cite{Yoshida:2020div}.


\subsubsection{Neutrino sources}\label{sssec:nusources}

The TeV--PeV astrophysical neutrino flux that IceCube has measured is approximately isotropic, and therefore is likely produced mainly by extragalactic sources~\cite{IceCube:2020wum}.  The sources are still almost entirely unresolved and pose a compelling mystery.  Two categories of analyses are used to search for the origins of the astrophysical neutrino signal: neutrino-only analyses and multi-messenger analyses.  The neutrino-only analyses search for clustering of neutrinos in direction and/or time.  They are generally model-independent, but a large trials factor is incurred by searching many directions (and times, for time-dependent studies), which degrades the sensitivity compared to correlation searches with modest numbers of multi-messenger source candidates.   Multi-messenger analyses search for correlation between neutrinos and known sources of other messengers, or at least with the directions of those messengers, including electromagnetic radiation, gravitational waves, and cosmic rays.  These analyses sometimes search for temporal as well as directional correlation.

Neutrino-only analyses (clustering searches) have established strong upper limits on the flux from point-like sources across the entire ($4\pi$) sky~\cite{IceCube:2019cia}.  Dividing the total neutrino flux by these upper limits provides a lower limit on the number of sources (regardless of their distance): there must be at least O($10^3$) sources.  Although the sources are very likely extragalactic, their distance distribution, and even their distance scale, is unknown.

There is evidence that at least one gamma-ray blazar, TXS 0506+056, contributes to the neutrino signal (and is therefore also a cosmic-ray source)~\cite{IceCube:2018dnn,IceCube:2018cha}.  However, the neutrino sky is remarkably different from the gamma-ray sky.  In particular, while gamma rays are an essential component of the evidence that TXS 0506+056 is a neutrino source, the relationship between gamma rays and neutrinos from this and other sources is not simple: the brightest gamma-ray sources are not the brightest neutrino sources~\cite{Silvestri:2009xb, Murase:2016gly, Ackermann:2019ows}.  Nevertheless, the neutrino sources could all be GeV and/or TeV gamma-ray sources, but without simple correlation between the neutrino and gamma-ray fluxes: a subset of the $\sim5 \times 10^3$ known $\gtrsim$GeV gamma-ray sources could be sufficient to explain the entire neutrino signal.  The lack of simple correlation between neutrino and gamma-ray fluxes underlines how complementary the two messengers are.  The lack of correlation could be explained by a combination of (1) the gamma-ray fluxes include leptonic as well as hadronic contributions, (2) the neutrino sources are more distant than the gamma-ray sources, and/or (3) the gamma-ray signals from neutrino sources are attenuated within or near the sources (though in this case gamma rays may be reprocessed to lower energies~\cite{Murase:2015xka}).

In addition to cross-correlation with gamma-ray sources~\cite{IceCube:2016xci,IceCube:2019scr,AMONTeam:2020otr}, analyses have searched for neutrino emission from known sources across the electromagnetic spectrum, including X-ray, UV/optical/IR, and radio sources.  These include blazars as well as the cores of active galactic nuclei (AGN), ultra luminous infrared galaxies (ULIRGs), and radio galaxies~\cite{IceCube:2021waz}.  Time-domain (variable or transient) searches have also been performed for neutrino emission from known sources including blazars, gamma-ray bursts (GRBs), fast radio bursts (FRBs), tidal disruption events (TDEs), supernovae, and novae~\cite{IceCube:2016qvd,Fermi-LAT:2019hte,IceCube:2019acm,Stein:2020xhk,IceCube:2017amx,IceCube:2015jsn}.  In addition to the electromagnetic spectrum, analyses have searched for coincidence between neutrinos and gravitational waves or cosmic rays~\cite{IceCube:2020xks,ANTARES:2022pdr}.

In addition to the predominant extragalactic signal, there could be a significant Galactic neutrino flux~\cite{Ahlers:2013xia, Ahlers:2015moa}.  Models predict both a diffuse Galactic signal (from cosmic-ray interactions producing pions) and emission by discrete sources~\cite{2012Fermi,Gaggero:2014xla}.  Discrete source candidates include the highest-energy gamma-ray sources identified by IACTs as well as wide-field instruments including HAWC and LHAASO, including those that have been detected above 100~TeV~\cite{Vargas:2016hcp,LHAASO:2021zta}.  As in the case of extragalactic sources, detecting neutrino emission from Galactic sources would identify them as cosmic-ray sources.  Detecting diffuse Galactic emission would also provide essential insight into cosmic-ray propagation and interaction.  Current upper limits on the diffuse Galactic flux are comparable to model predictions based on the related gamma-ray signal, which is well measured but at lower energies~\cite{IceCube:2017trr, ANTARES:2018nyb, ANTARES:2020leh}.

Although limits on the spatial clustering of the neutrino signal indicate that there must be at least O($10^3$) sources, the actual number may be close to this or arbitrarily larger.  A more sophisticated analysis than the simple comparison of the total flux to the point-source limits accounts for the source-to-source variation not only in flux but also in redshift and luminosity.  Observational constraints can be set in a 2D phase space (source luminosity, source number density) and compared to known source classes characterized with other messengers.  As has been established for other messengers and wavebands, there are likely multiple diverse source classes, and they likely vary substantially in contribution to the total flux, such that one or two source classes dominate the total astrophysical neutrino flux.


\subsubsection{Multi-messenger modeling of point sources}\label{sssec:mmmodeling} 

The coincident observations of a high-energy neutrino event, IceCube-170922A, with X-rays and  $\gamma$-rays from a blazar TXS 0506+056~\cite{IceCube:2018cha,IceCube:2018dnn} demonstrated the power of the multi-messenger approach. Nevertheless, there are still open questions in this case as we still lack a concordance picture of the multi-messenger data. Neutrino production must be accompanied by the production of gamma rays, which, after cascading to lower energies due to interactions on the CMB and EBL, eventually appear in the X-ray and/or gamma-ray energy range. The resulting flux is predicted to be comparable to the neutrino flux. For the 2017 multi-messenger flare from TXS 0506+056, the electromagnetic spectrum was observed all the way from radio, optical, and X-ray bands to the gamma-ray band, and it showed a valley in X-rays observed by {\it Swift} and {\it NuSTAR}~\cite{Keivani:2018rnh} that is especially constraining. The cascade constraints lead to upper limits on the neutrino flux which are in tension with the observation of IceCube-170922A~\cite{Keivani:2018rnh,Gao:2018mnu,MAGIC:2018sak,Cerruti:2018tmc}. The situation is more challenging for the 2014-2015 neutrino flare from TXS 0506+056~\cite{Murase:2018iyl,Reimer:2018vvw,Rodrigues:2018tku,Petropoulou:2019zqp}, for which various possibilities have been discussed to explain the missing electromagnetic energy~\cite{Xue:2019txw,Zhang:2019htg}. That these results are not sensitive to details of the modeling challenges the single-zone model approach for neutrino and gamma-ray emission.  Other coincidences with flares from supermassive black holes have been reported~\cite{Kadler:2016ygj,Giommi:2020viy,Petropoulou:2020pqh,Rodrigues:2020fbu,Oikonomou:2021akf}. In particular, coincidences with optically-detected tidal disruption events~\cite{Stein:2020xhk} with radio and X-ray counterparts are of special interest~\cite{Murase:2020lnu,Winter:2020ptf}. The connection among different messengers remains clouded. 

Multi-messenger analyses have begun to help our understanding of the astrophysical environments where cosmic rays are accelerated. The ten-year point-source searches with IceCube have indicated that NGC 1068 is the most significant steady source of neutrinos at a significance of $\sim3\sigma$ \cite{IceCube:2019cia}. Intriguingly, NGC 1068 is known to be not only a type-II Seyfert galaxy but also a starburst galaxy.  In the starburst model, NGC 1068 has been considered to be one of the most promising sources~\cite{Yoast-Hull:2013qfa,Murase:2016gly} in the Northern sky (given that M82 may not be ideal for confining cosmic rays with $\sim10-100$~PeV). However, gamma-ray limits provided by MAGIC and HESS observations clearly show that it has to be a hidden neutrino source in the sense that sub-TeV gamma-rays are attenuated or cascaded down to lower energies~\cite{MAGIC:2019fvw}. 
High-energy neutrinos may be produced in the vicinity of the supermassive black hole, and particles may be accelerated in hot coronae around the accretion disk~\cite{Murase:2019vdl,Inoue:2019yfs} or by disk-driven winds~\cite{Tamborra:2014xia,Wang:2016oid,Liu:2017bjr}. 
In particular, given that NGC 1068 is among the brightest sources in intrinsic X-rays, the AGN corona model predicts that NGC 1068 is the most promising neutrino source in the Northern sky~\cite{Murase:2019vdl,Kheirandish:2021wkm}. 
The IceCube flux can be explained if a fraction of the accretion power goes to protons, and gamma rays are predicted to appear in the MeV range without violating the observational data by {\it Fermi}, MAGIC, and HESS.   


\subsubsection{Multi-messenger alerts} \label{sssec:mmalerts}

The conclusive identification of astrophysical neutrino sources and their electromagnetic counterparts will not only enable the study of hadronic processes in cosmic-ray accelerators but also determine the baselines over which cosmic neutrinos propagate, a critical ingredient in many of the already-mentioned studies of fundamental physics. 

The hunt for neutrino sources is typically performed by searching for an excess of neutrinos in space and time (see Sec.~\ref{sssec:nusources}) with respect to the dominant atmospheric background. However, at energies above $\sim100$~TeV, the atmospheric background becomes subdominant with respect to the diffuse astrophysical neutrino flux discovered by IceCube~\cite{Aartsen:2013jdh}, which allows to search for sources by correlating in space and time individual neutrino events with a high probability of being astrophysical in origin with other multi-messenger signals. This concept has been realized in current-generation neutrino telescopes such as IceCube and ANTARES, where the detection of single high-energy neutrinos is promptly communicated to the astronomical community so that targeted observations can be collected to identify an electromagnetic (EM) counterpart~\cite{IceCube:2016cqr,Blaufuss:2019fgv,Dornic:2021kko}. The blazar TXS 0506+056 was identified as a candidate neutrino using this technique as it was observed in a flaring state in gamma rays in coincidence with the detection of the high-energy neutrino IceCube-170922A~\cite{IceCube:2018dnn}. The prompt alert streams maintained by current-generation neutrino observatories have mostly focused on the selection of muon-track events as their positional uncertainty regions (of order 1$^{\circ}$ radius or better) can be covered by pointed EM telescopes, although more recently particle shower events (or ``cascades'') have been added which have a higher astrophysical purity given the lower atmospheric backgrounds for this selection although at the cost of a poorer angular resolution (median error radius of $\sim7^{\circ}$ for IceCube)~\cite{IceCube:2019lzm}. These neutrino alerts are currently communicated via prompt Gamma-ray Coordination Network (GCN) notices, circulars and Astronomer's Telegrams and include the information required to facilitate their follow-up, such as sky position (with uncertainty), energy, and false alert rate for each event. The all-sky coverage of neutrino telescopes can also be used to search for correlated emission with multi-messenger alerts of interest, such as gravitational-wave events or the observation of a transient or variable EM sources that are potential neutrino emitters. Real-time programs such as AMON follow up on these external triggers and alert the community if a coincidence of interest is identified~\cite{AyalaSolares:2019iiy}.


\section{Goals for the next decade}
\label{sec:goals_future}


\subsection{In particle physics}
\label{subsec:future_particle}

\noindent
{\bf\color{header_color}The landscape of new-physics models:}
New physics may affect various features of high-energy cosmic neutrinos: their energy spectrum, arrival directions, flavor composition, and arrival times, as Fig. \ref{fig:models} illustrates for a representative part of the landscape of BSM models. 
Some of the outstanding questions to address today and in coming years are:
How do neutrino cross sections behave at high energies?
\cite{Berezinsky:1974kz, Cornet:2001gy, Kusenko:2001gj, AlvarezMuniz:2001mk, Anchordoqui:2001cg, Kowalski:2002gb, AlvarezMuniz:2002ga, Hooper:2002yq, Friess:2002cc, Anchordoqui:2002vb, Anchordoqui:2005pn, PalomaresRuiz:2005xw, Anchordoqui:2006ta, Hussain:2006wg, Borriello:2007cs, Hussain:2007ba, Barger:2013pla, Marfatia:2015hva, Ellis:2016dgb, Dev:2016uxj, IceCube:2017roe, Bustamante:2017xuy, Bertone:2018dse, Anchordoqui:2019ufu, IceCube:2018pgc, Zhou:2019vxt, Zhou:2019frk, Klein:2019nbu, Mack:2019bps, IceCube:2020rnc, Denton:2020jft, Huang:2021mki, Valera:2022ylt, Esteban:2022uuw}
How do flavors mix at high energies?
\cite{Learned:1994wg, Beacom:2003nh, Beacom:2003eu, Pakvasa:2007dc, Esmaili:2009fk, Bustamante:2010nq, Bustamante:2010bf, Mehta:2011qb, Esmaili:2012ac, Arguelles:2015dca, Bustamante:2015waa, Shoemaker:2015qul, Ahn:2016hhq, Gonzalez-Garcia:2016gpq, Rasmussen:2017ert, Ahlers:2018yom, Bustamante:2019sdb, Song:2020nfh}
What are the fundamental symmetries of Nature?
\cite{AmelinoCamelia:1997gz, Colladay:1998fq, Kostelecky:2003cr, GonzalezGarcia:2005xw, Hooper:2005jp, Anchordoqui:2005gj, Anchordoqui:2006wc, Kostelecky:2008ts, Kostelecky:2011gq, Gorham:2012qs, Borriello:2013ala, Diaz:2013wia, Stecker:2014xja, Anchordoqui:2014hua, Stecker:2014oxa, Amelino-Camelia:2015nqa, Tomar:2015fha, Liao:2017yuy, Ellis:2018ogq, Addazi:2021xuf}
Are neutrinos stable?~\cite{Beacom:2002vi, Anchordoqui:2005ey, Baerwald:2012kc, Pagliaroli:2015rca, Shoemaker:2015qul, Bustamante:2016ciw, Denton:2018aml,  Bustamante:2020niz, Abdullahi:2020rge} Is there evidence of dark matter in the flux of high-energy neutrinos?~\cite{Feldstein:2013kka, Esmaili:2013gha, Bai:2013nga, Ema:2013nda, Bhattacharya:2014vwa, Zavala:2014dla, Rott:2014kfa, Esmaili:2014rma, Fong:2014bsa, Murase:2015gea, Ahlers:2015moa, Aisati:2015vma, Anchordoqui:2015lqa, Troitsky:2015cnk, Chianese:2016opp, Chianese:2016kpu, Arguelles:2017atb, Bhattacharya:2017jaw, Chianese:2017nwe, Hiroshima:2017hmy, Aartsen:2018mxl,  Sui:2018bbh,  Bhattacharya:2019ucd, Arguelles:2019boy, Chianese:2019kyl, Aartsen:2020tdl}
Are there hidden interactions with cosmic backgrounds?
\cite{Talk:Weiler_TeVPA2006, Hooper:2007jr, Lykken:2007kp, Anchordoqui:2007iw, Ando:2009ts, Barranco:2010xt, Miranda:2013wla, Ioka:2014kca, Ng:2014pca, Ibe:2014pja, Blum:2014ewa, DiFranzo:2015qea, Shoemaker:2015qul, deSalas:2016svi, Cherry:2016jol, Reynoso:2016hjr, Arguelles:2017atb, Klop:2017dim, Bustamante:2018mzu, Kelly:2018tyg, Farzan:2018pnk, Alvey:2019jzx, Barenboim:2019tux, Choi:2019ixb, Koren:2019wwi, Murase:2019xqi, Bustamante:2020mep, Creque-Sarbinowski:2020qhz, Esteban:2021tub}
What is the origin of the anomalous EeV events seen by ANITA?~\cite{ANITA:2016vrp, Cherry:2018rxj, ANITA:2018sgj, Anchordoqui:2018ucj, Huang:2018als, Collins:2018jpg, Chauhan:2018lnq, Anchordoqui:2018ssd, Heurtier:2019git, deVries:2019gzs, Hooper:2019ytr, Cline:2019snp, Shoemaker:2019xlt, Esteban:2019hcm, Heurtier:2019rkz, Chipman:2019vjm, Borah:2019ciw, Abdullah:2019ofw, Esmaili:2019pcy}
Neutrino telescopes may also probe new neutrino interactions and hypothesized particles, such as magnetic monopoles~\cite{Dirac:1931kp, Abbasi:2012eda, Aartsen:2014awd, Aartsen:2015exf, Albert:2017fud}, by looking for exotic signatures~\cite{Ahlers:2018mkf}.
\newline
\newline
\noindent
{\bf\color{header_color}Synergies with sub-TeV neutrino experiments:}
Upcoming sub-TeV experiments will reduce uncertainties that limit the sensitivity of fundamental-physics searches using high-energy cosmic neutrinos.
First, the IceCube Upgrade~\cite{Ishihara:2019aao}, though geared towards low-energy neutrinos, will reduce systematic errors that affect the detection of high-energy neutrinos.  Second, oscillation experiments---DUNE\ \cite{DUNE:2020jqi}, JUNO\ \cite{Wu:2019maq}, Hyper-Kamiokande\ \cite{Walker:2019cxi}, the IceCube Upgrade\ \cite{Ishihara:2019aao}, KM3NeT/ORCA\ \cite{KM3Net:2016zxf}---will reduce the uncertainties on lepton mixing parameters\ \cite{Ellis:2020hus}, permitting more precise tests of new physics via the flavor composition of high-energy cosmic neutrinos \cite{Bustamante:2015waa, Song:2020nfh}.  Third, FASER$\nu$ will reduce systematic uncertainties in charm production at parton momentum fraction $x$ close to $1$ by measuring the high-energy neutrino flux from the LHC in the forward direction~\cite{Abreu:2019yak, FASER:2020gpr,Anchordoqui:2021ghd,Feng:2022inv}.  This will improve predictions of the undiscovered background of prompt atmospheric neutrinos that muddle searches of new physics with high-energy cosmic neutrinos.

Given the unique potential and rich experimental outlook of high-energy cosmic neutrinos to extend our view of fundamental physics,
this topic should feature prominently in the high-energy physics program, as it pushes the boundaries for the neutrino, cosmic, energy, theory, and instrumentation frontiers. Companion Snowmass Whitepapers study in detail dark matter physics~\cite{Carney:2022gse,Aramaki:2022zpw,Berti:2022rwn,Chakrabarti:2022cbu}, Beyond-the-Standard Model physics~\cite{Arguelles:2022xxa}, the Forward Physics Facility~\cite{Anchordoqui:2021ghd}, prompt neutrinos facilities~\cite{Bai:2022jcs}, tau neutrinos~\cite{Abraham:2022jse}, multi-messenger facilities and capabilities~\cite{Coleman:2022abf, Engel:2022bgx, Berti:2022wzk, Engel:2022yig}. 


\subsubsection{Cross sections}\label{subsec:future_crosssection}

The last decade yielded the first measurements of the deep-inelastic neutrino nucleon interaction (DIS) 
cross section in the HE range. Future experiments, both at accelerators and with cosmic neutrinos, aim to significantly increase event rates and study neutrino interactions in a new energy region. 
In particular,
both astrophysical and cosmogenic neutrinos provide
neutrino beams at energies far beyond the reach of human-made accelerators,
which allows us to explore particle physics in the phase space 
we could not access to otherwise. 

Improvements in cross section measurements by neutrino telescopes at the HE scale will largely be due to increased statistics, as future optical experiments increase their sensitivities and exposures. This means that they will not expand their energy reach by a large amount. At the lower energy threshold, systematic uncertainties will dominate how well small absorption features can be resolved and at the higher energy the cross section measurement will be limited by the low cosmic neutrino flux. 

To access the higher energies, UHE experiments plan to instrument much larger volumes. Indeed, detector volumes of 100 km$^3$ of Antarctic ice are required to detect a large enough sample of UHE neutrino interactions to probe PDFs with Bjorken$-x$ below $10^{-4}$. Other experiments that rely on Earth-skimming tau neutrinos have larger effective volumes and are sensitive to even higher energy neutrinos, in part through neutrino flux attenuation in the Earth. Much progress has been made in recent years in the modeling of neutrino propagation in the Earth \cite{Niess:2018opy,NuSpaceSim:2021hgs,Safa:2021ghs,Garcia:2020jwr}. These simulation codes will be useful tools to test the standard model and BSM neutrino cross sections inputs.

Beyond DIS cross sections,
the first detection of $W$-boson production via the Glashow resonance  \cite{IceCube:2021rpz} is promising. Improved experimental sensitivity in the PeV band with IceCube-Gen2 will further enhance the science gains in this regard. The $\bar\nu_e+e$ Glashow resonant production of real $W$-bosons mainly yields showers in the high-energy neutrino detectors, so Glashow resonant events could be detected via the shower spectrum. Ref.~\cite{Zhou:2019frk} shows that, on top of the shower events from neutrino deep-inelastic scattering with nuclei, $W$-boson production can be identified with $1\sigma$ and $3\sigma$ in the conservative and optimistic cases, respectively, with ten years of current IceCube data or with one year of IceCube-Gen2 data.

Through higher-order electroweak processes \cite{Seckel:1997kk, Alikhanov:2015kla, Garcia:2020jwr,Zhou:2019vxt, Zhou:2019frk},
$\nu_\mu$ induced $W$-boson production can produce dimuon events~\cite{Zhou:2019frk, Zhou:2021xuh}. One muon comes from the leptonic vertex while the other from the $W$ muonic decay. 
In Ref.~\cite{Zhou:2021xuh}, the authors proposed that dimuons can be detected using the small vertical spacing between the digital optical modules of high-energy neutrino telescopes like IceCube and IceCube-Gen2.

Another source of dimuon events are neutrino charged-current scattering with nuclei that produces a charm hadron. One muon comes from the charged-current scattering, and the second muon comes from the semi-leptonic charm hadron decay. This source of dimuons (DIS dimuons) typically has associated showers.  With realistic shower cuts, the DIS dimuons produced in the detector can be almost completely removed, leaving the yield of dimuons from $\nu_\mu$-induced $W$-bosons produced in the detector largely background free  \cite{Zhou:2021xuh}.
IceCube-Gen2 could detect $>6$  such $W$-boson dimuons in ten years \cite{Zhou:2021xuh}, and other signatures can be used for detection $\nu$ induced $W$-boson production, including pure electromagnetic showers and showerless single tracks ~\cite{Zhou:2019frk}.
DIS dimuons event analyses have the potential to measure the strange-quark PDF. In the optimistic case, it is predicted that IceCube can detect $\simeq 130$ DIS dimuons and IceCube-Gen2 can detect $\simeq 620$ DIS dimuons, most from atmospheric neutrinos \cite{Zhou:2021xuh} (see also Ref.~\cite{Sun:2022lti}).

UHE measurements of the cross section can constrain several proposed models of BSM physics involving the virtual exchange
of the Kaluza-Klein graviton~\cite{Jain:2000pu}, microscopic black hole production~\cite{Anchordoqui:2002vb},
or neutrino-induced sphaleron transition~\cite{Ellis:2016dgb}.  At these energies, again the neutrino-nucleon cross section can be measured by exploiting the attenuation of the flux as neutrinos cross the Earth, particularly because most events occur near the horizon. These experiments can constrain the cross section using the angular distribution of the events if the UHE flux is high, such that more than one event per year is detected~\cite{Connolly:2011vc, Denton:2020jft, Huang:2021mki, Valera:2022ylt, Esteban:2022uuw}. Since the event rate of UHE neutrinos by large neutrino telescopes
is essentially a product of neutrino flux and the neutrino-nucleon interaction
cross section, the upper limit of UHE neutrino fluxes are turned into
constraints on these BSM physics, assuming the cosmogenic neutrino flux
from proton-dominated UHECRs~\cite{Yoshida:2010kp}. 
Any new interactions enhancing the cross section
by more than 100 are disfavored under this flux assumption~\cite{Connolly:2011vc}. 


\subsubsection{Inelasticity}\label{subsub:future_inelasticity}

Future HE neutrino detectors, particularly optical Cherenkov detectors, are expected to enhance their flux sensitivity -- and therefore statistics -- by factors of two to four over the current measurements~\cite{IceCube:2018pgc, IceCube-Gen2:2020qha}. This will improve precision substantially in that energy range. 

At UHE energies, inelasticity measurements are yet to come, but may be possible with future radio-detection experiments~\cite{Klein:2019nbu}. Detectors must be able to detect and separate outgoing leptons and the hadronic cascade produced by neutrino interactions. At energies above about $10^{19-20}$ eV, the Landau-Pomeranchuk-Migdal (LPM) effect lengthens electromagnetic showers, giving them a longer distance scale than hadronic showers \cite{Gerhardt:2010bj}.  At still higher energies, the LPM effect will break electromagnetic showers into multiple subshowers, from a single charge-current induced electron. It may be possible to separate the hadronic and electromagnetic cascades on the basis of this length scale, which leads to different radio-emission spectra, allowing for inelasticity measurements in $\nu_e$ charged current interactions.   Alternately, if multiple subshowers are detected as separate showers, this would also allow for inelasticity measurements.  This measurement is most straightforward at extremely high energies ($10^{20}$~eV), promising recent results suggest that hadronic and electromagnetic cascades may be separated on the basis of their different radio-emission spectra ~\cite{Stjarnholm:2021xpj, Alvarez-Muniz:1999qlp}. 


\subsubsection{Neutrino decay}\label{subsub:future_neutrinodecay}

Neutrino decay constraints in the future may evolve with measurements of the diffuse supernova neutrino background, cosmological measurements, and possibly with a measurement of the cosmic neutrino background.
At shorter lifetimes, high-energy astrophysical neutrino constraints will continue to improve.
These probes also have a benefit over the others in that it is easier to discover neutrino decay given astrophysical uncertainties due to the fact that all three flavors can be differentiated at high energies but not at low energies.
The best improvement will come from IceCube augmented by KM3NeT and Baikal-GVD.
Experiments sensitive to higher energy neutrinos are less sensitive to neutrino decay since the neutrinos are more boosted and experience less proper time.
The increased statistics of events collected by IceCube-Gen2 will help, as well as the presumably increased $\nu_\tau$ data set which is a key data set for neutrino decay.


\subsubsection{Dark matter} \label{sec:dm_future}

IceCube and ANTARES have provided the principal constraints on DM annihilation to neutrinos for DM with a mass larger than a TeV. Future experiments will boost the sensitivity of the neutrino telescope in the search for very heavy dark matter. The sensitivity of the future experiments to probe very-heavy DM are shown in Fig. \ref{fig:nuDM_annihilation}. For masses between 1--100 TeV, KM3NeT will provide the leading constraints on the annihilation cross section, closing in on the parameter space near the thermal relic abundance. Above 1 PeV, improved constraints will be provided by IceCube-Gen2, radio arrays like RNO-G and GRAND, and other planned experiments. Experiments operating in the mid-range between 1 PeV to 1 EeV, such as TAMBO, Trinity, or RET-N, can provide a crucial overlap between the PeV scale and the EeV scale experiments.  
Beyond the EeV scale, POEMMA's fluorescence observation mode could probe DM masses beyond $10^{11}$ GeV \cite{Guepin:2021ljb}. 

Interactions of ultra-light dark matter with neutrinos can also impact the observed flavor ratios.  The nearly even expectation for the flavor ratio at the Earth is very robust against varying the condition at the source and even invoking new physics \cite{Bustamante:2015waa}.  In fact, Lorentz symmetry implies that neutrinos arrive at the Earth as incoherent combinations of different neutrino mass eigenstates. However, as shown in \cite{Arguelles:2015dca,Arguelles:2019rbn,deSalas:2016svi}, in the presence of Lorentz violation, the standard expectation of even flavor ratios can be relaxed because the flavor state arriving at the Earth can remain as a coherent combination of the mass eigenstates. In the background of ultralight dark matter, $\phi$, neutrinos can obtain a Lorentz violating effective mass of form $(m_{eff})_{\alpha \beta}\nu_\alpha^\dagger \nu_\beta$ originating from vector vector interaction of the following form with ultra light complex dark matter denoted by $\phi$ \cite{Farzan:2018pnk,Farzan:2021gbx,Farzan:2021slf}
\begin{equation}\frac{g_{\alpha\beta}}{\Lambda^2}(\phi^*\partial_\mu\phi-\phi \partial_\mu \phi^*)(\bar{\nu}_\alpha \gamma^\mu \nu_\beta).\label{vec-vec}\end{equation}
Then, $(m_{eff})_{\alpha \beta}=(\rho_{DM}/m_{DM})(g_{\alpha\beta}/\Lambda^2)$. In the halo of dark matter where the dark matter density, $\rho_{DM}$, is relatively large,
$(m_{eff})_{\alpha \beta}$ may dominate over $(\Delta m^2/E_\nu)$. As a result, the energy eigenstates will correspond to the eigenvectors of $(m_{eff})_{\alpha \beta}$ rather than the mass eigenstates in vacuum. 
Since the DM density variation across the route of neutrinos is smooth, the transition will be adiabatic. 

As shown in \cite{Farzan:2018pnk}, interaction of form (\ref{vec-vec}) can be obtained in a UV complete model based on gauging the $L_\mu-L_\tau$ symmetry. 
In this case, $g_{\alpha \beta}$ will be diagonal with a remarkable result that the original flavor ratio at the source located in a dark matter halo will be maintained up to the Earth. That is, if at the source $F_{\nu_\tau}^0\ll F_{\nu_\mu}^0, F_{\nu_e}^0$, the $\nu_\tau$ flux at the Earth will remain negligible. The observation of two $\nu_\tau$ events by IceCube \cite{IceCube:2020abv} already puts an upper bound on $g_\alpha/\Lambda^2$. However, as discussed in \cite{Farzan:2021gbx},  it is  still possible that higher energy neutrinos ($E_\nu>10$~PeV) originating from sources located inside DM halos maintain their original flavor ratio. 
This implies that  while detectors designed to detect air showers from Earth-skimming $\nu_\tau$ may report null results, the radio telescopes designed to detect the Askaryan radiation from all neutrino flavors may report a significant flux \cite{Farzan:2021gbx}. Cosmogenic neutrinos, having originated typically outside DM halos will still have the canonical democratic flavor ratios \cite{Farzan:2021gbx}.


\subsubsection{Secret neutrino interactions}

Neutrinos may interact with the cosmic neutrino background through non-standard neutrino interactions~\cite{Kolb:1987qy, Ioka:2014kca,Ng:2014pca,Blum:2014ewa}.
Upcoming experiments will allow for precision measurements of the energy spectrum of high-energy astrophysical neutrinos.  The most salient feature imprinted by secret interactions on the high-energy neutrino spectrum are the dips around the resonance energy of the neutrino-neutrino cross section.  Thus, previous searches and forecasts have centered predominantly on looking for evidence of spectral dips.  

There are two main limitations to these searches; future detectors may try to improve on them.  First, in optical Cherenkov neutrino telescopes, the energy $E$ of a neutrino-initiated event is typically reconstructed to within, roughly, 10\% in $\log_{10}(E/{\rm GeV})$.  The limitation stems from uncertainties in the medium properties, in the position of the interaction vertex, and other detector characteristics.  The detector energy resolution sets the size of the smallest dip feature that we can expect to detect: if the energy width of the dip is smaller than the energy resolution, the dip will not be detected.  Second, when looking for the presence of the dips in the energy spectrum---or other unusual features---searches use samples of contained events, where the neutrino interaction occurs inside the detector volume.  In contained events, most or all of the neutrino energy can be inferred from the event energy, and so the shape of the energy spectrum can be inferred with relatively little bias.  Unfortunately, contained events are rare; after ten years, IceCube has observed only approximately one hundred of them.  Future optical detectors, with better knowledge of the properties of the detector medium, and of larger size, should overcome both of the above limitations.  

Further, the couplings to the new mediator need not be flavor-universal: different neutrino flavors experience secret interactions with different strengths. This expands the available parameter space where effects may be present and require searches.  Specifically, couplings to $\nu_\tau$ are the least constrained.  By including the capability to identify neutrino flavor, future searches could start probing flavor-dependent couplings.  However, when doing so, the significant uncertainties in the flavor composition with which neutrinos are produced need to be factored in, and will weaken the probing power. Refs.~\cite{Shoemaker:2015qul,Esteban:2021tub} explored future prospects in detail, highlighting the capability of IceCube-Gen2 to improve constraints on $\nu_\tau$-coupled mediators by over two orders of magnitude with ten years of data collection, for mediator masses up to 100~MeV.  Ultra-high-energy neutrino telescopes could extend the range to tens of GeV~\cite{POEMMA:2020ykm}.  

Additionally, measuring neutrinos from identified sources with multi-messenger observations will allow for further constraints on secret neutrino interactions by constraining their mean free path and any possible neutrino echoes or halos~\cite{Kelly:2018tyg,Murase:2019xqi}.


\subsubsection{Prompt neutrinos}

The extension of IceCube, the proposed IceCube-Gen2, will be able to enhance the detectability of prompt atmospheric neutrinos with higher statistics~\cite{IceCube-Gen2:2020qha}. It will additionally provide a unique laboratory to study PeV prompt muons as produced in air showers, which are strongly linked to the production of prompt neutrinos using an approach that is complementary to accelerator studies of prompt neutrinos~\cite{IceCube-Gen2:2021aek}. 
In the meantime, upcoming forward experiments at the LHC, FASER$\nu$ \cite{FASER:2020gpr} and SND@LHC \cite{SHiP:2020sos}, will measure the prompt neutrinos and reduce the uncertainties in heavy flavor production.
In addition, further developing experiments with upgrades of FASER$\nu$ and SND@LHC, and additional experiments for the next stage, the so-called Forward Physics Facility (FPF) are under study \cite{Anchordoqui:2021ghd,Feng:2022inv}.
The FPF will be designed to cover rapidity ranges beyond those probed by LHCb and by forward experiments deployed during Run 3 of the LHC. The FPF will play a crucial role through measurements of high energy neutrinos from the decays of the charm mesons produced in $pp$ collisions at the LHC at $\sqrt{s}=14$ TeV. These FPF measurements will then be related to improved theoretical predictions of prompt atmospheric neutrino fluxes
\cite{Feng:2022inv}. Likewise, measurements of the prompt atmospheric neutrino flux may lead to a better understanding of structure of the proton through the parton distribution functions and, for example, the role of intrinsic charm~\cite{Halzen:2016thi,Laha:2016dri,Giannini:2018utr,Bhattacharya:2018tbc,Goncalves:2021yvw}.


\subsubsection{Fundamental symmetries}

\paragraph{$\mathbf{U(1)^\prime}$ extensions to the Standard Model}

\begin{figure*}[t!]
	\centering
	\includegraphics[width=0.33\linewidth]{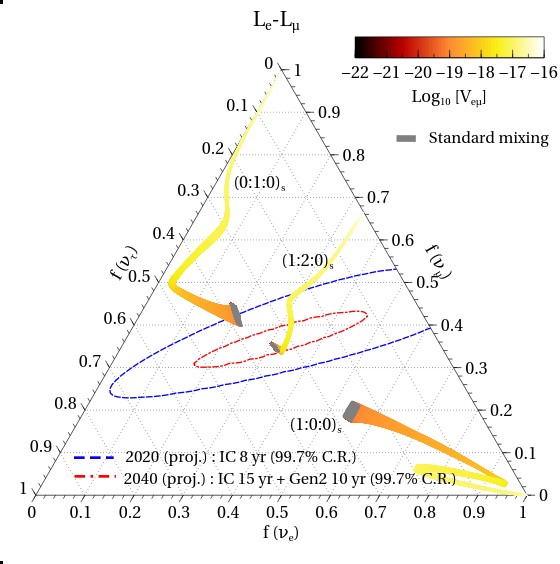}\hfil
	\includegraphics[width=0.33\linewidth]{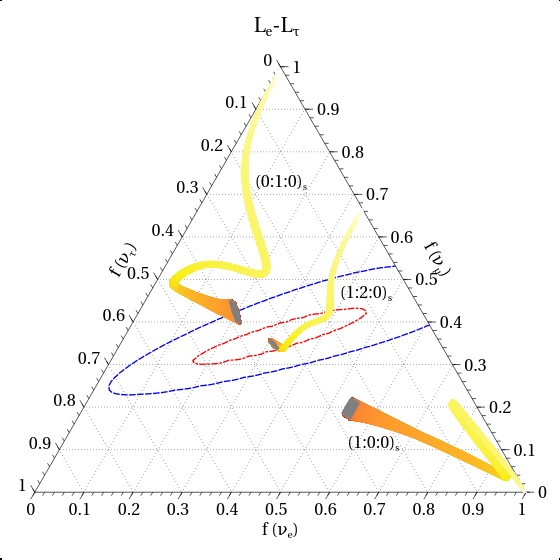}\hfil
	\includegraphics[width=0.33\linewidth]{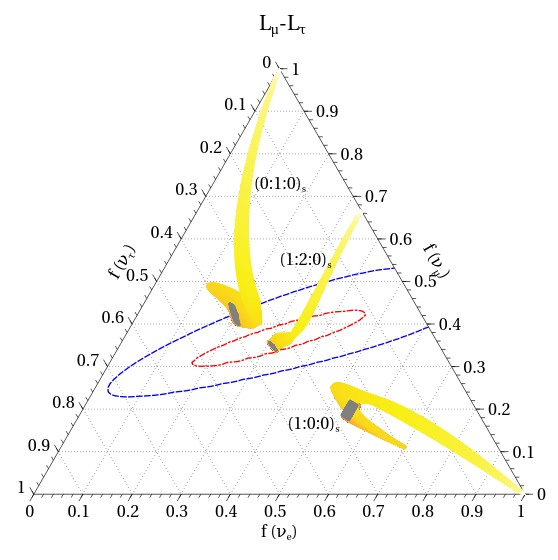}
	\caption{Neutrino flavor ratio at Earth as a function of long-range potential $V_{\alpha\beta}$ associated with $L_e-L_\mu$ ({\it left}), $L_e-L_\tau$ ({\it middle}), and $L_\mu-L_\tau$ ({\it right}) for the three choices of the flavor ratio at the sources ($f_{e,S}$:$f_{\mu,S}$:$f_{\tau,S}$) =($\frac{1}{3}$:$\frac{2}{3}$:0), (0:1:0), and (1:0:0). Equal contribution from neutrino and antineutrino is considered.  All the oscillation parameters are varied in $1\sigma$ allowed range under the normal mass ordering (NMO) scenario as given in Ref.~\cite{deSalas:2017kay}. Neutrino energy is fixed at $E_{\nu}=100$ TeV. Dashed blue and dot-dashed red contours respectively show projected IceCube flavor ratios for 8 years runtime and 35 years runtime (15 years IceCube and 10 years IceCube +Gen2). Left and middle ternary diagrams are adapted from Ref.~\cite{Bustamante:2018mzu}.}
	\label{Fig-1}
\end{figure*}

By measuring the flavor composition of astrophysical neutrinos, next generation flavor-sensitive experiments will significantly improve our sensitivity to possible $U(1)^\prime$ extensions in the Standard Model. 
Fig.~\ref{Fig-1}, shows the expected neutrino flavor composition at Earth in the SM and in the presence of $L_{\alpha}-L_{\beta}$ interactions~\cite{Bustamante:2018mzu}. Dashed-blue and dot-dashed-red contours, respectively, show projected IceCube-measured flavor ratios for an 8-year runtime and a 35-year runtime (15 years IceCube and 10 years IceCube +Gen2). Three benchmark scenarios for flux compositions at the source, namely, full pion decay (1:2:0),  muon damping (0:1:0), and neutron decay (1:0:0) are considered. The respective flavor composition at Earth is shown by grey blobs in the standard mixing scenario and by yellow-orange strips in the presence of $L_e-L_{\beta}$ long-range force. For all the assumed flavor ratios at source, the flavor ratio at Earth deviates from SM predictions. For ($V_{\alpha \beta}>>(H_{vac})_{ee}$), the original flavor ratio is preserved and not modified during propagation. Therefore, using the IceCube flavor ratio data, it is possible to constrain the mass and the coupling of the new interactions associated with $L_\alpha-L_\beta$ symmetries. With upcoming experiments like IceCube-Gen2 and the IceCube Upgrade, the flavor composition will be measured with high precision, which will allow for tighter constraints on the coupling and mediator mass of these new interactions. 


\subsection{In astrophysics}
\label{subsec:future_astro}


\subsubsection{Diffuse flux of TeV--100 PeV astrophysical neutrinos}

IceCube has measured, for the first time, the spectrum and flavor composition of cosmic neutrinos over three orders of magnitude in energy (10~TeV -- 10~PeV). The apparent isotropic distribution of the observed neutrinos on the sky points to an extragalactic origin. This is also corroborated by the first observational evidence for a neutrino source, the blazar TXS~0506+056. 

Over the next decade, IceCube's measurement will be complemented by new neutrino observatories currently under construction in the Northern Hemisphere: the KM3NeT (~\cite{KM3Net:2016zxf}, see also Sec. \ref{subsec:km3net}) neutrino telescope in the Mediterranean Sea, and the Gigaton Volume Detector (GVD) (\cite{Allakhverdyan:2021}, see also Sec. \ref{subsec:baikal}) in Lake Baikal. The sky coverage of these telescopes is complimentary to that of IceCube, due to their geographic locations. Comparable in size and sensitivity to the latter, they will enhance the precision of the current measurements, and, more importantly, should be able to probe the existence of a Galactic component in the diffuse neutrino flux. The observation of such a diffuse Galactic component, originating from the interactions of TeV-PeV cosmic rays with interstellar gas, and the measurement of its spectrum would be a unique input for understanding the production and propagation of the highest-energy cosmic rays in the Milky Way~(e.g., \cite{Gabici:2019jvz}).   
Joint spectral and flavor measurements combining complementary data from all operating neutrino telescopes will be feasible, and may even be necessary to solve the apparent slight tensions currently seen in IceCube data.

\begin{figure*}[t]
  \centering
  \includegraphics[width=0.7\textwidth]{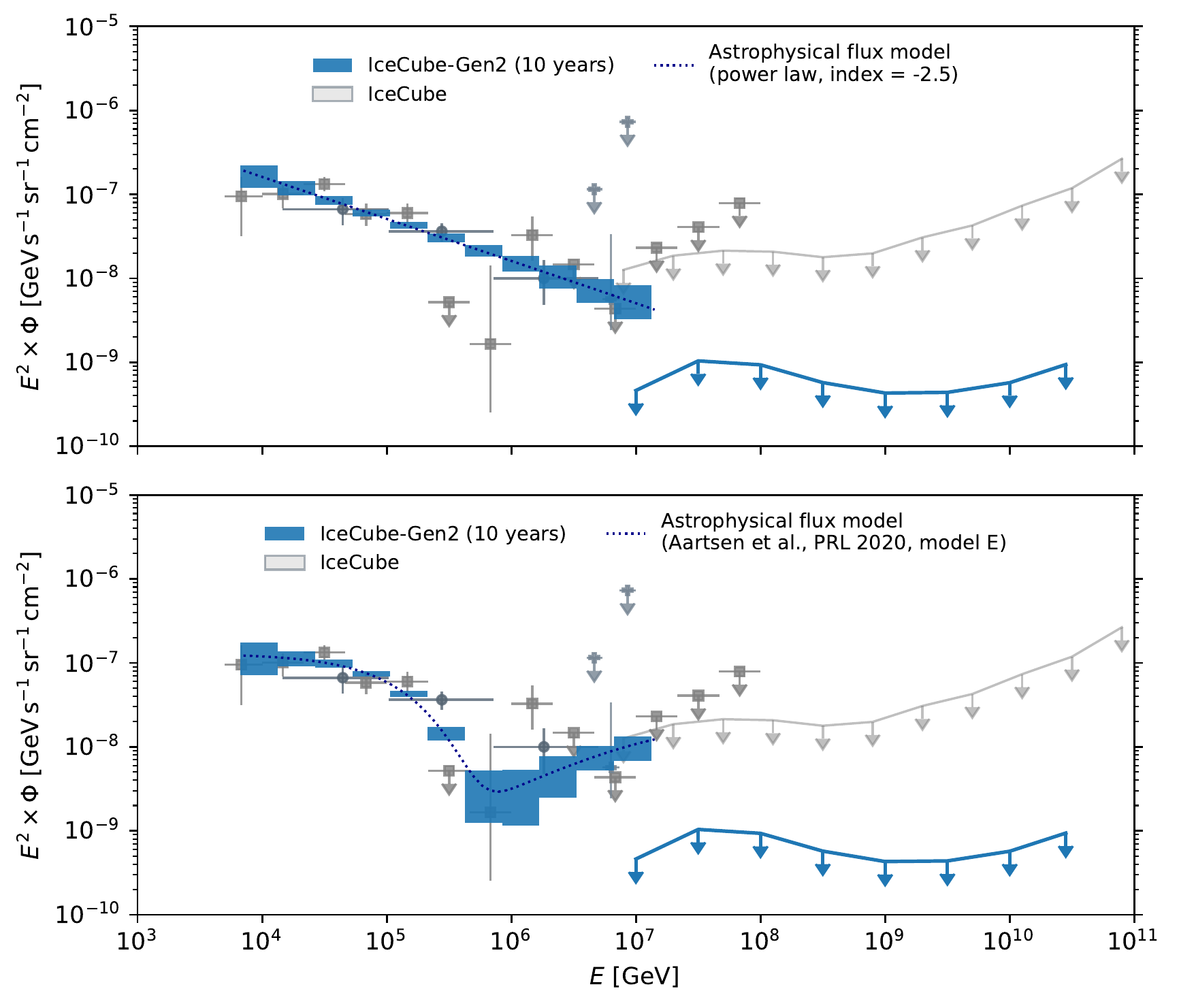}
  \caption{A comparison of the IceCube measurements of the diffuse astrophysical neutrino flux with projections for the future IceCube-Gen2 detector. The IceCube measurements shown are from high-energy cascades~(\cite{IceCube:2020acn}, squares), tracks~(\cite{IceCube:2021uhz}, circles), a Glashow resonance event~(\cite{IceCube:2021rpz}, +), and a search for EHE events~(\cite{IceCube:2018fhm}, line with arrows). The predicted unfolded diffuse spectrum assuming 10 years of IceCube-Gen2 data is shown in blue with error bands representing 68\% confidence intervals up to 10\,PeV. The upper panel assumes a single power law model with a spectral index of -2.5. The lower panel assumes a two component spectrum, consisting of a power law with an exponential cutoff at approximately 100~TeV, and a component reflecting a model of the $p\gamma$ emission of BLLac which peaks at tens of PeV. Both models are based on fits to the cascade data~\cite{IceCube:2020acn}. With current data from IceCube, it is not statistically significant to reject either hypothesis. IceCube-Gen2 will provide key measurements to reveal spectral features in the astrophysical neutrino component. Above 10\,PeV, the expected differential 90\%~C.L. sensitivities for the combined optical and radio instrumentation for IceCube-Gen2 are shown in solid blue line.}
  \label{fig:unfolded_spectrum_astro}\small 
\end{figure*}

A major step forward in measurement precision for spectrum and flavor composition at TeV and PeV energies will arise from the proposed IceCube-Gen2 observatory (\cite{IceCube-Gen2:2020qha}, see also section \ref{subsec:gen2}).
Figure~\ref{fig:unfolded_spectrum_astro} shows a comparison of current IceCube measurements and projections for 10 years of IceCube-Gen2 data. The IceCube-Gen2 projections \cite{IceCube-Gen2:2020qha}, with an updated astrophysical flux model and improved characterizations of the surface and radio array efficiencies, give an impression of the expected unprecedented precision and energy coverage of the measurement of the diffuse neutrino spectrum.
Together with significantly stronger constraints of the flavor composition of the cosmic neutrinos (cf. Fig.~\ref{Fig-1}), these data can be used to study the properties of the source populations that dominate the cosmic neutrino flux. Spectral features allow for study of cosmic-ray interaction processes and maximum acceleration energies. Investigations of the resonant production of neutrinos at the Glashow resonance~\cite{Glashow:1960zz} can be used to study the $\nu$/$\bar{\nu}$ ratio of cosmic neutrinos, and derive constraints on the production environments and targets~\cite{Biehl:2016psj}. Differential measurements of the flavor composition in various energy ranges can probe magnetic field conditions in the neutrino production regions around the sources~\cite{IceCube-Gen2:2020qha}. 
For flavor measurements, tau neutrino identification capabilities are of particular importance. See Ref.~\cite{Abraham:2022jse} for a detailed discussion of tau neutrino identification prospects with IceCube-Gen2, KM3NeT, and Baikal-GVD.


\subsubsection{Diffuse flux of ultra-high-energy neutrinos: cosmogenic and from sources}

The flux of cosmogenic neutrinos is expected to encode complimentary and unique information about the nature and flux of UHECR accelerators, including their redshift evolution, their chemical composition, and maximum accelerating energy.
The cosmogenic neutrino flux measurements can also be used for probing the Galactic-to-extragalatic transition~\cite{Takami:2007pp}.
Interestingly, there is a ``sweet spot"~\cite{vanVliet:2019nse} near 1~EeV where the flux of UHE neutrinos is predicted to depend mostly on two parameters: the redshift evolution of sources and the chemical composition of the cosmic rays, in particular the fraction of protons above $\sim$~40~EeV. For example, a source's redshift evolution is often parameterized as being distributed in $z$ according to $(1+z)^{m}$; IceCube data already disfavors $m\geq 3.5$~\cite{IceCube:2016uab} for purely power law sources, constraining otherwise promising sources such as radio-loud FR-II AGN. 
With regards to composition, there is a long standing discrepancy between two measurements of the UHECR composition, with the Telescope Array experiment favoring a ``light" (proton-dominated) composition~\cite{TelescopeArray:2018xyi}, while the Pierre Auger observatory favors a ``heavy" (iron-rich) composition~\cite{PierreAuger:2016use}. Despite these different interpretations, the results are approximately compatible to each other within uncertainties~\cite{deSouza:2017wgx, Hanlon:2018dvd}. These two scenarios have markedly different signatures in the cosmogenic neutrino flux however, with a light proton-dominated UHECR flux giving rise to a much higher cosmogenic neutrino flux than a heavy iron-rich one. 
The cosmogenic neutrino intensity, therefore, is a completely orthogonal probe of the UHECR composition~\cite{Ahlers:2009rf, vanVliet:2019nse} with different systematic uncertainties.

The IceCube, Auger, and ANITA experiments already constrain the cosmogenic neutrino parameter space, and a major goal for the next generation of observatories is to definitively detect this flux, or improve the constraints by at least two orders of magnitude, reaching a flux sensitivity near $10^{-10}$ GeV~cm$^{-2}$~s$^{-1}$~sr$^{-1}$ at 1~EeV. Such sensitivity will probe even ``pessimistic" cosmogenic models where the UHECR flux is relatively heavy~\cite{Murase:2010gj, AlvesBatista:2018zui}, as suggested by the Pierre Auger Observatory. A cosmic-ray composition of even 10-20\% protons will lead to a few detected neutrinos per year for typical source evolution models~\cite{vanVliet:2019nse}. If the flux is in fact even lighter, as suggested by the Telescope Array, experiments with this level of sensitivity could detect dozens or hundreds of events events per year.

\begin{figure*}[t]
  \centering
  \includegraphics[width=0.9\textwidth]{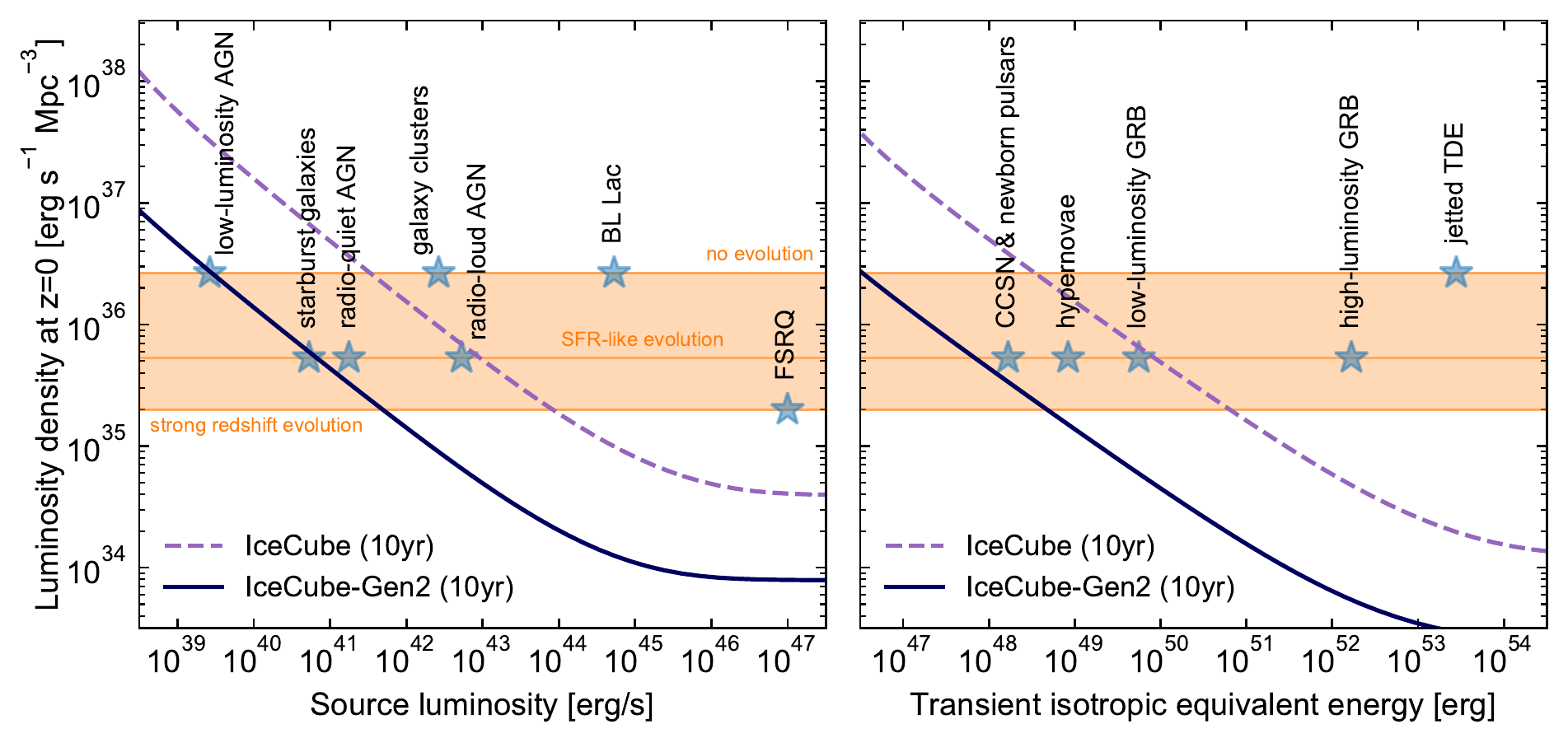}
  \caption{\label{fig:pop_sensitivity_astro}
  {\it Left:} Comparison of the effective local density and luminosity density of extragalactic neutrino source populations to the discovery potential of IceCube and IceCube-Gen2 (optical array only). We indicate several candidate populations by the required neutrino luminosity density to account for the full diffuse flux~\cite{Murase:2016gly} observed by IceCube. The orange band indicates the luminosity / density range that is compatible with the total observed diffuse neutrino flux. The lower (upper) edge of the band assumes rapid (no) redshift evolution. The lines indicate the parameter space (above the lines) for which IceCube and IceCube-Gen2 are able to discover one or more sources of the population. {\it Right:} Same comparison for transient neutrino sources parametrized by their local rate density~\cite{Murase:2018utn}. \label{fig:pointGen2}}
\end{figure*}


\subsubsection{Neutrino point sources}

IceCube has found the first compelling evidence for an extragalactic neutrino source, and several more hints that point to a rich extragalactic neutrino sky that may be composed of various source populations  contributing to the cosmic neutrino flux (cf. Sec.s \ref{sssec:nusources} and \ref{sssec:mmmodeling}). Ultimately, complementary and more sensitive instruments are needed for revealing the ``dominant'' population for the observed diffuse neutrino flux (among various source populations) and a comprehensive study of the neutrino sky.  

To date, IceCube has found no evidence for Galactic neutrino sources nor, therefore, Galactic cosmic-ray accelerators. Northern Hemisphere observatories (such as the under-construction Baikal-GVD and KM3NeT) will have a much better sensitivity of the inner galactic plane and the Galactic center region, where the non-thermal emission from the Milky Way is brightest and several candidate cosmic-ray accelerators are located, in particular some of the recently established Galactic Pevatrons (sources with observed gamma ray emission at $O$(100~TeV))~\cite{HESS:2016pst,HAWC:2019tcx,Amenomori:2019rjd,2021Natur.594...33C}. The detection of neutrinos is an unambiguous signature of the acceleration of protons or nuclei. Therefore, the combined data of IceCube and the new generation of Northern Hemisphere detectors may finally resolve the puzzle of where the highest-energy cosmic rays in the Milky Way originate. 

An order of magnitude improvement in detector volume, corresponding to a five-fold increase in sensitivity, will be critical for testing the dominant origin of the extragalactic diffuse TeV-PeV neutrinos and investigating the properties of their primary populations with reasonable statistics~\cite{Murase:2016gly}. 
Next-generation detectors such as IceCube-Gen2 will be able to identify the brightest sources and transients of populations currently considered as potential cosmic-ray accelerators. 
This is the case for both cosmic-ray reservoirs that allow the production escape of UHECRs and cosmic-ray accelerators in which cosmic rays are confined and/or depleted~\cite{Murase:2016gly} (see Fig.~\ref{fig:pointGen2}).  
It will also enable high-significance observations of the brightest sources and flares similar to the one observed for TXS+0506+056~\cite{IceCube-Gen2:2020qha}, allowing a precise measurement of the spectral properties of such sources/transients. 

The energy generation rate density of 10-100~TeV neutrinos, indicated by the IceCube shower data~\cite{IceCube:2020acn}, is surprisingly large. The neutrino flux seems to violate the Waxman-Bahcall bound and the corresponding gamma-ray flux violates the IGRB in the transparent limit~\cite{Murase:2015xka}. Revealing the sources in the 10~TeV range is important especially if the diffuse flux consists of multiple source populations. This will also enable investigation of particle acceleration and related plasma processes in dense environments that cannot be directly observed by electromagnetic observations, and successful detection will also be useful for probing new physics beyond the Standard Model. The candidate sources include the vicinity of supermassive black holes of AGN~\cite{Stecker:2005hn,Kalashev:2015cma,Murase:2019vdl} and choked jets~\cite{Murase:2013ffa,Tamborra:2015fzv,Denton:2018tdj,Carpio:2020app}.
For this purpose, neutrino detectors that are sensitive to 10~TeV energies or lower are important. KM3NeT and Baikal-GVD, which has good angular resolutions even for shower topographies, will also help to identify bright sources in the southern sky~\cite{Kheirandish:2021wkm}.  

For the purpose of identifying UHECR sources, it is important to reveal the origin of neutrinos 
$\gtrsim 10-100$~PeV. Here, novel detector concepts that will be able detect neutrino-induced showers in Earth's natural ice shields or in the atmosphere with effective volumes of hundreds to thousands of km$^3$ (see Sec.~\ref{sec:experiments} for an overview) will play key roles. Having such large effective volumes, these detectors will be able to identify astrophysical sources and transients that are directly linked to UHECR production in a wide range of scenarios.

\begin{figure*}[t]
  \centering
  \includegraphics[width=\textwidth]{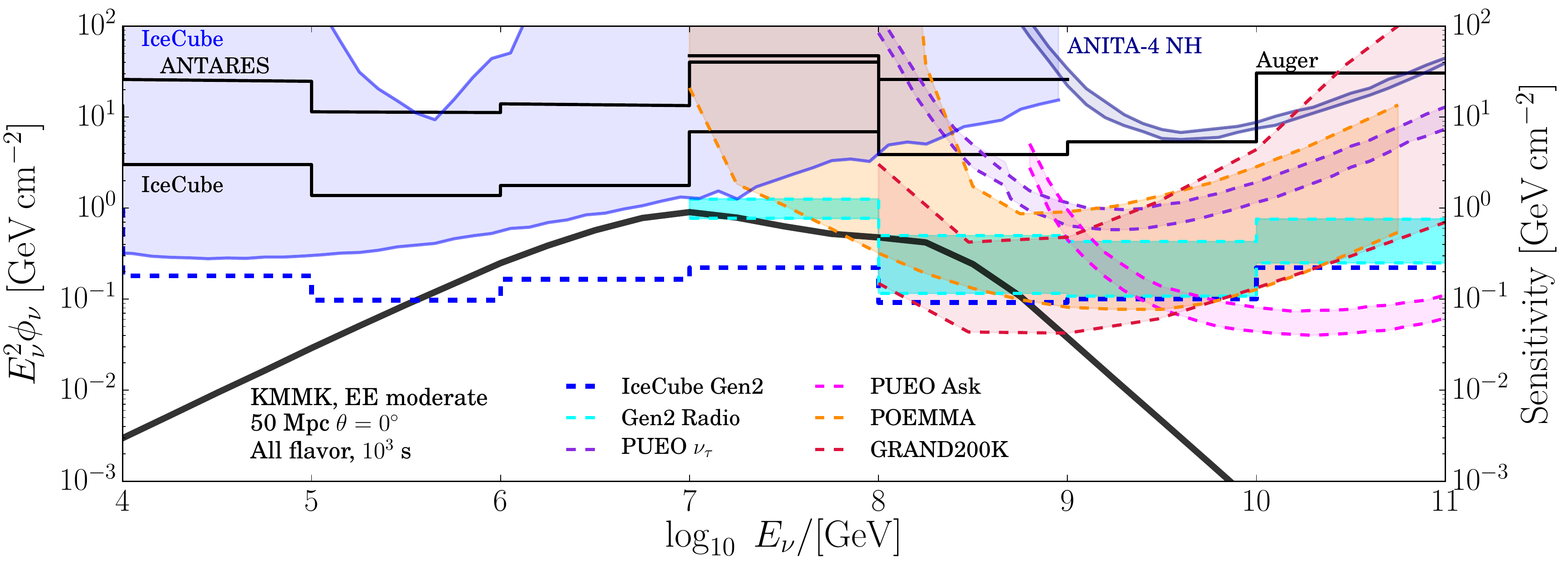}
  \caption{\label{fig:short_burst_nu}
  The sensitivity to short (1000 s) burst all-flavor neutrino plus antineutrino spectral fluence for some current and future detectors. The ANTARES, IceCube and Auger limits are 90\% confidence level limits in a $\pm$ 500 s window around the gravitational wave event from GW170817 
  \cite{ANTARES:2017bia}. The dashed blue histogram shows IceCube-Gen2's projected sensitivity for such an event \cite{IceCube-Gen2:2020qha}, including IceCube-Gen2 radio \cite{IceCube-Gen2:2021rkf} (shown for a range of declination with cyan).   The blue shaded band comes from IceCube's 
   all-sky  point-source  effective  area  values tabulated for 2012 with 86 strings \cite{IceCubePointSource,IceCube:2016tpw}.
  The ANITA-4 NH limit is shown with the solid purple curves, and
  projected all-flavor sensitivities for PUEO \cite{PUEO:2020bnn} from $\nu_\tau$-sourced and Askaryan signals are shown with purple and magenta dashed curves,
  for POEMMA (orange)\cite{Venters:2019xwi} and for GRAND200K (red) \cite{GRAND:2018iaj} updated according to \cite{grandupdate}. The Kimura et al.~\cite{Kimura:2017kan} extended emission short gamma ray burst fluence for on-axis viewing ($\theta=0^\circ$)\cite{Kimura:2017kan} from 50 Mpc is shown with the solid black curve. 
  }
\end{figure*}


\subsubsection{Multi-messenger astronomy} 

The combination of neutrinos with multi-wavelength observations across the electromagnetic spectrum is key for the identification of astrophysical neutrino sources, as well as for the modeling of the physical processes, conditions and environments that enable particle acceleration in such sources (cf. Sec. \ref{sssec:mmmodeling} and Ref.~\cite{Engel:2022yig}).

One of the most important results from the multi-messenger observations is that energy generation densities (or luminosity densities) of high-energy neutrinos, sub-TeV gamma rays, and UHECRs are all comparable~\cite{Murase:2013rfa,Katz:2013ooa,Murase:2018utn}. This may indicate that all three messengers are physically connected. 
The simplest scenario is that all three messengers come from a single source class. While such a grand-unification scenario is appealing~\cite{Fang:2017zjf}, the multi-messenger analysis has indicated other populations, particularly below 100~TeV energies~\cite{Murase:2015xka}. 
Even if the physical connection exists, it may be clouded by different source components and/or different emission regions. Indeed, luminosity densities of various sources including AGN and stellar deaths such as supernovae are not much different~\cite{Murase:2018utn}, and dissipation at multi-scales has been observed in these sources. Detailed information from electromagnetic observations from radio, optical, X-ray, and gamma-ray bands will be crucial, and next-generation gamma-ray detectors such as the Cherenkov Telescope Array will provide critical tests for the unification picture. 

One of the most successful strategies of multi-messenger astronomy has been the real-time distribution of neutrino alerts (cf. Sec. \ref{sssec:mmalerts}). Over the coming decade, the quality and quantity of neutrino alerts will improve significantly thanks to the construction and operation of additional gigaton-scale neutrino telescopes such as KM3NeT, Baikal-GVD, and the recently-proposed P-ONE. Their distribution around the world enables a constant monitoring of the entire sky with good sensitivity in the TeV--PeV range. The distribution mechanisms for these alerts will also benefit from current efforts to upgrade the communication infrastructure between astrophysical facilities to prepare for the massive alert rate expected from Rubin observatory and other time-domain instruments. 

After completion, the proposed IceCube-Gen2 detector will increase the rate of individual high-energy neutrinos by a factor of five that in combination with improved pointing, and more sensitive follow-up instruments will boost the number of observed correlations with candidate sources \cite{IceCube-Gen2:2020qha} by an order of magnitude and enhance their significance. 
For example, Fig.~\ref{fig:short_burst_nu} gives an illustration of the sensitivities to short (1000~s) bursts of neutrinos currently achieved for the gravitational wave event GW170817 from IceCube, ANTARES and Auger  \cite{ANTARES:2017bia}, and short burst sensitivities of IceCube \cite{IceCubePointSource,IceCube:2016tpw}
and ANITA-4 NH, as projected sensitivities from IceCube-Gen2 \cite{IceCube-Gen2:2020qha}, including IceCube-Gen2 radio \cite{IceCube-Gen2:2021rkf}, PUEO \cite{PUEO:2020bnn}, POEMMA \cite{Venters:2019xwi}, and GRAND200K \cite{GRAND:2018iaj}.

The construction of neutrino observatories operating in the extremely-high-energy range (beyond PeV) will also provide a new source of neutrino alerts that is yet largely unexplored, but would be of interest given the very low backgrounds expected in this energy range.


\section{Experimental landscape}
\label{sec:experiments}


\subsection{Overview}

\begin{figure}[t]
    \centering
    \includegraphics[width=\textwidth]{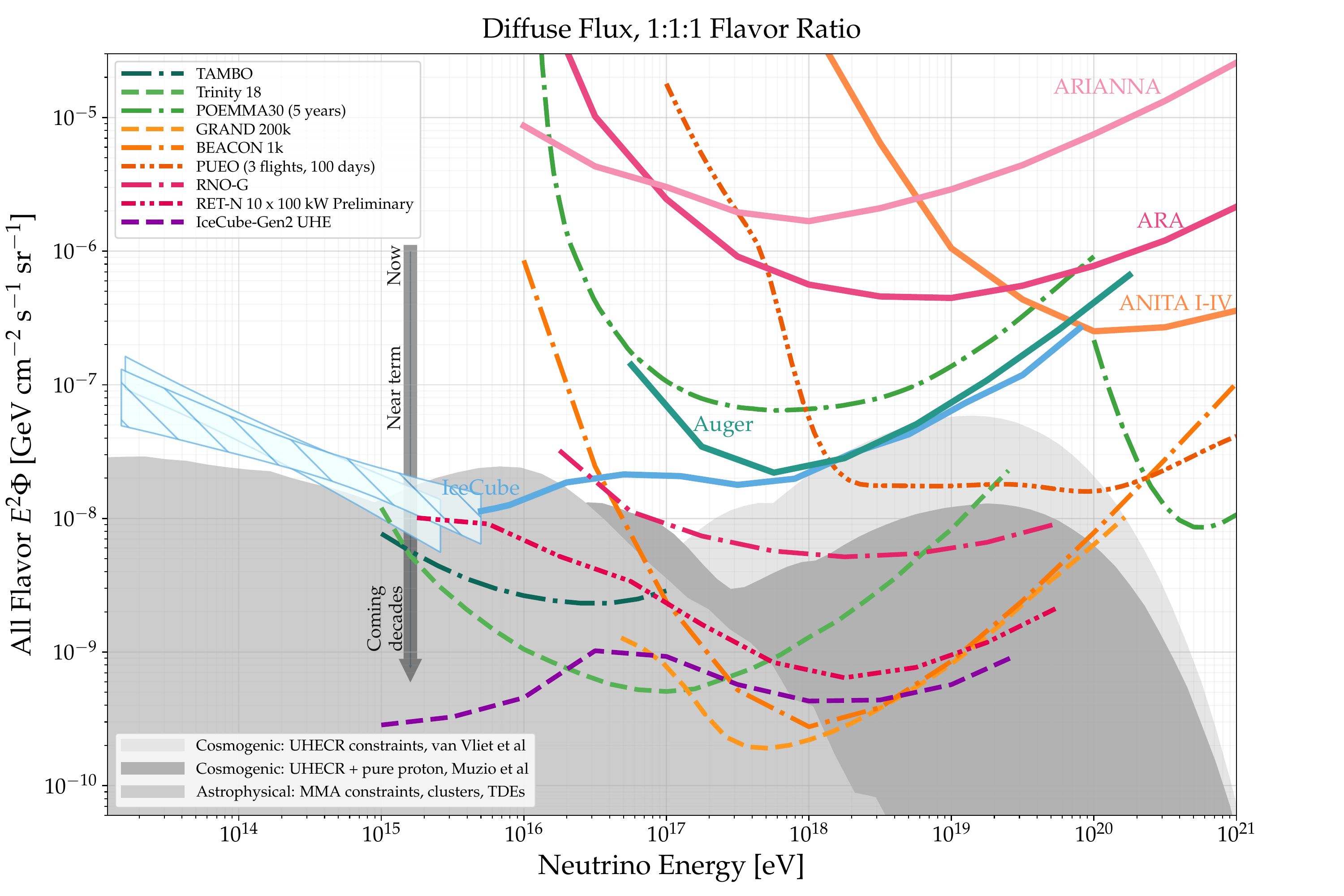}
    \caption{The expected differential 90\%~C.L. sensitivities for a variety of experiments to an all-flavor diffuse neutrino flux computed in decade-wide energy bins and assuming a ten-year integration (unless otherwise noted in the legend). The measurements and sensitivities are compared with a range of cosmogenic neutrino models~\cite{vanVliet:2019nse, Muzio:2021zud} and astrophysical neutrino models~\cite{Muzio:2021zud, Fang:2017zjf, Biehl:2017hnb}.  The blue bordered bands show the astrophysical neutrino flux measured by IceCube using tracks ($\nu_\mu$~\cite{IceCube:2021uhz}) in hatch and using cascade-like events ($\nu_e$ and $\nu_\tau$~\cite{IceCube:2020acn}) in solid band. The solid lines show experimental upper limits at higher energies from the Pierre Auger Observatory~\cite{PierreAuger:2019ens}, ARA~\cite{ARA:2019wcf}, ARIANNA~\cite{Anker:2019rzo}, ANITA I-IV~\cite{Gorham:2019guw}, and IceCube~\cite{IceCube:2016zyt}. The dashed lines show the sensitivities of a selection of proposed experiments currently in various design and prototyping stages (GRAND with 200,000 stations~\cite{GRAND:2018iaj}, BEACON with 1000 stations~\cite{Wissel:2020sec}, TAMBO with 22,000 detectors~\cite{Romero-Wolf:2020pzh}, Trinity with 18 telescopes~(updated from \cite{Brown:2021ane}), RET-N with 10 stations each with a 100~kW transmitter ~\cite{Prohira:2019glh}, POEMMA30~\cite{POEMMA:2020ykm}) and under construction (RNO-G~\cite{RNO-G:2020rmc,RNO-G:2021hfx}, PUEO~\cite{PUEO:2020bnn}). Experiments using the same detection technique are grouped into similar colors (orange, Earth-skimming radio (GRAND, BEACON); dark teal, particle showers (TAMBO); light green, Earth-skimming optical Cherenkov and fluorescence (Trinity, POEMMA30); pink, in-ice radio (ARIANNA, ARA, PUEO, RNO-G, RET-N (radar); blue, in-ice optical Cherenkov (IceCube)). Sensitivity to UHE neutrinos from IceCube-Gen2 (dashed purple) is computed using radio and optical for 10 years (see Fig.~\ref{fig:unfolded_spectrum_astro}), and PUEO (dashed orange) uses both in-ice and Earth-skimming radio techniques (shown here for 3 flights, 100 days). Auger (teal) uses particle showers and fluorescence and its upgrade, AugerPrime, will employ radio.  
    The grey downward-pointing arrow is a reminder that experimental sensitivities improve not only as exposure increases with time, but also as new experimental techniques and analysis methods are both demonstrated and scaled to larger detection volumes. }
    \label{fig:diffuse_sensitivity}
\end{figure}

In this section, we give a brief overview of the experimental landscape, neutrino detection strategies at the highest energies, and current and planned experiments.


\subsubsection{The broad experimental landscape}

Because the neutrino-nucleon cross section grows with energy, the Earth is opaque to high-energy neutrinos, reducing the angular coverage for experiments to well below $4\pi$.
Charged-current interactions (CC, $\nu_l + N \to l + X$, where $l = e, \mu, \tau$ and $X$ are final-state hadrons) remove neutrinos from the flux.  Neutral-current interactions (NC, $\nu_l + N \to \nu_l + X$) redistribute high-energy neutrinos to lower energies.  

In the TeV--PeV range, the opacity is high~\cite{Kusenko:2001gj, Hooper:2002yq, Anchordoqui:2005pn, Anchordoqui:2006ta, Hussain:2006wg, Borriello:2007cs, Hussain:2007ba}, but some neutrinos still traverse up to thousands of km inside the Earth before interacting\ \cite{IceCube:2017roe, Bustamante:2017xuy, IceCube:2020rnc}.  HE neutrinos are detected by instrumenting a cubic-kilometer of water or ice with photon detectors, which capture the optical Cherenkov light from the particle shower following a neutrino interaction. The in-ice optical Cherenkov technique is mature and led to the discovery of HE astrophysical neutrinos. In the near term, we anticipate broader sky coverage as IceCube continues to monitor the Northern sky, while KM3NeT, Baikal-GVD, and P-ONE will open the Southern HE-neutrino sky.

For UHE neutrinos, the interaction length is so short~\cite{Gandhi:1995tf, Gandhi:1998ri, Connolly:2011vc, Garcia:2020jwr} that instruments must either be sensitive to neutrinos interacting in the atmosphere or to neutrinos interacting in rock, ice, or water outside of the detector volume~\cite{PalomaresRuiz:2005xw,Reno:2019jtr,Venters:2019xwi}. A variety of suitable UHE-neutrino detection techniques have been developed and are currently being implemented. These techniques include detecting neutrino interactions in dense media (ice or the lunar regolith) and detecting Earth-skimming tau neutrinos from the air showers they generate; we describe them below.  While some UHE experiments are sensitive to all neutrino flavors, other are sensitive predominantly or only to tau neutrinos.

The sensitivity of proposed experiments is shown in Fig.~\ref{fig:diffuse_sensitivity}. In the TeV--PeV energy range, large optical Cherenkov detectors (IceCube~\cite{Halzen:2010yj,Gaisser:2014foa}, KM3NeT~\cite{KM3Net:2016zxf}, Baikal-GVD~\cite{Avrorin:2019dli}, P-ONE~\cite{P-ONE:2020ljt}) are expected to survey the full HE neutrino sky with improved statistics in the coming decade. In the 1--100~PeV energy range, IceCube and several proposed experiments (IceCube-Gen2~\cite{IceCube-Gen2:2020qha}, RET-N~\cite{Prohira:2019glh},TAMBO~\cite{Romero-Wolf:2020pzh}, Trinity~\cite{Otte:2019aaf}) may observe neutrinos from astrophysical sources and study the spectral shape of the diffuse astrophysical neutrinos by extending measurements to higher energies or constraining a spectral cutoff. At even higher energies, we eventually expect a transition from astrophysical neutrinos to cosmogenic neutrinos (or a combination of cosmogenic and astrophysical neutrinos). Several experiments are sensitive to these energies (IceCube~\cite{Halzen:2010yj,Gaisser:2014foa}, the Pierre Auger Observatory~\cite{PierreAuger:2019ens}, ARIANNA~\cite{ARIANNA:2019scz}, ARA~\cite{Allison:2011wk,ARA:2019wcf,ARA:2022rwq}) or are under construction (RNO-G~\cite{RNO-G:2020rmc,RNO-G:2021hfx}, PUEO~\cite{PUEO:2020bnn}). The most constraining limits on UHE neutrinos come from IceCube (in-ice optical), Pierre Auger Observatory (particle showers and fluorescence)~\cite{PierreAuger:2019ens}, and ANITA experiments (in-ice radio~\cite{Gorham:2019guw}). 

Several mid-scale UHE experiments currently operating (ARA~\cite{Allison:2011wk,ARA:2019wcf}, ARIANNA~\cite{ARIANNA:2019scz}, TAROGE-M~\cite{Nam:2020hng}) or under construction (PUEO~\cite{PUEO:2020bnn}, RNO-G~\cite{RNO-G:2020rmc,RNO-G:2021hfx}, EUSO-SPB2~\cite{Scotti:2020}) have the sensitivity to constrain the proton fraction in UHECR sources~\cite{vanVliet:2019nse}. Still other experiments in the design or prototyping phase have promising projected sensitivities (BEACON~\cite{Wissel:2020sec}, GRANDProto300~\cite{Decoene:2019sgx}, RET~\cite{RadarEchoTelescope:2021rca}, POEMMA~\cite{POEMMA:2020ykm}). The full array of experiments inform our understanding of fundamental physics and cosmic ray sources, but individual instruments also serve as prototypes for larger experiments targeting flux sensitivities consistent with the lowest expectations from UHE neutrino flux models (Trinity-18~\cite{Otte:2019knb}, BEACON-1K~\cite{Wissel:2020sec}, GRAND-200k~\cite{GRAND:2018iaj}, IceCube-Gen2~\cite{IceCube-Gen2:2020qha}) or broad-sky coverage and transient phenomena (POEMMA30~\cite{POEMMA:2020ykm}) planned in the coming decades. 


\subsubsection{Detection strategies} 
\medskip
\noindent
{\color{header_color}\bf Neutrino detection in ice and in water:} Signatures of HE and UHE neutrinos can be detected via optical Cherenkov emission in ice and water. UHE neutrinos can also be detected using their radio signatures. 

\begin{itemize}[label=\textcolor{header_color}{\textbullet}]
  \item 
  {\color{header_color}\bf Optical detection in ice and water:} IceCube~\cite{Halzen:2010yj,Gaisser:2014foa} and, until very recently, ANTARES~\cite{Antares:2011nsa} search for the optical Cherenkov emission from neutrino-induced showers and tracks from all flavors of neutrinos. Optical Cherenkov experiments are sensitive to HE and UHE neutrinos that interact near or inside the detector volume~\cite{IceCube:2016uab}. 
  Experiments of this type typically look for tracks of secondary leptons produced by charged-current neutrino interactions or cascades produced by  neutral-current and charged-current neutrino interactions. Track events have excellent angular resolution due to the long lever arm left by a final-state muon track, while cascade events have superior energy resolution. Typically, astrophysical purity increases with energy as the backgrounds from atmospheric neutrinos and muons fall more steeply with energy than the astrophysical neutrino flux.
  At UHE energies, these detectors are also sensitive to neutrinos that cascaded from EeV $\nu_\tau$ to PeV energies~\cite{Safa:2019ege}. In the future, IceCube-Gen2~\cite{IceCube-Gen2:2020qha} will have a much improved HE sensitivity and a broader energy coverage. Broader HE sky coverage is achieved by expanding the detector volumes of optical arrays in the Northern Hemisphere, e.g., KM3NeT\ \cite{KM3Net:2016zxf}, Baikal-GVD\ \cite{Avrorin:2019dli}, and P-ONE\ \cite{P-ONE:2020ljt}.
  \item
  {\color{header_color}\bf Radio detection in ice:}  In dense media like ice, compact electromagnetic showers generated after UHE neutrino interactions emit coherent Askaryan radiation at the Cherenkov angle. Askaryan radiation---fast, coherent radio-frequency impulses---is due to the excess negative charge in the shower~\cite{Askaryan:1961pfb}. The long attenuation length at radio frequencies allows the signal to propagate over kilometer-long distances. In-ice radio experiments are sensitive to all three flavors~\cite{Anchordoqui:2019omw,Garcia-Fernandez:2020dhb}, and may have the power to discriminate flavors based on different event topologies~\cite{Garcia-Fernandez:2020dhb,Lai:2013kja}, such as the stretching of electromagnetic showers due to the Landau-Pomeranchuk-Migdal (LPM) effect~\cite{Gerhardt:2010bj}. In ARA \cite{Allison:2014kha,ARA:2022rwq} and ARIANNA~\cite{ARIANNA:2019scz, Anker:2020lre}, radio antennas are buried in the Antarctic ice. RNO-G, based on a similar concept, is a radio detector under construction in Greenland~\cite{RNO-G:2020rmc}. The experience gained in these experiments will directly feed into the design of the expansive, sparse radio array of IceCube-Gen2~\cite{IceCube-Gen2:2020qha}. Radar signals reflecting off in-ice showers is also being explored as a detection method~ \cite{deVries:2013qwa, Prohira:2017nyr, Prohira:2019glh}.
  \item
  {\color{header_color}\bf Radio detection from the upper atmosphere or in space:} ANITA is a balloon-borne radio detector that searches for in-ice UHE neutrino interactions via the Askaryan radio emission that refracts out of the ice ~\cite{Gorham:2008dv}. The high altitude provides ANITA with a large effective area at the highest energies~\cite{Gorham:2019guw}. PUEO~\cite{Deaconu:2019rdx,PUEO:2020bnn} is an upgrade to ANITA currently under construction. NuMoon searches for radio emission from neutrino interactions in the lunar regolith~\cite{Buitink:2010qn}.
\end{itemize}
Neutrino detection via acoustic signals generated by UHE neutrino interactions in water or solids \cite{Askaryan:1977,Bowen:1977} have also been pursued (see e.g., \cite{Nahnhauer:2010uv,Lahmann:2019unc} and references therein).

\medskip
\noindent
{\color{header_color}\bf Air-shower detection techniques of UHE $\nu_\tau$:} The lifetime of the tau and $\nu_\tau$ regeneration \cite{Halzen:1998be,Dutta:2002zc,Bigas:2008sw,Alvarez-Muniz:2017mpk,Safa:2019ege} allow for unique detection techniques. The CC interaction of a $\nu_\tau$ produces a tau lepton that, at UHE energies, typically decays over a distance of tens to hundreds of kilometers. If the geometry of the experimental setting is right, the neutrino interacts inside the Earth and the tau emerges and decays in the atmosphere. The decay initiates an extensive air shower, which can be detected with particle, air-shower imaging, or radio detectors~\cite{Fargion:2000iz}. But even in cases where a $\nu_\tau$ interacts deep underground and the tau decays before reaching the surface, a new $\nu_\tau$ is produced in the decay which can again interact in the Earth and generate a tau emerging from the surface.  This ``$\nu_\tau$ regeneration" increases the chances of Earth-skimming $\nu_\tau$ reaching the detector.  Although with a lower probability, Earth-emerging muons from tau neutrinos and muon neutrinos (and electron antineutrinos in the PeV energy range due to Glashow Resonance) can undergo catastrophic energy losses in air, initiating deeply penetrating cascades that can increase the sensitivity to neutrinos with energies below 10 PeV for optical experiments \cite{Cummings:2020ycz}. For $\nu_e$, the final-state electrons showers immediately after the CC interaction, i.e., before reaching the surface.)  Earth-skimming neutrino fluxes have the added advantage of being unaffected by atmospheric neutrinos and only mildly affected by atmospheric muons~\cite{Garcia-Fernandez:2020dhb}.   

\begin{itemize}[label=\textcolor{header_color}{\textbullet}]
  \item
  {\color{header_color}\bf Air-shower particle detection:} The Pierre Auger Observatory (Auger) is a long-running large-scale array of surface water tanks that detect the Cherenkov light from air-shower particles passing through them.  Auger is designed to detect UHECRs, but it has been used to search for horizontal showers initiated by UHE neutrinos in the atmosphere~\cite{PierreAuger:2019ens}. Planned experiments using particle detectors combined with radio detectors include the upgrade to Auger, AugerPrime~\cite{Aab:2016vlz}, and GCOS~\cite{Horandel:2021prj}.  The Telescope Array (TA)~\cite{Abbasi:2019fmh} and  HAWC~\cite{Vargas:2016hcp} experiments have used a similar detection strategy, but have a more limited sensitivity.  TAMBO is a planned array of water tanks to be located on one side of an Andean canyon, designed to detect the showers initiated by UHE taus emerging from the opposite side~\cite{Romero-Wolf:2020pzh}.
  \item
  {\color{header_color}\bf Air-shower radio detection:}  Radio-detection of Earth-skimming tau neutrinos is a promising method due to the long attenuation lengths of radio waves in air. As with in-ice radio detection, sparse arrays can be used to instrument large areas.  Radio emission is generated in air showers initiated by tau decays, via the geomagnetic effect, due to charge separation in the magnetic field of Earth as the air showers progress through the atmosphere. Moreover, the narrow Cherenkov cone and fast radio imaging enables sub-degree angular resolution. BEACON~\cite{Wissel:2020sec}, in its prototype phase, TAROGE~\cite{Nam:2016cib}, and TAROGE-M~\cite{Nam:2020hng} are compact antenna arrays in elevated locations that aim to detect UHE $\nu_\tau$ emerging upwards via the radio emission of the air showers that they trigger.  ANITA~\cite{Deaconu:2019rdx} and PUEO~\cite{Deaconu:2019rdx,PUEO:2020bnn} are also sensitive to upgoing $\nu_\tau$, from a higher elevation. GRAND~\cite{GRAND:2018iaj} is a planned experiment that will cover large areas with a sparse antenna array to detect the radio emission from air showers triggered by UHE $\nu_\tau$, cosmic rays, and gamma rays.
  \item
  {\color{header_color}\bf Air-shower imaging:} Several air-shower imaging instruments, although optimized for cosmic-ray and gamma-ray detection, have demonstrated that the imaging of air showers via the Cherenkov and fluorescence light radiated by shower particles is a viable detection method of UHE $\nu_\tau$~\cite{Aramo:2004pr,  Gora:2014lya, Gora:2016mmy, Gora:2016mmy, MAGIC:2018gza, Fiorillo:2020xst}. Air-shower imaging allows the reconstruction of the air-shower arrival direction with arcminute resolution and the shower energy within a few tens of percent of uncertainty. These excellent reconstruction characteristics are why the very-high-energy gamma-ray and UHECR communities have been using air-shower imaging for quite some time~\cite{Weekes:1989tc}. Two planned instruments optimized for the detection of UHE neutrinos from the ground are Trinity~\cite{Otte:2018uxj} and Ashra NTA~\cite{Sasaki:2014mwa}.
  POEMMA is designed to detect the Cherenkov light of UHE $\nu_\tau$-initiated showers from a satellite. A unique feature of POEMMA is its ability to rapidly reposition to target transient multi-messenger events~\cite{Venters:2019xwi, POEMMA:2020ykm}. EUSO-SPB2 is a telescope mounted on a super-pressure balloon which will fly at high altitudes and serve as a pathfinder for POEMMA~\cite{Adams:2017fjh}.
\end{itemize}
\noindent
{\color{header_color}\bf UHE neutrino detection in lunar regolith:} The Moon, with its radio-transparent lunar regolith, provides a 19~million km$^2$ target for observing UHE neutrinos with Earth-based radio telescopes~\cite{Dagkesamanskii}. The radio emission generated by showers following a neutrino interaction in the rim of the Moon, are searched for by forming beams on various regions on the moon~\cite{Bray:2014loa}.  The technical requirements on radio telescopes are access to full time-domain voltages of the receiver and multiple beams on- and off-Moon, making not all telescopes suitable for these studies. Past observations have been performed most notably by the Parkes~\cite{Bray:2015lda} and Westerbork telescopes~\cite{terVeen:2010gb} and the EVLA~\cite{Jaeger:2009whi}, with past and on-going work being performed with LOFAR~\cite{Buitink:2012awa, Krampah:2021ysn}. The next step in sensitivity will be obtainable with the Square-Kilometre Array (SKA)~\cite{Bray:2014loa, James:2015cav, Bray:2015ota, James:2017pvr}. While have being able to observe huge target volumes, the technique suffers from an intrinsically high energy threshold beyond $10^{20}$~eV~\cite{Bray:2016xrn}, thus targeting more exotic models beyond the Standard Model. It is complimentary to the energy range of dedicated neutrino experiments and uses existing radio telescope infrastructure without much additional effort.


\subsection{High-energy range}
\label{subsec:landscape_tev_pev}

\subsubsection{IceCube}

\label{subsec:landscape_tev_pev_icecube}

\begin{figure}[t]
\centering
\begin{subfigure}
  \centering
  \includegraphics[width=0.35\textwidth]{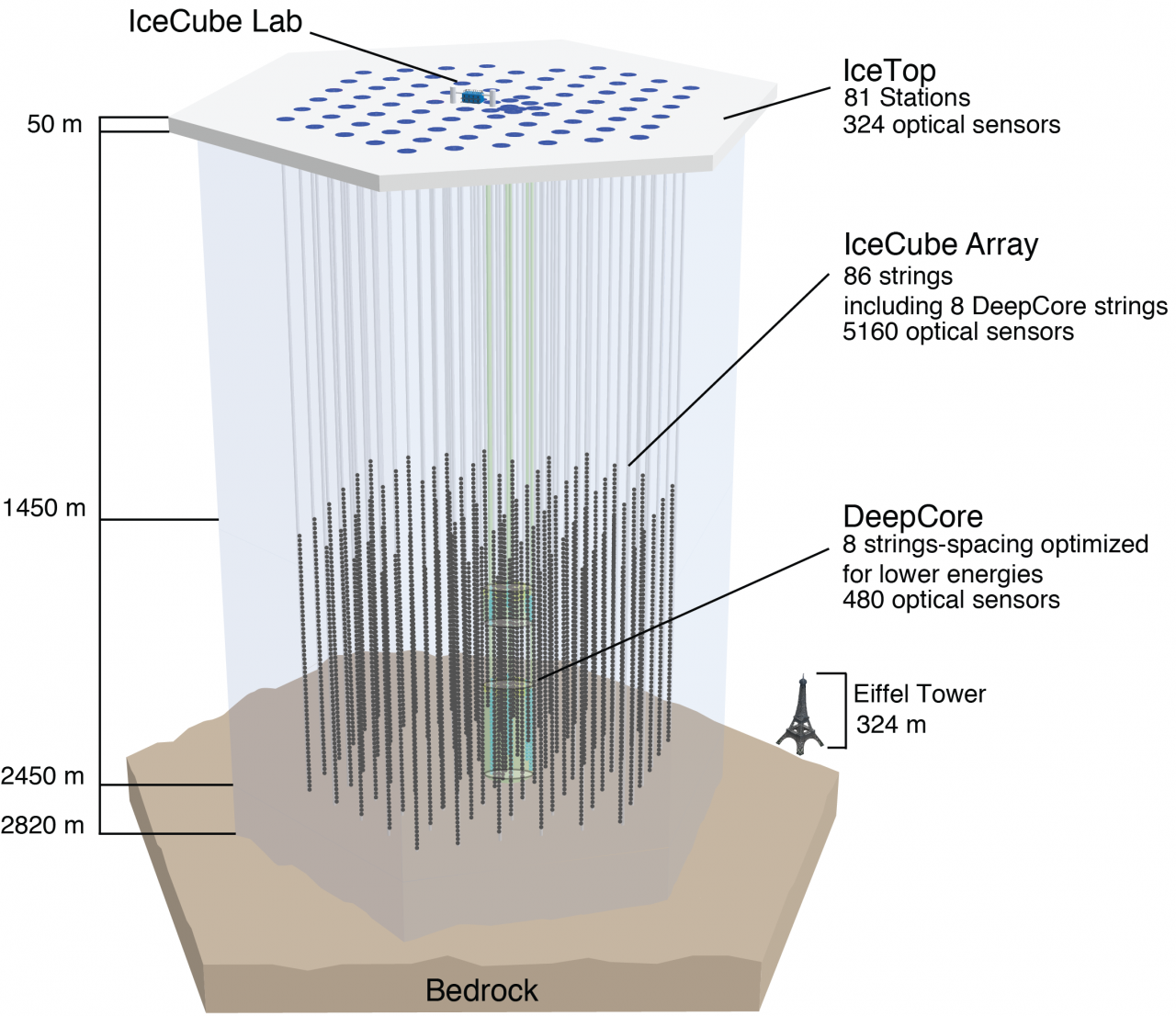}
\end{subfigure}
\begin{subfigure}
  \centering
  \includegraphics[width=0.5\linewidth]{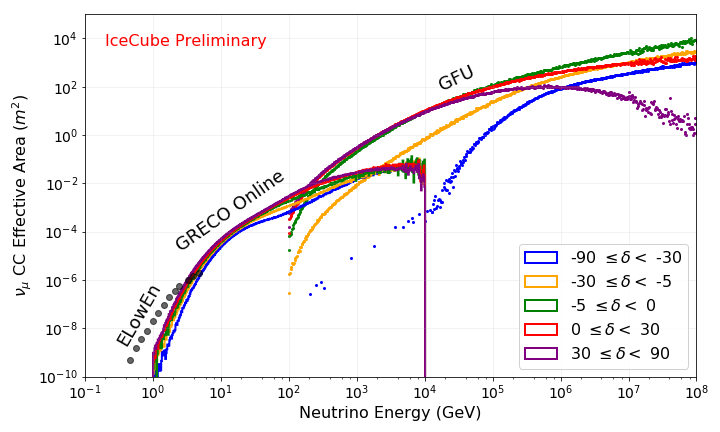}
  \end{subfigure}
\caption{
       {\it Left:} The IceCube detector, showing the three subdetectors described in \ref{subsec:landscape_tev_pev_icecube}. The IceCube array is sensitive to interactions above a few hundred GeV. DeepCore lowers the detection threshold to around 5~GeV. The IceTop surface array allows for vetoing of down-going events and for study of atmospheric air showers.  Figure reproduced from Ref.~\cite{IceCube:2016zyt}.
       {\it Right:} The effective area for various charged-current muon neutrino data sets used by IceCube~\cite{IceCube:2021oos}, including the ``ELowEn'' data set for sub-GeV transients \cite{IceCube:2021ddq}, the ``GRECO Online'' data set for sub-TeV transients \cite{IceCube:2021oos}, and the gamma-ray follow-up data set (GFU) used for real-time follow-up analyses~\cite{IceCube:2016cqr}.  Figure reproduced from Ref.~\cite{IceCube:2021oos}.
    }
\label{fig:IceCube-geometry-aeff}
\end{figure}

The IceCube Neutrino Observatory~\cite{Gaisser:2014foa}, shown in Fig.~\ref{fig:IceCube-geometry-aeff}, is the first gigaton neutrino detector. Situated at the geographic South Pole, it consists of the IceTop surface array used for cosmic-ray physics and the IceCube detector, an in-ice Cherenkov detector deployed between 1450 and 2450 meters below the surface of the Antarctic glacier. The IceCube detector contains 5160 Digital Optical Modules (DOMs) arranged in 86 strings. Each DOM contains a single 10-inch-diameter PMT, LEDs used for calibration measurements, and associated electronics housed in a pressure sphere. Seventy-eight strings are arranged in a regular hexagonal grid with strings 125~m apart and DOMs separated by 17~m along each string. Using this configuration, IceCube has measured neutrino fluxes from 100~GeV to several PeV, although its sensitivity to high-energy neutrinos extends beyond 1 EeV \cite{IceCube:2018fhm, IceCube:2021rpz}. The remaining eight strings are deployed as an infill array at the center of the detector, exploiting the clear ice below 1750~m to lower the IceCube threshold to 10~GeV and to study neutrino oscillations and sub-TeV dark matter. Using specialized studies of detector-wide noise rates, IceCube is also capable of studying nearby MeV~scale core-collapse supernova neutrinos.

\begin{figure}[t]
\centering
\begin{subfigure}
  \centering
  \includegraphics[width=0.47\linewidth]{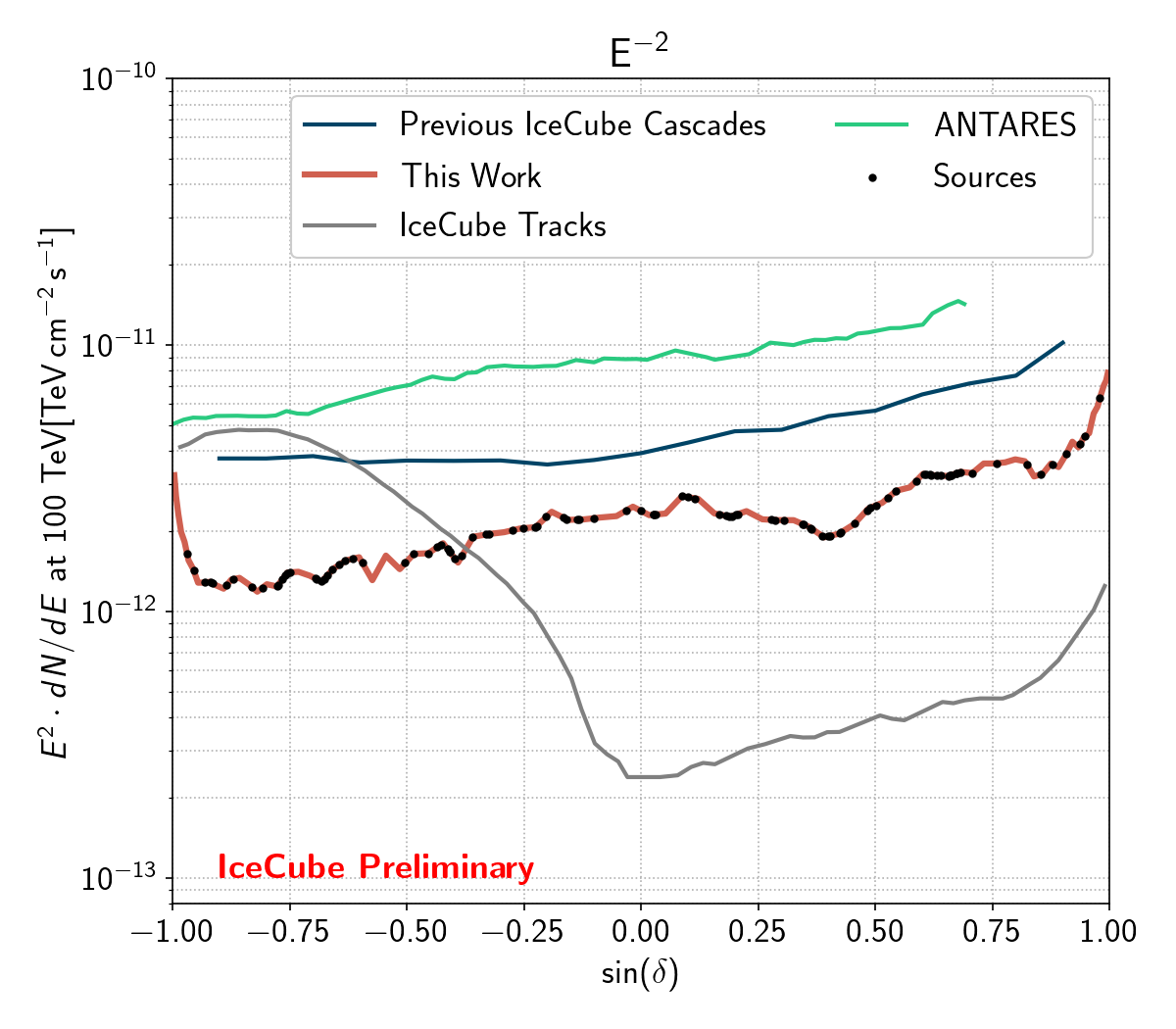}
\end{subfigure}
\begin{subfigure}
  \centering
  \includegraphics[width=0.47\linewidth]{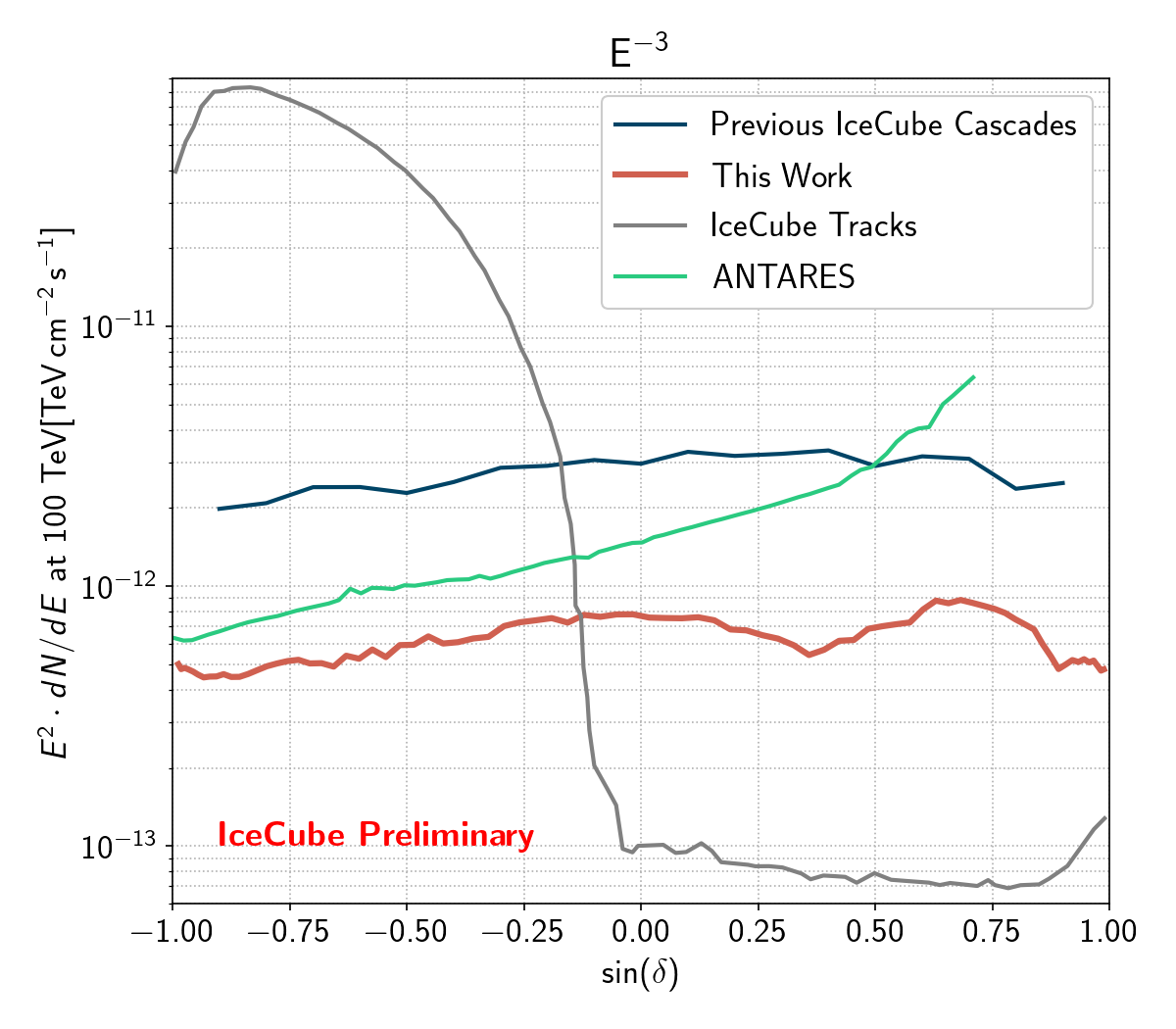}
  \end{subfigure}
\caption{IceCube expected 90\% sensitivity for an $E^{-2}$ ({\it left}) and $E^{-3}$ ({\it right}) spectrum~\cite{IceCube:2021mbf}. Sensitivities are shown for 10~years of IceCube tracks~\cite{IceCube:2019cia} (grey), IceCube cascades (10~years~\cite{IceCube:2021mbf}, red; 7~years~\cite{IceCube:2019lzm}, blue), and for 11~years of ANTARES data~\cite{Aublin:2019zzn} (teal). The IceCube track sensitivity dominates the Northern sky while cascade events allow IceCube to remain competitive in the Southern sky for soft spectra.  Figures reproduced from Ref.~\cite{IceCube:2021mbf}}.
\label{fig:icecube_sensitivities}
\end{figure}

Able to reconstruct muon tracks in the detector to $0.25^{\circ}$ at 1~PeV~\cite{IceCube:2019cia} and shower events to about $7^{\circ}$ at 1~PeV~\cite{IceCube:2021mbf}, IceCube has provided continuous full-sky coverage since detector completion in 2010. The unique location of IceCube at the geographic South Pole results in a clean sample of TeV neutrinos from the Northern sky with world-leading sensitivity to astrophysical neutrino sources, as shown in Fig.~\ref{fig:icecube_sensitivities}. These events have been used to test for emission from many source populations, including blazars~\cite{IceCube:2020nig}, diffuse Galactic emission~\cite{IceCube:2017trr,Aartsen:2020tdl}, gamma~ray bursts~\cite{IceCube:2017amx}, fast radio bursts~\cite{IceCube:2017fpg,IceCube:2019acm}, and more. IceCube tracks from the Northern sky are also used in broader physics analyses ranging from dark matter searches, flavor physics~\cite{IceCube:2020abv}, Lorentz invariance~\cite{IceCube:2017qyp}, neutrino cross sections~\cite{IceCube:2017roe,Bustamante:2017xuy,IceCube:2020rnc}, and sterile neutrinos~\cite{IceCube:2017ivd, IceCube:2016rnb,IceCube:2020phf,IceCube:2020tka}.

IceCube's view of the Southern sky is dominated by muons produced in atmospheric air showers. The harsh cuts necessary to remove atmospheric muons lead to a large reduction in effective area, shown in Fig.~\ref{fig:IceCube-geometry-aeff}, limiting the Southern sky sensitivity with tracks. To mitigate the impact of atmospheric backgrounds, the highest-energy tracks, cascades and starting events may be used to perform searches. In 2013, using a sample of these high-energy events starting inside of the detector~\cite{IceCube:2014stg}, IceCube reported the first evidence of unresolved flux of astrophysical neutrinos, opening a new window into high-energy physics in the Universe. While no individual sources have yet been identified in the high energy data set~\cite{Arguelles:2019boy}, characterization of the observed diffuse flux continues with improvements to data selections and analysis methods~\cite{IceCube:2015gsk}.

IceCube releases information on events likely to be of astrophysical origin to the community in real time. as described in Sec.~\ref{sssec:mmalerts}, allowing partners to search for potential sources of neutrinos within minutes~\cite{IceCube:2016cqr}. IceCube also regularly performs follow-up searches for neutrino emission in spatial and temporal coincidence with flares observed with both electromagnetic observatories and gravitational-wave detectors~\cite{IceCube:2020xks}. In 2017, a neutrino released through the IceCube real-time event program, IceCube-170922A, was found to be in coincidence with a gamma-ray flare from blazar TXS~0506+056~\cite{IceCube:2018dnn}. A follow-up analysis unearthed a previously unobserved neutrino excess between September~2014 and March~2015, with $3.5\sigma$ evidence of high-energy astrophysical neutrino emission~\cite{IceCube:2018cha} from the direction of TXS 0506+056. 
Following the observation of neutrinos from the direction of TXS~0506+056, new analysis tools have been constructed to search the sky for similar emission across 10~years of data~\cite{IceCube:2021slf}. 

The long operation of IceCube provides more than a decade's worth of data. New data can provide additional sensitivity to TeV physics, but improvements based solely on accumulating data slow the pace of discovery. Instead, significant improvements must come also from progress in selection, reconstruction, and analysis. New studies using starting tracks and shower events reject atmospheric muons better, lowering the energy threshold for observations and dramatically improving the sensitivity in the Southern sky, as shown in Fig.~\ref{fig:icecube_sensitivities}, enabling IceCube to continue to perform analyses competitive with planned Northern observatories \cite{IceCube:2021ctg,IceCube:2021mbf}. Lower-energy data sets targeting astrophysical transients with all-flavor full-sky coverage have recently given new limits on sub-TeV astrophysical phenomena \cite{IceCube:2021ddq,IceCube:2020qls} and an expanded data set, shown in Fig.~\ref{fig:IceCube-geometry-aeff}, will soon open a new channel for multi-messenger analyses~\cite{IceCube:2021oos}. New event reconstructions relying on updated likelihood constructions~\cite{IceCube:2021oqo} or advanced convolutional neural networks~\cite{Abbasi:2021ryj} show promise in improving the angular and energy resolution  necessary for precise characterization of astrophysical fluxes. Recent updates in the point-source likelihood used by IceCube searches aim to improve modeling by utilizing a fuller description of its point spread function~\cite{IceCube:2021wbi}. With these updates and future development of machine learning tools, IceCube will continue to be a leader in neutrino astronomy.


\subsubsection{KM3NeT}\label{subsec:km3net}

\begin{figure*}[t]
  \centering
  \includegraphics[width=0.99\textwidth]{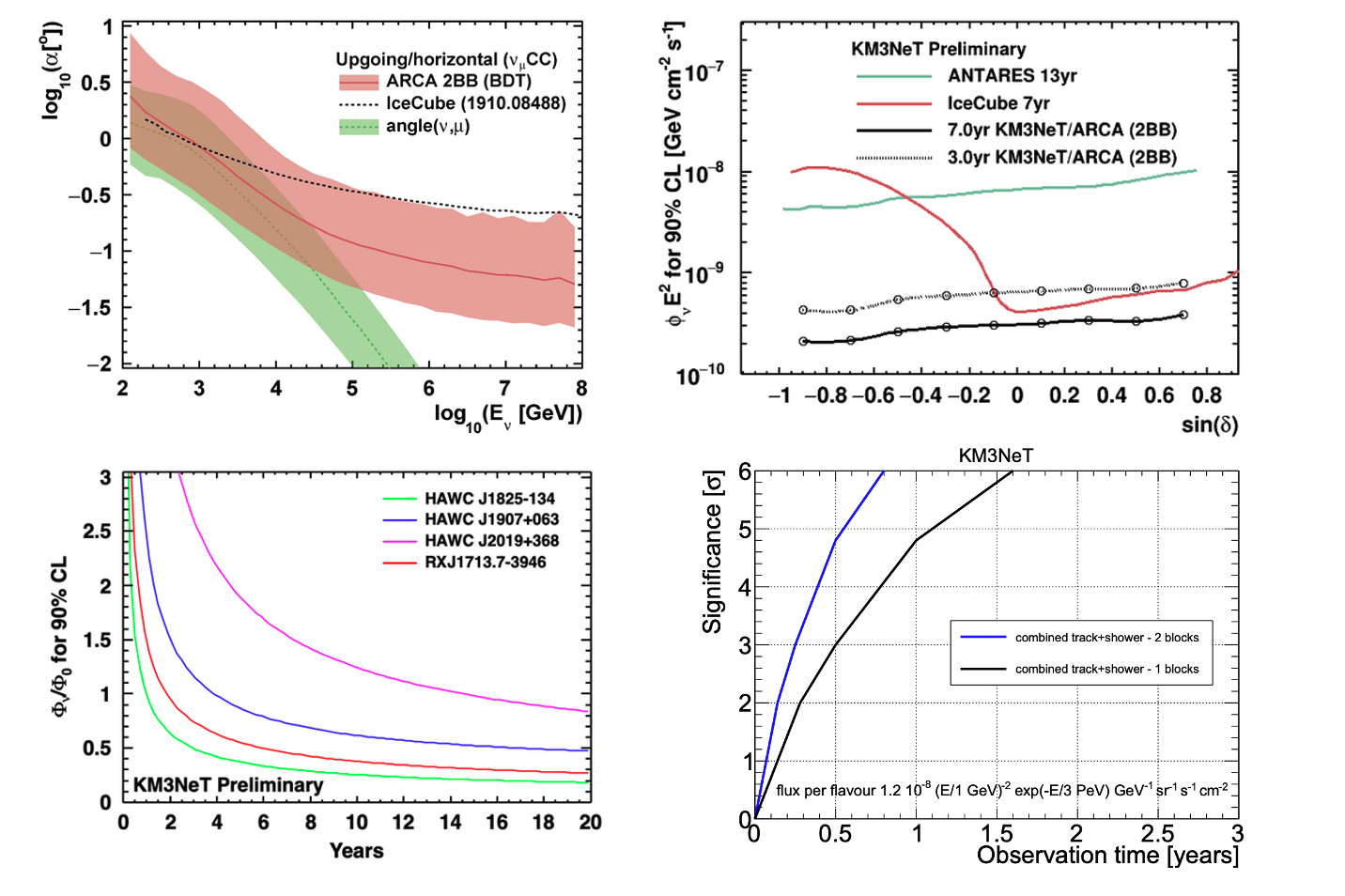}
  \caption{\label{fig:KM3NeT-figure}\small 
  {\it Top left:} ARCA angular resolutions as a function of the neutrino energy for $\nu_{\mu}$~CC events~\cite{ARCAproc}. {\it Top right:} Sensitivity, defined as the median upper limit at 90\%~CL, of the complete ARCA detector to point-like sources emitting with a $E^{-2}$ spectrum as a function of the source declination after three and seven years of data taking~\cite{ARCAproc}. For comparison, the ANTARES and IceCube sensitivities are also shown. {\it Bottom left:} Ratio between the 90\%~CL sensitivity and the expected neutrino flux normalization for different Galactic sources as a function of the data taking time of the complete ARCA detector~\cite{ARCAproc}. {\it Bottom right:} Expected significance in the detection of the cosmic neutrino flux reached by ARCA as a function of the data taking time in either half or full configuration~\cite{KM3Net:2016zxf, ARCADiffuseProc}.}
\end{figure*}

KM3NeT~\cite{KM3Net:2016zxf} is a multi-purpose cubic-kilometer water Cherenkov neutrino telescope currently being deployed at the bottom of the Mediterranean Sea. It consists of two sub-detectors: ARCA (for Astroparticle Research with Cosmics in the Abyss), located offshore from Capo Passero (Italy) at a depth of 3500~m, and ORCA (for Oscillation Research with Cosmics in the Abyss), placed 2450~m deep offshore from Toulon (France). A KM3NeT string, or Detection Unit (DU), holds 18 Digital Optical Modules (DOMs), each a 17-inch-diameter pressure-resistant glass sphere housing 31 PMTs together with the associated electronics. An array of 115 DUs will constitute a detector building block. Two blocks will be deployed at the ARCA site and one at the ORCA site. Besides the number of building blocks, a different granularity distinguishes ARCA and ORCA. With the DOMs being vertically spaced by 36~m and with a horizontal spacing of 90~m between DUs, ARCA will instrument a total volume of $1~\textrm{km}^3$ with the primary goal of detecting astrophysical neutrinos in the $100$--$10^8$~GeV energy range. ORCA, with 9~m (20~m) spacing between DOMs (DUs), is optimized for the study of fundamental neutrino properties using 1--100-GeV atmospheric neutrinos. The combination of these two detectors allows KM3NeT to exploit the full potential of neutrino astronomy, accessing a wide energy range, from MeV-scale core-collapse supernova neutrinos (CCSNs) to astrophysical neutrinos up to several PeV.

ARCA will have an excellent angular resolution: $0.1^{\circ}$ at 1~PeV for muon neutrinos, as shown in Fig.~\ref{fig:KM3NeT-figure}, and around $1^{\circ}$ for showers. It will be a powerful tool for identifying the sources of high-energy cosmic neutrinos. Being located in the Northern Hemisphere, ARCA will have a sensitivity to point-like sources on the Southern sky that will be over one order of magnitude better than the current detectors (see Fig.~\ref{fig:KM3NeT-figure})~\cite{ARCAproc}. 
Moreover, its geographical location will allow ARCA to test predictions of neutrino production in Galactic sources in the 1--10 TeV energy range, which are based on gamma-ray measurements. As shown in Fig.~\ref{fig:KM3NeT-figure}, ARCA will reach the sensitivity to probe the predicted fluxes for several Galactic sources after less than four years of data taking, assuming the observed gamma rays from these sources originate from the interactions of hadrons~\cite{ARCAproc}.

The IceCube observation of the diffuse astrophysical flux in 2013 set a milestone in the field of neutrino astronomy. As depicted in Fig.~\ref{fig:KM3NeT-figure}, ARCA will quickly confirm the cosmic diffuse neutrino flux with an expected significance of $5\sigma$ reached in 1 (0.5) year of operation with one (two) building blocks~\cite{KM3Net:2016zxf, ARCADiffuseProc}. It will provide valuable constraints on the neutrino spectrum and flavor composition~\cite{ARCAproc} and its superior angular resolution will enhance the power of multi-messenger follow-up studies for the detected neutrinos, reducing chance coincidences with potential counterpart sources.

ARCA will have a unique sensitivity to detect the diffuse high-energy neutrino emission from the Galactic Plane, especially from the vicinity of the Galactic Center. For certain cosmic-ray propagation models predicting a harder cosmic-ray spectrum in the inner Galaxy, the resulting neutrino flux (the KRA$\gamma$ models~\cite{Gaggero:2015xza}) is expected to be detected with $5\sigma$ significance in four years with the complete detector~\cite{ARCADiffuseProc}. Similarly, only three years of data taken by one ARCA building block would be needed to observe the expected neutrino flux from the region of the Fermi Bubbles, in case their gamma-ray emission originates from the decay of pions, and assuming a cutoff in the neutrino energy spectrum at 100~TeV~\cite{KM3NeT:2013dmm}.

The world-leading angular resolution, wide energy range, full-sky coverage and 100\% duty cycle will make KM3NeT a key player in the field of real-time multi-messenger astronomy, performing fast follow-up-studies of transient phenomena and distributing real-time alerts of interesting neutrino detection~\cite{KM3NeT-MMframework}.


\subsubsection{Baikal-GVD} 
\label{subsec:baikal}

\begin{figure*}[t]
  \centering
  \includegraphics[width=0.52\textwidth]{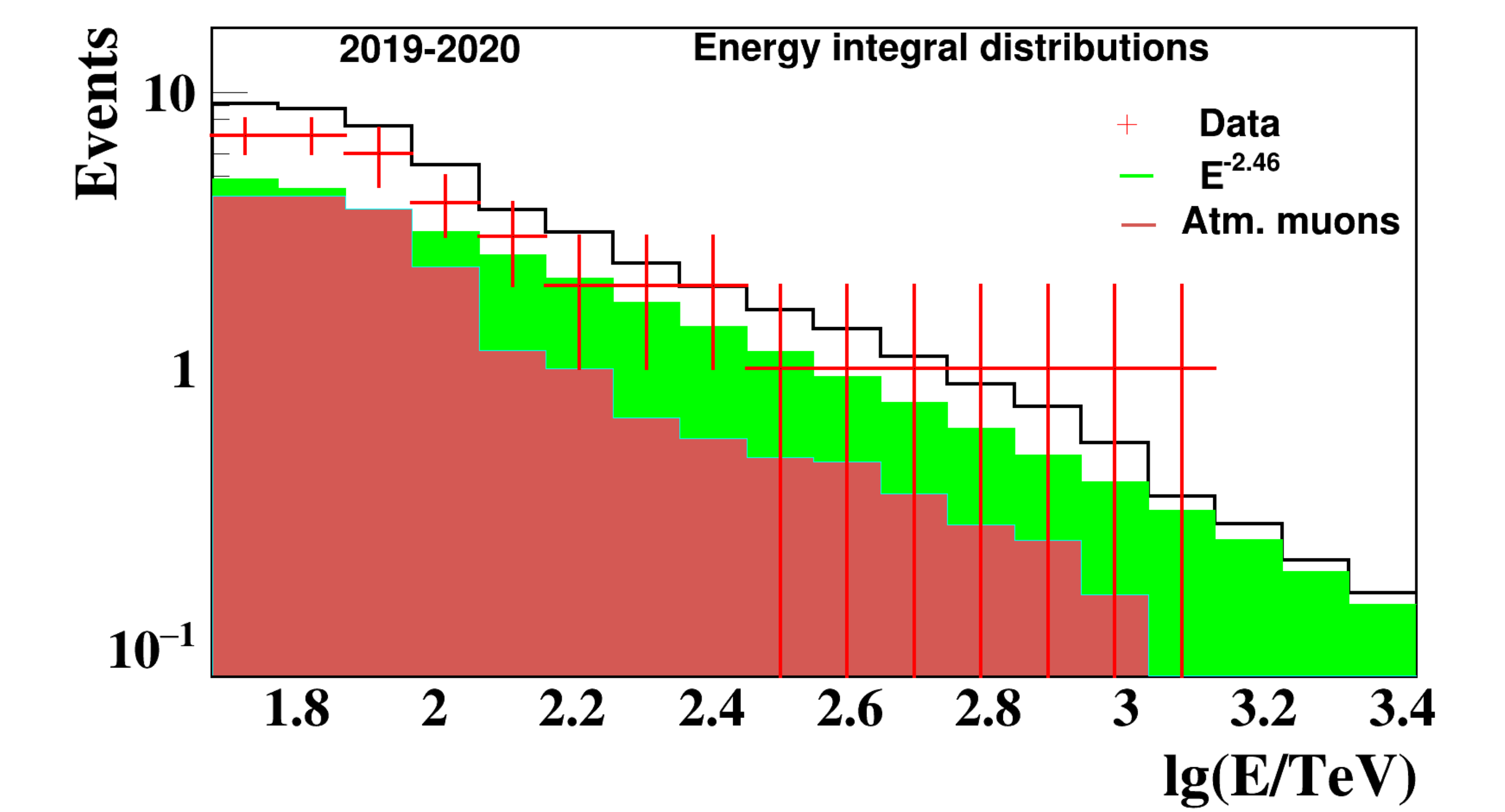}
  \includegraphics[width=0.47\textwidth]{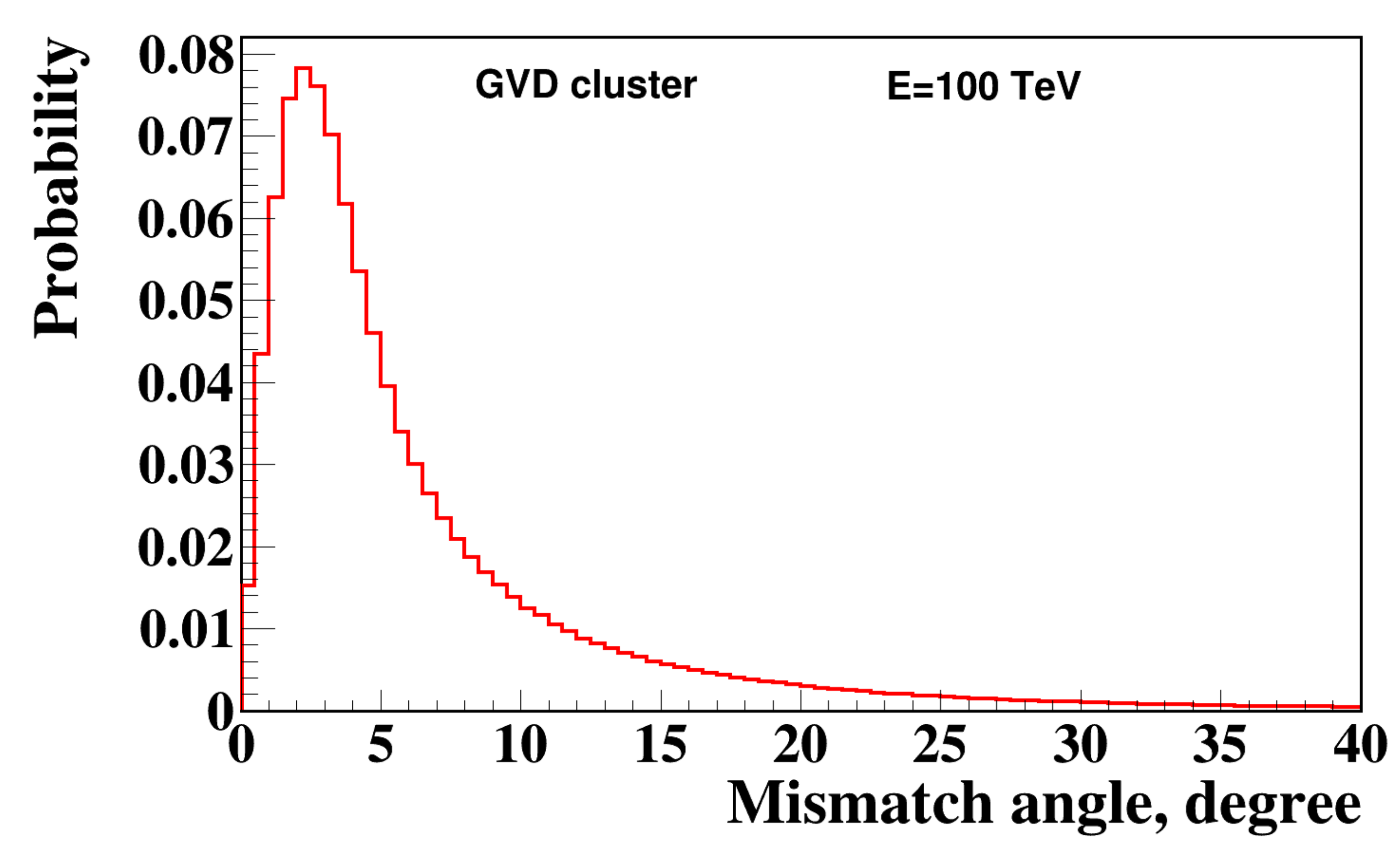}
  \newline
  \includegraphics[width=0.65\textwidth,trim=0cm 0cm 0cm -0.8cm, clip]{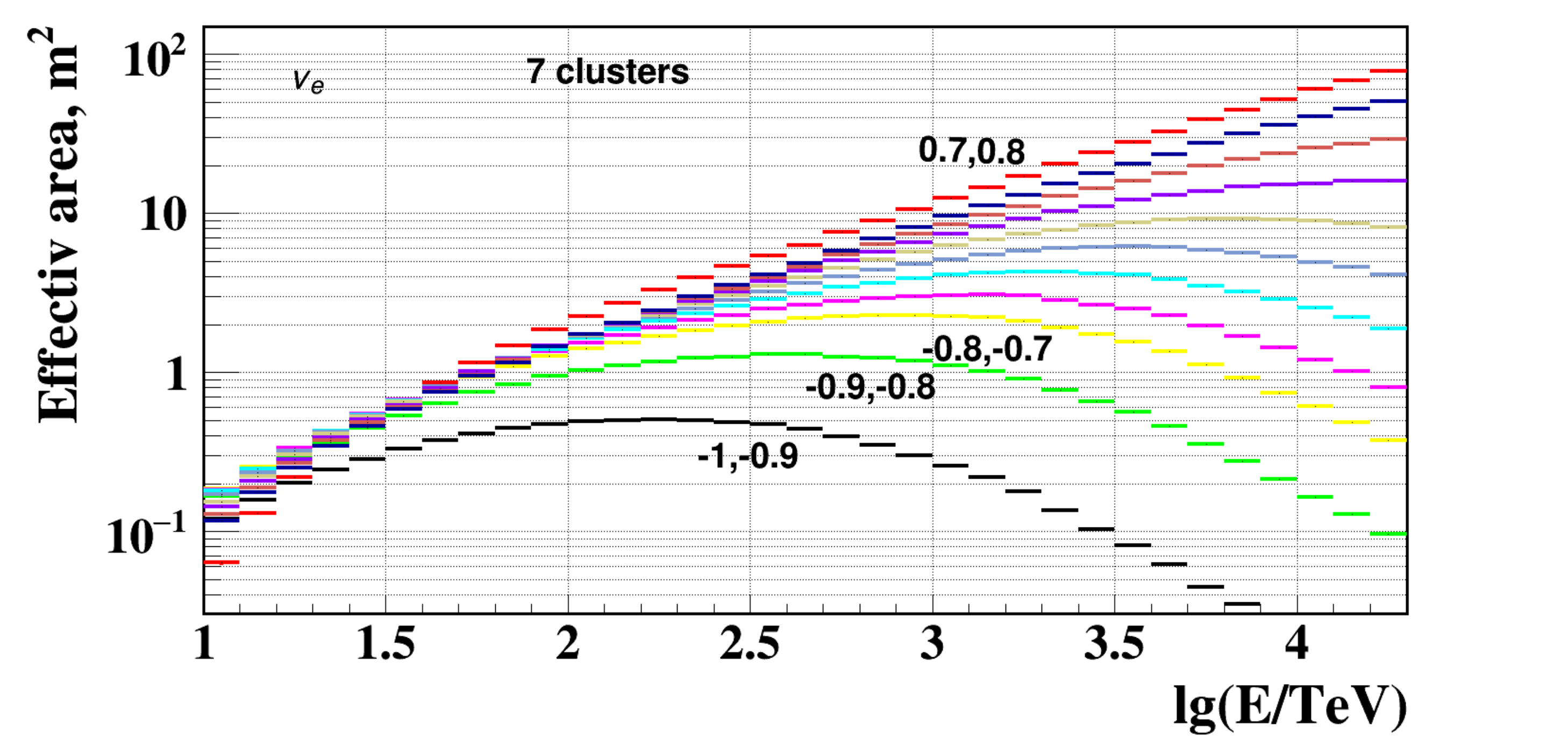}
  \includegraphics[width=0.34\textwidth,trim=0cm 0cm 0cm -0.8cm, clip]{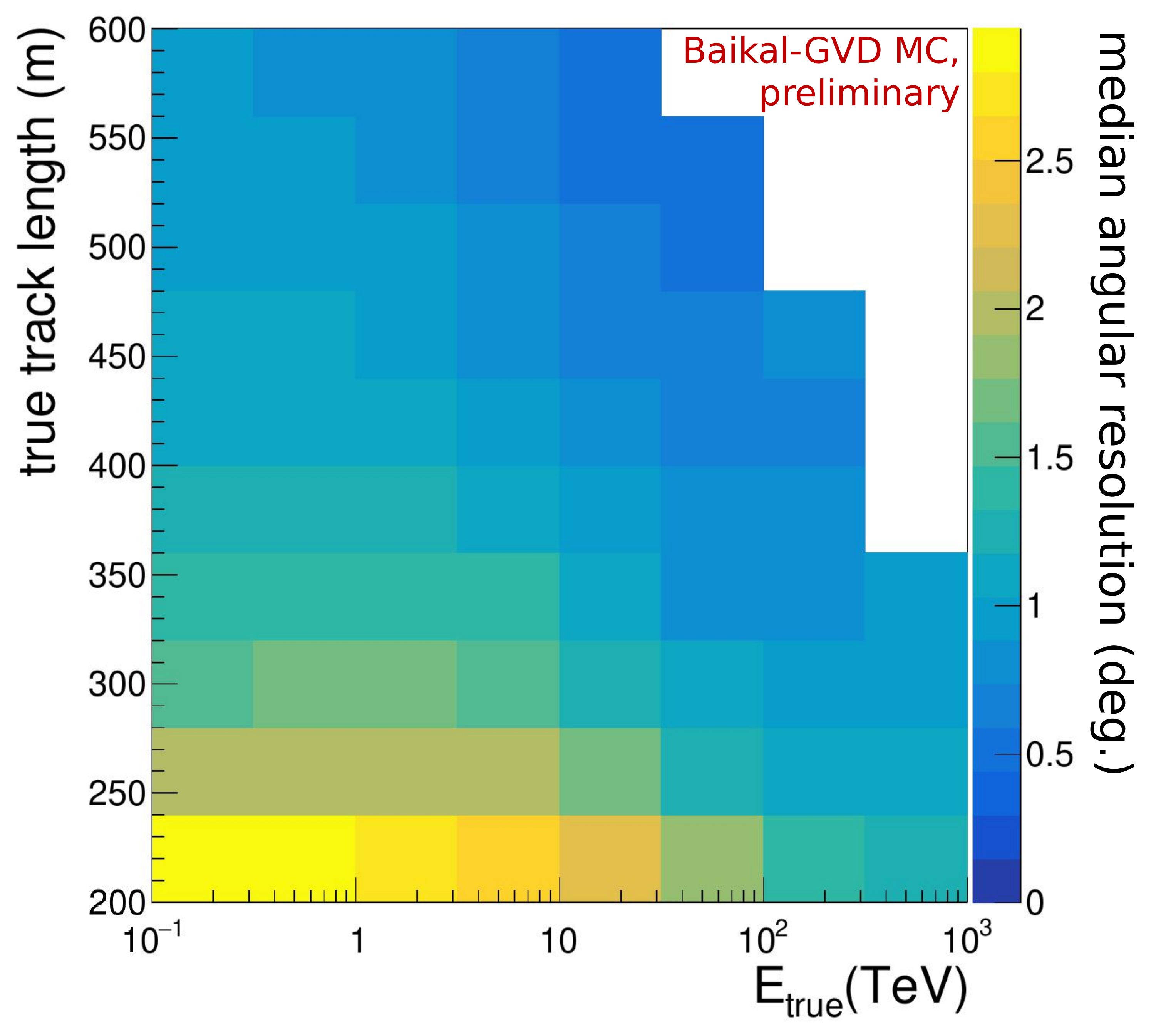}
  \caption{\label{fig:GVD-figure}\small {\it Top left:} Integral distributions of reconstructed events selected by quality cuts versus energy (crosses) and expected events from atmospheric muons (brown histogram) and from a diffuse flux of astrophysical neutrinos (green histogram)~\cite{Allakhverdyan:2022}. {\it Top right:} Accuracy of the angular reconstruction of events in cascade mode for a neutrino energy of 100~TeV. {\it Bottom left:}  Effective areas of the Baikal-GVD telescope with the 7 clusters deployed as of 2020 in the cascade reconstruction mode for electron neutrino interactions for different ranges of cos(zenith angle)~\cite{Baikal-GVD:2021civ}. {\it Bottom right:} Preliminary estimates for the single-cluster angular resolution of muon tracks produced in CC neutrino interactions~\cite{Baikal-GVD:2021krq}.}
\end{figure*}

The Baikal Gigaton Volume Detector (Baikal-GVD)~\cite{Allakhverdyan:2021} is currently the largest operating water Cherenkov neutrino telescope in the Northern Hemisphere, designed to search for high-energy neutrinos of astrophysical origin, whose sources are not yet reliably known.
The telescope has a modular structure and consists of clusters. Each cluster is a fully functional detector equipped with detection, triggering, calibration, positioning, and data acquisition systems. The first cluster of  Baikal-GVD was deployed on Lake Baikal at a depth of 1360 meters, 4 km from the shore, starting in 2016. In subsequent seasons, the effective volume of Baikal-GVD was gradually increased by adding one or two functionally independent clusters per season. Further volume expansion to one cubic kilometer by 2025 is ongoing.

Each cluster consists of eight strings 525~m long with optical modules (OM) placed on them to register the Cherenkov radiation in the lake water. The distance between the strings in a cluster is 60~m, where one string is in the center and seven are arranged in a circle around it. The distance between the clusters is 300~m. The OM~\cite{Avrorin:2016lxq} is a borosilicate glass sphere 42 cm in diameter that houses a PMT of 25.5~cm in diameter.
The electronics of the OM are mounted directly on the base of the PMT and include a LED calibration system. Eight GVD clusters have been successfully operating since April 2021 and receiving data from 2304 OMs placed on 64 strings. 

The modular structure of the telescope allowed the study of muons and neutrinos in the early stages of detector deployment. 
Ten (3+7) cascade events have been selected from data collected between 2018 and 2020
with energies above 60 TeV as the first astrophysical neutrino candidates~\cite{Allakhverdyan:2022}. The energy distributions are shown in  the upper left of Fig.~\ref{fig:GVD-figure} for the season 2019-2020 in comparison with expectations from atmospheric muon and diffuse astrophysical neutrino fluxes. 

The effective area of the Baikal-GVD telescope with 7 clusters for events from $\nu_e$-induced cascades in the lake water is shown in the lower left of Fig.~\ref{fig:GVD-figure}~\cite{Baikal-GVD:2021civ}. At energies above 100~TeV, there is a significant decrease in the effective area, due to neutrino absorption in the Earth.  

The angular resolution for cascade events is shown in the upper right of Fig.~\ref{fig:GVD-figure}. 
The median value is 2--4 degrees, depending also on neutrino energy and zenith angle. Preliminary estimates of the angular resolution of muon tracks produced in CC neutrino interactions for single GVD cluster are displayed in the bottom right of Fig.~\ref{fig:GVD-figure}, along with the its dependence on neutrino energy and the visible track length~\cite{Baikal-GVD:2021krq}. An accuracy of better than $1^\circ$ is achieved for energies above 10~TeV and track lengths greater than 300~m. This precision will allow for significant contributions to multi-messenger searches for neutrino sources via the transmission of neutrino alerts. GVD already participates in follow-up analyses~\cite{Avrorin:2021dwe}. Presently, the delay in issuing an alert takes on average 3--5 hours~\cite{Baikal-GVD:2021ypx}.  However, a fast mode has been recently implemented~\cite{Baikal-GVD:2021pyu} that promises low latency alerts on the order of several minutes, which will be useful for multi-messenger alerts. 


\subsubsection{P-ONE} 

\begin{figure*}[t]
  \centering
  \includegraphics[width=0.9\textwidth]{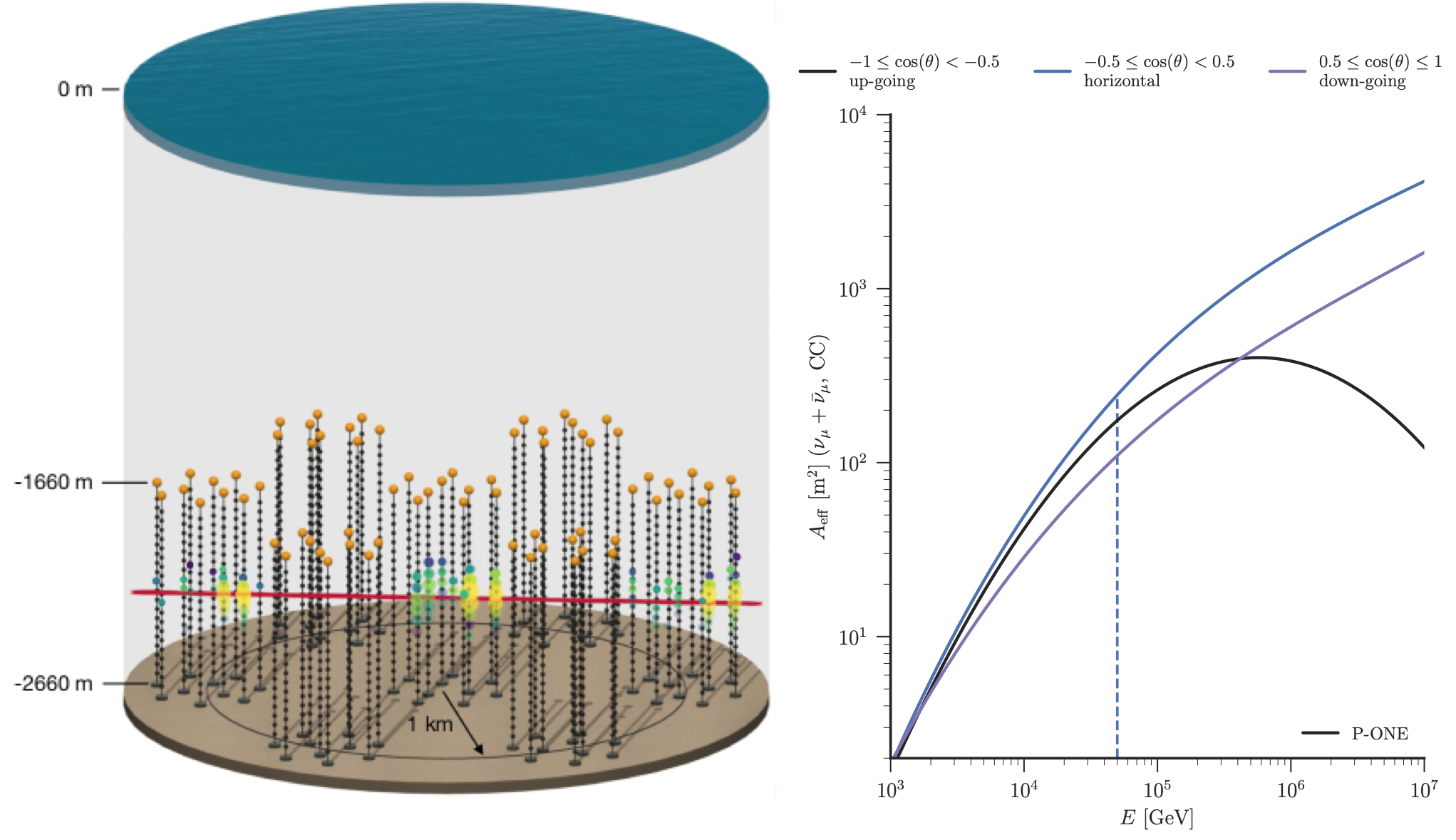}
  \caption{\label{fig:p-one-figure}\small 
  {\it Left:}  Design of the Pacific Ocean Neutrino Experiment (P-ONE) showing the strings of photomultipliers, and the light pattern of a 50-TeV neutrino.  {\it Right:}  Charged-current effective areas of $\nu_\mu + \bar{\nu}_\mu$ for different arrival directions.  Figures reproduced from Ref.~\cite{P-ONE:2020ljt}.}
\end{figure*}

The vision behind the Pacific Ocean Neutrino Experiment (P-ONE) \cite{P-ONE:2020ljt, Resconi:2021ezb}, 
shown schematically in Fig. \ref{fig:p-one-figure},
is to base a neutrino telescope within a pre-existing large-scale oceanographic infrastructure. As of 2018, the P-ONE collaboration, consisting of researchers in Germany, Canada, and United States, has obtained evidence that the NEPTUNE observatory, operated by Ocean Networks Canada (ONC) since 2009, is an ideal instrumented site to operate a large-volume neutrino telescope.

The 800-km loop of fibre-optic telecommunications cables comprising NEPTUNE provides a high-speed (up to 4~Gbit s$^{-1}$) and high-power (of 8 kW per node) data link to five nodes that serve as the local hub for the main observations and experiments. A total of 17 primary junction boxes are already wired to the nodes and used to connect hundreds of instruments. The connections are based on underwater mating connectors with field-proven reliability (less than 2\% failure in connector pairs deployed over ten years). The ONC success rate of offshore maintenance cruises has been of the order of 95\% over several years. 

Two pathfinder missions deployed in 2018 and 2020 by the P-ONE collaboration monitored continuously the optical properties of the 2.6 km deep instrumented site of Cascadia Basin in the Pacific Ocean over three years. The in-situ data delivered the baseline of ambient bioluminescence and $^{40}$K and an attenuation length of about 30~meters at 450~nm \cite{STRAW:2018osc, Bailly:2021dxn}. The successful pathfinder mission phase triggered a prototype phase in 2021 envisioning the deployment of at least three fully instrumented lines by 2026. The prototyping phase will lay the scientific groundwork and feasibility for an eventual 70-line multi-cubic kilometer telescope. The scientific goal of P-ONE is to extend IceCube's sky coverage in neutrinos at energies above TeV, explore the central region of the Galaxy and operate in real time in a complementary way to KM3NeT and GVD. Together, the four telescopes mentioned above can cover the entire sky, optimising possible reciprocal follow-ups and improving the sensitivity compared to that achieved by IceCube by up to three orders of magnitude as recently calculated by the PLE$\nu$M group~\cite{Schumacher:2021Pv}. 


\subsubsection{IceCube-Gen2 optical}\label{subsec:gen2}

\begin{figure}[t]
    \centering
    \includegraphics[width=.75\textwidth]{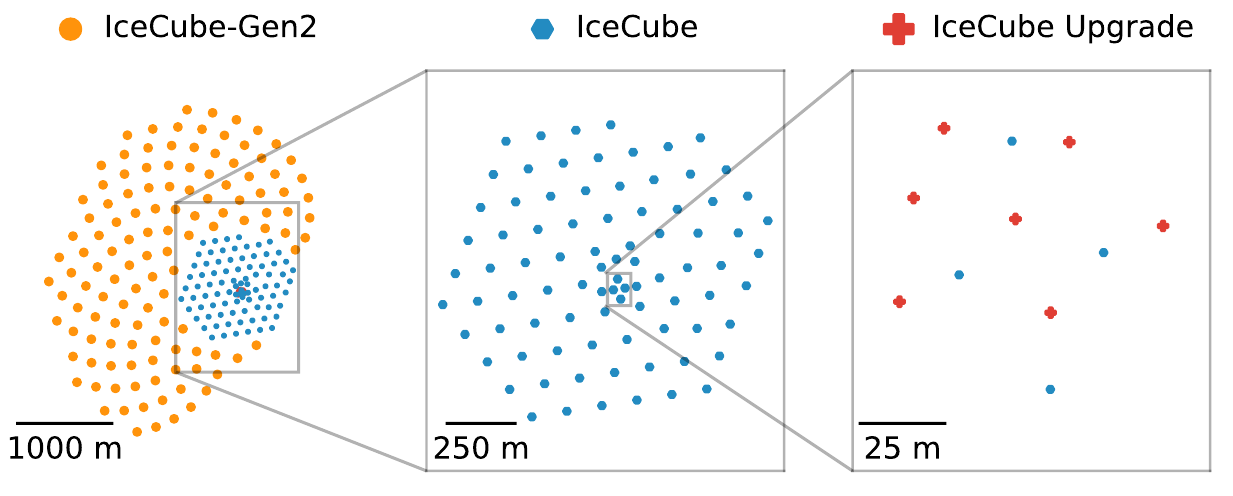}
    \caption{Top view of the IceCube optical detector arrays. {\it Left:} the IceCube-Gen2 optical array, in the reference ``sunflower'' design instrumenting 7.9~$\mathrm{km}^3$. {\it Center:} the current $\mathrm{km}^3$ IceCube array. {\it Right:} the low-energy IceCube Upgrade infill, focusing on neutrino oscillation properties \cite{Ishihara:2019aao}. The IceCube-Gen2 radio component (not shown) sparsely instruments an additional 500~$\mathrm{km}^2$ and is discussed in Sec.~\ref{sec:icecube_gen2_radio}.  Figure reproduced from Ref.~\cite{IceCube-Gen2:2020qha}.}
    \label{fig:ic_gen2_optical}
\end{figure}

IceCube-Gen2~\cite{Aartsen:2014njl, IceCube-Gen2:2020qha} is a planned next-generation expansion of the IceCube Neutrino Observatory at the South Pole. IceCube-Gen2 will observe the neutrino sky from TeV to EeV energies and consists of three sub-components: an optical Cherenkov detector focusing on the HE-neutrino energy regime; a large, sparse radio array extending neutrino detection to the UHE regime; and a hybrid surface detector for cosmic-ray air-shower detection and veto. We focus in this section on the design and capabilities of the optical array; the radio array is described in Sec.~\ref{sec:icecube_gen2_radio}. 

The IceCube-Gen2 optical array will instrument 8~$\mathrm{km}^3$ of ultra-clear ice near the geographic South Pole and include the existing IceCube optical array (Fig.~\ref{fig:ic_gen2_optical}). DOMs detect the Cherenkov light emitted from secondary charged particles created in neutrino interactions in the ice and bedrock. The IceCube-Gen2 DOMs will be deployed along 120 cables or ``strings'', with 80 DOMs on each string deployed between 1344~m and 2689~m below the surface. The string spacing, increased to 240~m from IceCube's 125~m, balances the increase in instrumented volume with the associated increase in energy threshold, while ensuring that angular resolution and calibration accuracy satisfy design requirements. The strings are arranged in a ``sunflower'' pattern to improve azimuthal homogeneity (removing lower-detection-efficiency corridors between straight rows of strings). Each IceCube-Gen2 DOM has approximately three times the photon collection of the original IceCube DOM, by employing a multi-photomultiplier-tube design.

IceCube-Gen2 will map the neutrino sky by observing an order of magnitude more neutrino events per year as IceCube, and with a sensitivity to individual neutrino sources at least 5 times better~\cite{IceCube-Gen2:2020qha}. The angular resolution of the optical array for track-like events is approximately 10~arcmin at PeV energies. IceCube-Gen2 can detect neutrino transients from over two times larger distances than IceCube, leading to a factor of 10 source detection volume increase for transient sources. Multi-messenger observations are an ongoing key science goal and will greatly profit from the up to five times higher neutrino alert rate and improved angular resolution when compared to IceCube. IceCube-Gen2's observation of the sky is energy-dependent, shifting from the Northern Hemisphere for energies below 100~TeV to the celestial equator and the Southern Hemisphere for PeV and higher energies. 


\subsubsection{Trinity}

Trinity is a proposed system of 18 air-shower Cherenkov telescopes optimized for detecting Earth-skimming neutrinos with energies between $10$\,PeV and $1000$\,PeV~\cite{Brown:2021lef, Wang:2021zkm}. Trinity is an evolution of the original Earth-skimming concept, which proposed monitoring nearby mountains with Cherenkov telescopes~\cite{Bertou:2001vm,Fargion:2003kn,Feng:2001ue} and has been tested and validated by several groups, for example, by NTA~\cite{Ogawa:2021irx} and MAGIC~\cite{MAGIC:2018gza}. 

\begin{figure}[t]
    \centering
    \includegraphics[width=.59\textwidth]{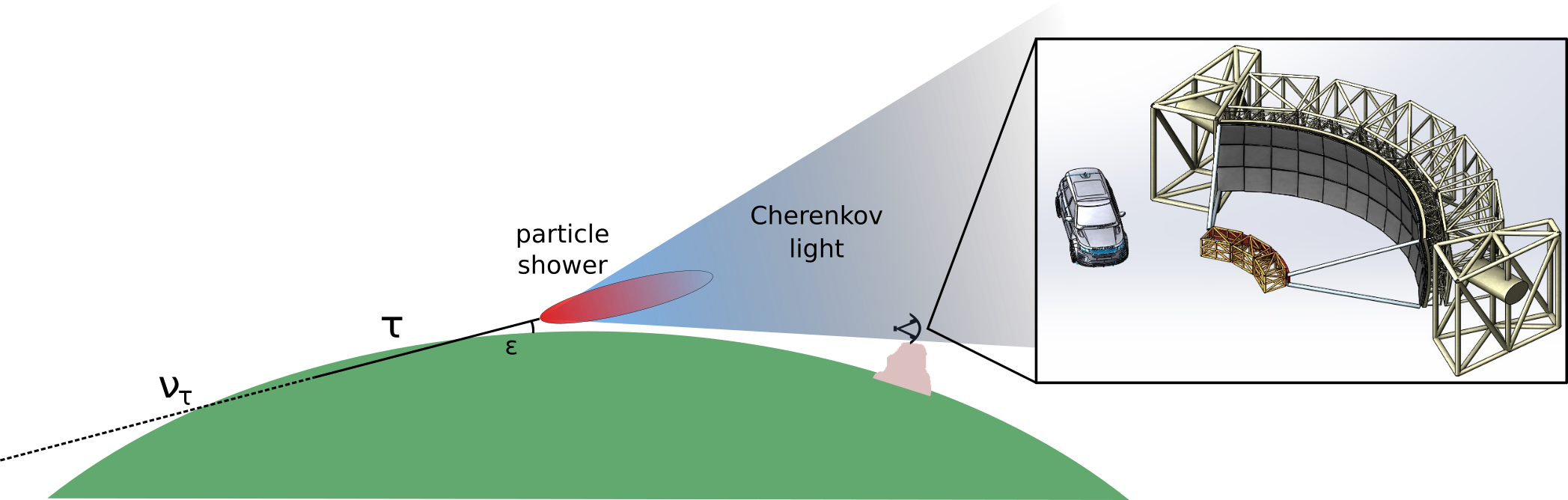}
     \includegraphics[width=.40\textwidth]{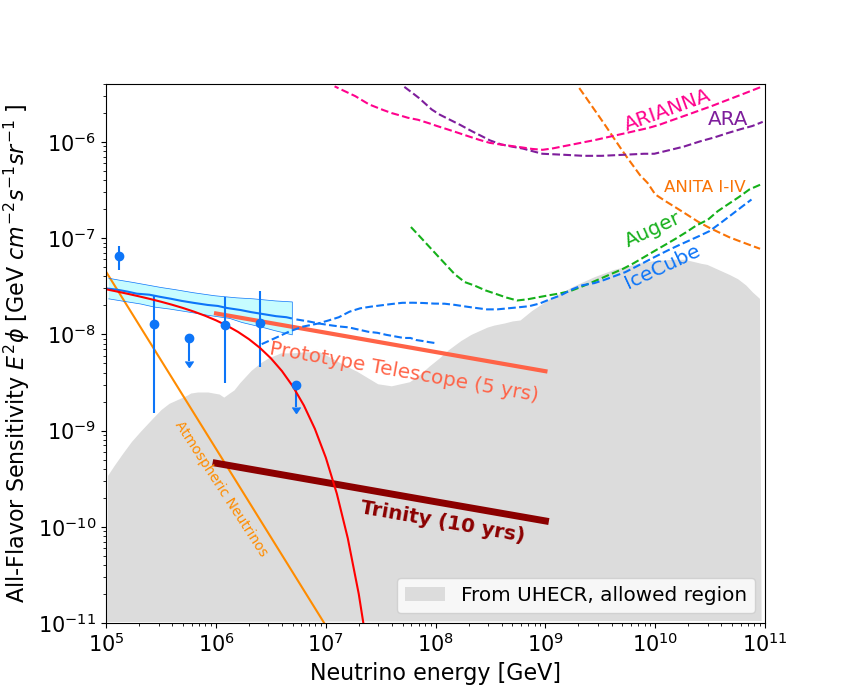}
    \caption{{\it Left:} Earth-skimming technique. The Trinity telescope images the air-shower, which develops in the atmosphere after the interaction of an Earth-skimming tau neutrino inside Earth. {\it Right:} The integral sensitivity of Trinity to diffuse neutrino fluxes for the prototype telescope in five years and 18 telescopes in ten years, respectively. Updated from Ref. \cite{Brown:2021ane}.}
    \label{fig:trinity}
\end{figure}   

Trinity's energy range overlaps with IceCube's due to Trinity's PeV threshold. The overlap enables studies of the astrophysical neutrino spectrum in regions of the sky not accessible with in-ice or atmospheric radio experiments (declinations from $-75^\circ$ to $55^\circ$~\cite{Wang:2021zkm}). That is because atmospheric radio becomes sensitive at $\sim10^{8}$\,GeV, and in-ice radio experiments, while sensitive down to $\sim10^{7}$\,GeV, have a limited sky acceptance due to their locations close to the poles. Trinity closes an important observational gap.
        
Trinity's telescopes will be located on mountains 2-3\,km above the surrounding ground and point at the horizon. At these altitudes, the telescopes can detect air-showers developing as far away as 200\,km~\cite{Otte:2018uxj}. The possibility of detecting these very distant showers compensates for the low 20\% duty cycle and therefore boosts Trinity's acceptance below $10^8$\,GeV when compared to other instruments. 
        
Trinity's telescopes are optimized to deliver the best possible detection sensitivity per cost \cite{Otte:2019aaf}. A key feature of the instrument is the extreme and unique 60-degree wide-field optics  (see Fig. \ref{fig:trinity})~\cite{Cortina:2015xra}. The telescopes use demonstrated technologies lowering costs and improving performance. The mirror technology, for example, has been demonstrated by the Cherenkov Telescope Array Consortium. High efficient, mechanical, and optical robust silicon PMTs populate the focal plane outperforming classical PMTs. The digitizer system has been developed for high-energy physics experiments focusing on low cost and high-channel density. These technologies make it possible to build a high-performance system for a fraction of the cost of a conventional Cherenkov telescope.

Trinity's sensitivity improves inverse-proportionally with the number of telescopes. One telescope alone can detect astrophysical neutrinos within five years if the spectrum extrapolates from IceCube energies without a cut-off (see Fig.~\ref{fig:trinity}).
        
Trinity is currently in its demonstrator phase~\cite{Brown:2021lef}, which constitutes building a $1\,$m$^2$ air-shower Cherenkov telescope that will be deployed on Frisco Peak, UT in 2022. The demonstrator will demonstrate the concept and camera technologies planned for a full Trinity telescope.


\subsubsection{RET}

The Radar Echo Telescope (RET) is a forthcoming observatory for detecting neutrinos with energies above $10^{16}$~eV using the radar echo method, where radio waves are reflected from ionization deposits left in the wake of high-energy particle cascades in dense material, such as ice. The concept has recently been demonstrated in the laboratory during experiment T576 at SLAC~\cite{Prohira:2019glh}, where a dense beam of high energy electrons was directed into a plastic target, to simulate a high-energy neutrino interaction in ice. The next step, testing the method in the field, is imminent, and will be discussed below.

\begin{figure}[t]
    \centering
    \includegraphics[width=.75\textwidth]{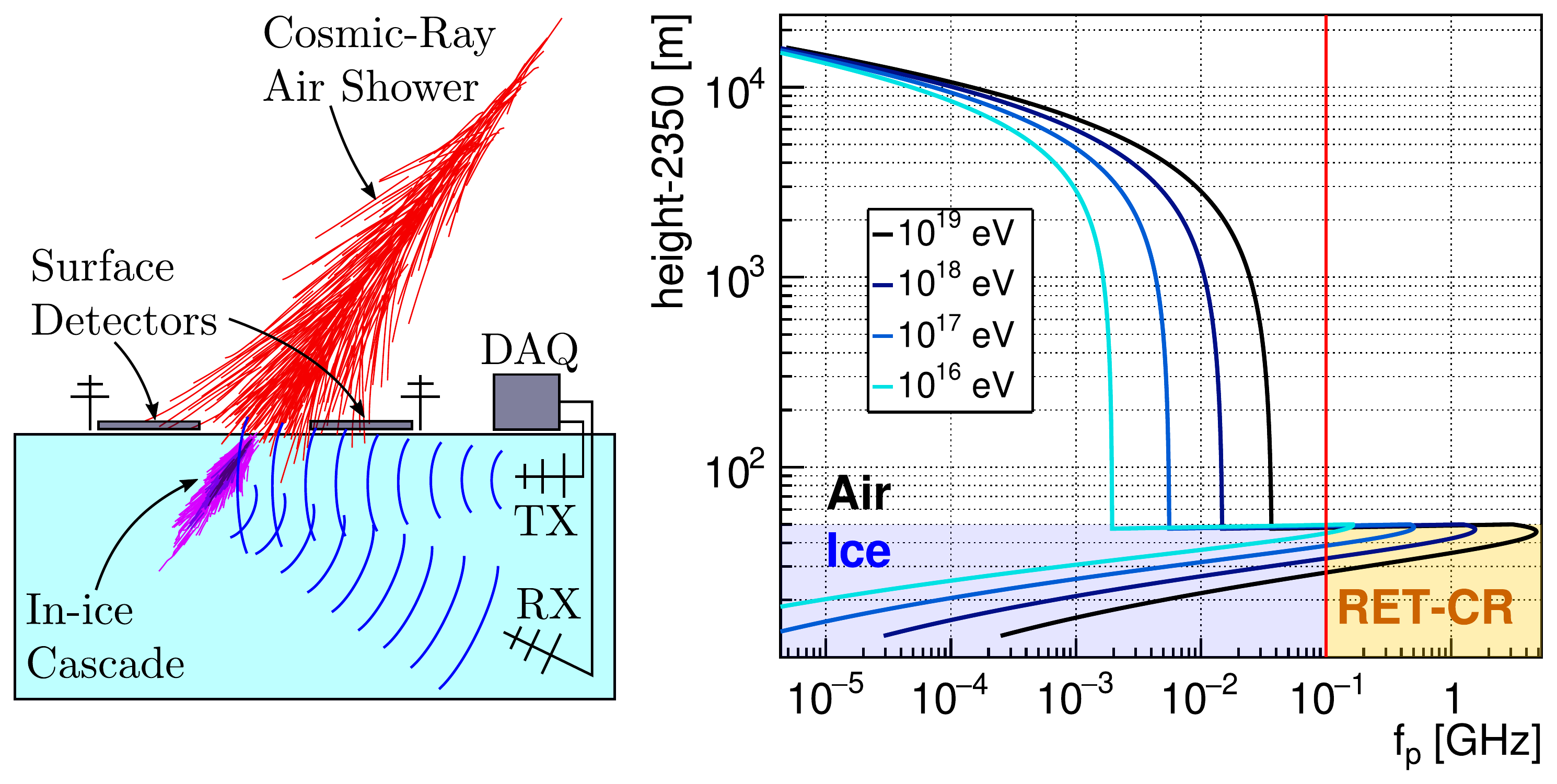}
    \caption{{\it Left:} The RET-CR concept. The signal from an in-ice transmitter is reflected from the dense in-ice cascade produced by a cosmic ray air shower, and detected by a receiving antenna. {\it Right:} The plasma frequency $f_p$ (a proxy for cascade ionization density) versus height for air showers of various energy. Here the 2.4\,km surface height is shown at 50\,m on the y axis, for clarity. The frequency range of interest to RET-CR is indicated by the vertical red line. }
    \label{fig:ret-cr}
\end{figure}

RET is an umbrella project for two instruments. One, the Radar Echo Telescope for Cosmic Rays (RET-CR)~\cite{RadarEchoTelescope:2021rca} is a prototype system designed to test the radar echo method in nature, shown diagrammatically in Fig.~\ref{fig:ret-cr}. To do this, RET-CR will use the ionized core of a cosmic ray shower---after it has impacted the ice---as a proxy for a neutrino-induced cascade. 
On a high-elevation ice sheet, 10\% or more of the energy of a primary cosmic ray will remain as the cascade impacts the ice. This energy is tightly collimated around the cascade axis, and produces a dense cascade just beneath the surface of the ice. RET-CR will deploy a phased transmitter and several phased receivers (see Refs.~\cite{Vieregg:2015baa, Allison:2018ynt} for more information on in-ice phased arrays) to detect the ionization left in the wake of these englacial cascades. A detected event rate of roughly 1 event per day with a primary energy at or above approximately $10^{17}$\,eV is expected. RET-CR is under construction at time of writing. 

The second system under the RET umbrella is the Radar Echo Telescope for Neutrinos (RET-N), a large-scale, in-ice neutrino detector capable of detecting neutrinos just above the reach of IceCube optical. A RET-N station is comprised of a central transmitter and an array of receivers on baselines of hundreds of meters, all buried $\sim$1.5\,km deep in the polar ice. RET-N is projected to have good sensitivity down to $10^{16}$\,eV. RET-N is particularly well-suited to flavor studies of UHE neutrinos. Because of the wide angular acceptance of the radar signal, so-called double-bang $\nu_{\tau}$ events~\cite{Learned:1994wg} are easily detected---even by a single receiving antenna---leading to an increase in acceptance for such events, providing a promising handle on flavor identification. Furthermore, as the only {\it active} method for detecting UHE neutrinos, it provides a different method of measuring properties of the cascade, such as direction, energy, and topology. As such, RET-N is complementary to the other radio- and optical-based techniques being explored to bring neutrino astronomy to the UHE regime. 


\subsubsection{TAMBO}

\begin{figure}[t]
    \centering
    \includegraphics[width=0.8\textwidth]{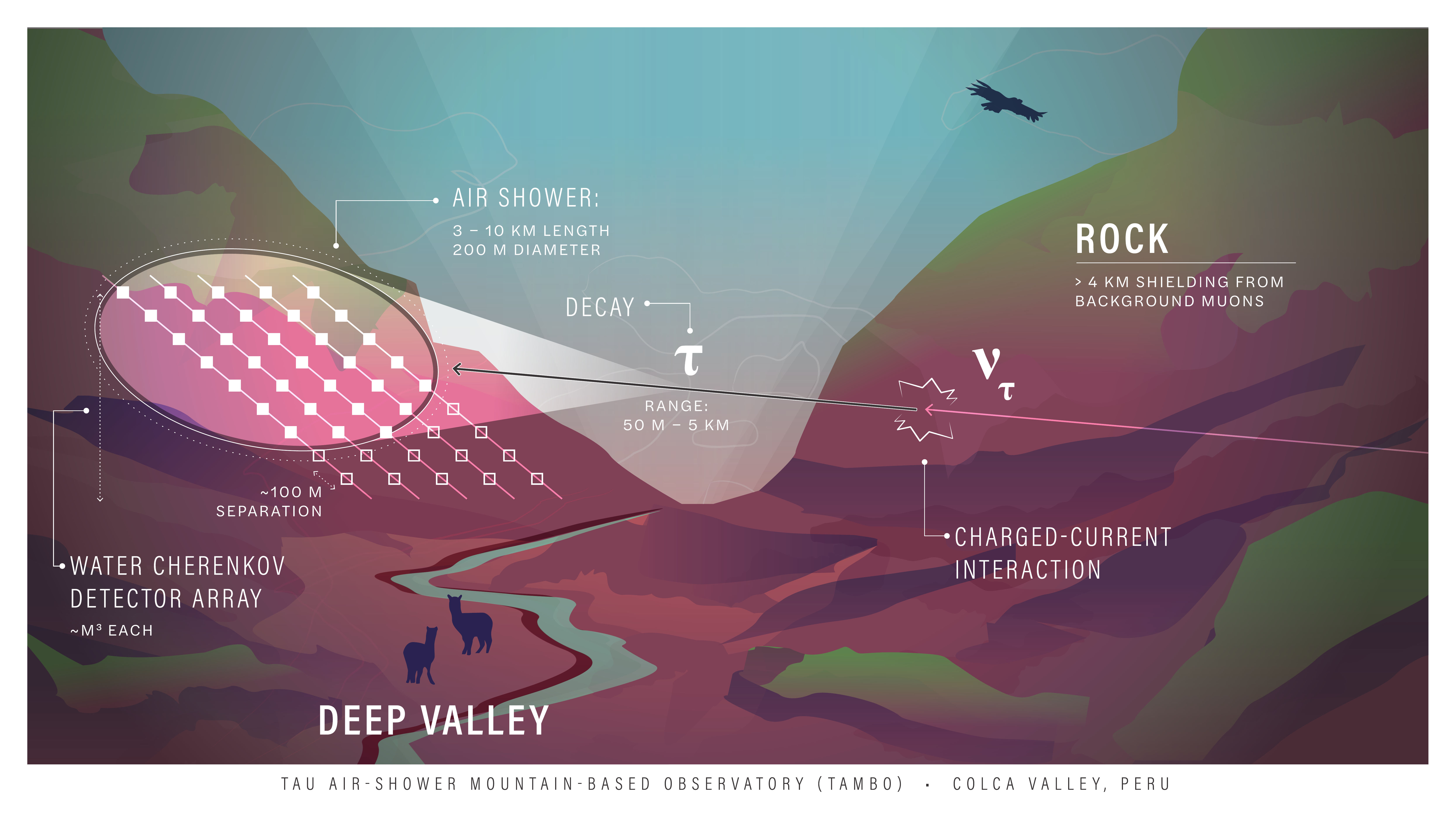}
    \caption{Detector concept schematic for TAMBO.
    A tau neutrino propagates through rock in the mountain.
    If a charged current interaction occurs, a tau is produced with a range of 50~m--5~km before it decays to produce an air shower.
    The electromagnetic component of the air shower extends 3--10~km , with a diameter of about 200~m.
    Such an air shower is detectable by an array of water-Cherenkov tanks, each a $m^{3}$ in volume, with 100~m separation. Figure reproduced from Ref.~\cite{Romero-Wolf:2020pzh}
    \label{fig:TAMBO}}
\end{figure}

TAMBO (Tau Air-Shower Mountain-Based Observatory) is a proposed detector in the Colca Valley in Peru to measure Earth-skimming astrophysical tau neutrinos in the 1~PeV to 100~PeV energy range 
(see Fig. \ref{fig:TAMBO}).
Tau neutrinos in that energy range can propagate through Earth with an interaction length of hundreds kilometers~\cite{Gandhi:1998ri}, which can produce tau particles via charged-current interactions.
The produced charged leptons decay within the range of 50~m to 5~km.
If the tau neutrino interaction occurs within this distance from the valley, the tau is likely to exit into the air, where it decays producing an air shower with a range of 3--10 km, and a diameter of around 200~m.
The air shower can then be detected by water Cherenkov tanks, plastic scintillator panels, or air-Cherenkov telescopes.
The preliminary design of TAMBO is based on water Cherenkov detectors, but plastic scintillators are also being studied as a complementary detection medium.

The geographical structure of the Colca valley makes it an ideal place for high-energy tau-neutrino detectors, such as TAMBO.
The separation between the slopes of the valley has a length comparable to the decay length scale of a tau.
This provides sufficient time for the tau particle to decay to form an extensive air-shower particles, which can eventually be detected by placing a ground array of small water-Cherenkov detectors, each of approximately of volume 1~m$^3$ and separated by $\sim$~100~m.
With this configuration and deploying 22,000 tanks, we expect that TAMBO will have an effective area ten times larger than IceCube's current tau neutrino effective area at $\sim$~3~PeV.
Compared to ongoing Earth-skimming experiments~\cite{Neto:2020zgu}, the deep-valley topography also provides a significant increase to the geometric acceptance compared to a flat ground array.
Similarly, the detector location is in the Southern hemisphere, allowing for observation of the Galactic Center and a wide variety of standard astrophysical sources~\cite{IceCube:2017trr, Kheirandish:2020zll} which also holds the prospect of detecting neutrinos from dark matter annihilation~\cite{Arguelles:2019ouk}.
Additionally, the smaller tank-to-tank distance significantly lowers the energy threshold, thereby increasing the rate of detected events.
For an $E^{-2.5}$ spectra with the current IceCube best-fit normalization, we expect to detect 7 tau-neutrino events per year over a small background predominantly due to coincident cosmic-ray air showers.

TAMBO will characterize the flux of astrophysical 1--10~PeV neutrinos by measuring the contribution from the tau neutrino component.
This would allow a better understanding of high-energy neutrino production~\cite{Ackermann:2019ows, Bustamante:2019sdb} and to more precisely test high-energy neutrino physics~\cite{Anchordoqui:2004eb, Arguelles:2015dca, Bustamante:2015waa, Rasmussen:2017ert, Arguelles:2019rbn, Arguelles:2019tum}.
Similarly, it would also help determine whether high-energy neutrino sources continue to accelerate particles above 10~PeV, testing some studies~\cite{IceCube:2017zho} that question whether sources of high-energy neutrinos have a cutoff at $\sim$~6~PeV.


\subsection{Ultra-high-energy range ($>100$ PeV)}
\label{subsec:landscape_100_pev}

Several experiments search for radio emission from neutrino interactions using embedded radio arrays. Taken together, they serve as complementary test benches for the future radio array of the planned IceCube-Gen2. 

\begin{figure}[t]
    \centering
    \includegraphics[width=0.8\textwidth]{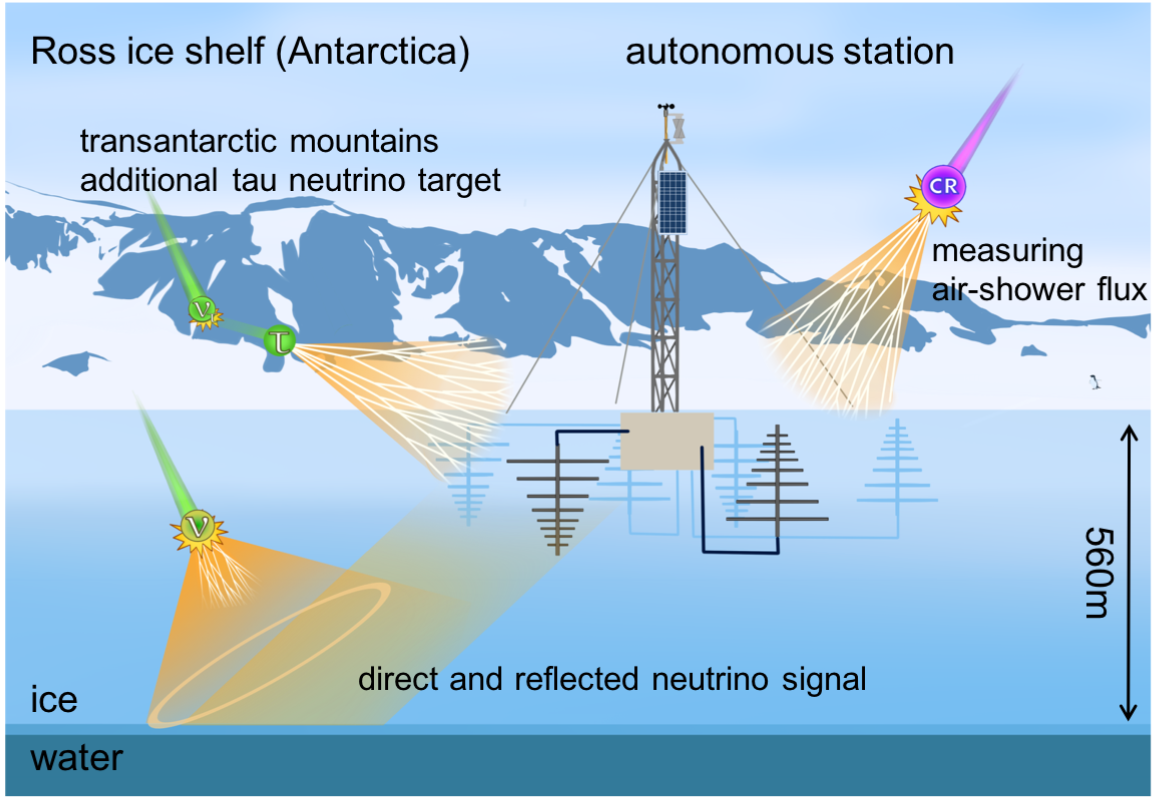}
    \caption{Schematic of ARIANNA stations at the Ross Ice Shelf. Figure reproduced from Ref.~\cite{ARIANNA:2019scz}}.
    \label{fig:arianna}
\end{figure}


\subsubsection{ARIANNA} 

The ARIANNA experiment uses the in-ice radio technique to search for UHE neutrinos~\cite{ARIANNA:2019scz}. In a uniquely radio quiet area on the Ross Ice Shelf in Antarctica, a hexagonal array of pilot-stations has been taking data for several years. In addition, two detector stations have been installed at the South Pole. The station design is shown schematically in Fig.~\ref{fig:arianna}.  
It deployed its first prototype station in 2009  \cite{Gerhardt:2010js} and has been operating a demonstration array since 2014~\cite{ARIANNA:2019scz}. Each station uses high gain antennas in both a cosmic ray and veto mode (upward pointing) and in neutrino mode (downward pointing)  a few meters below the ice surface. Limits on the UHE neutrino flux from ARIANNA demonstrate the feasibility of the in-ice radio detection technique~\cite{Arianna:2021vcx}. The stations provide a continuous test bench for optimization of future larger arrays like IceCube-Gen2~\cite{Glaser:2020pot, Arianna:2021vcx} as well as reconstruction and veto methods~\cite{Barwick:2016mxm, Anker:2019zcx, Glaser:2019rxw, Welling:2019scz, Gaswint:2020mmx,  Nelles:2020fow, ARIANNA:2020zrg, Glaser:2021jss, Stjarnholm:2021xpj, Gaswint:2021smu, ARIANNA:2021pzm, Arianna:2021lnr}. 


\subsubsection{ARA}

\begin{figure}[t]
    \centering
    \includegraphics[width=0.9\textwidth]{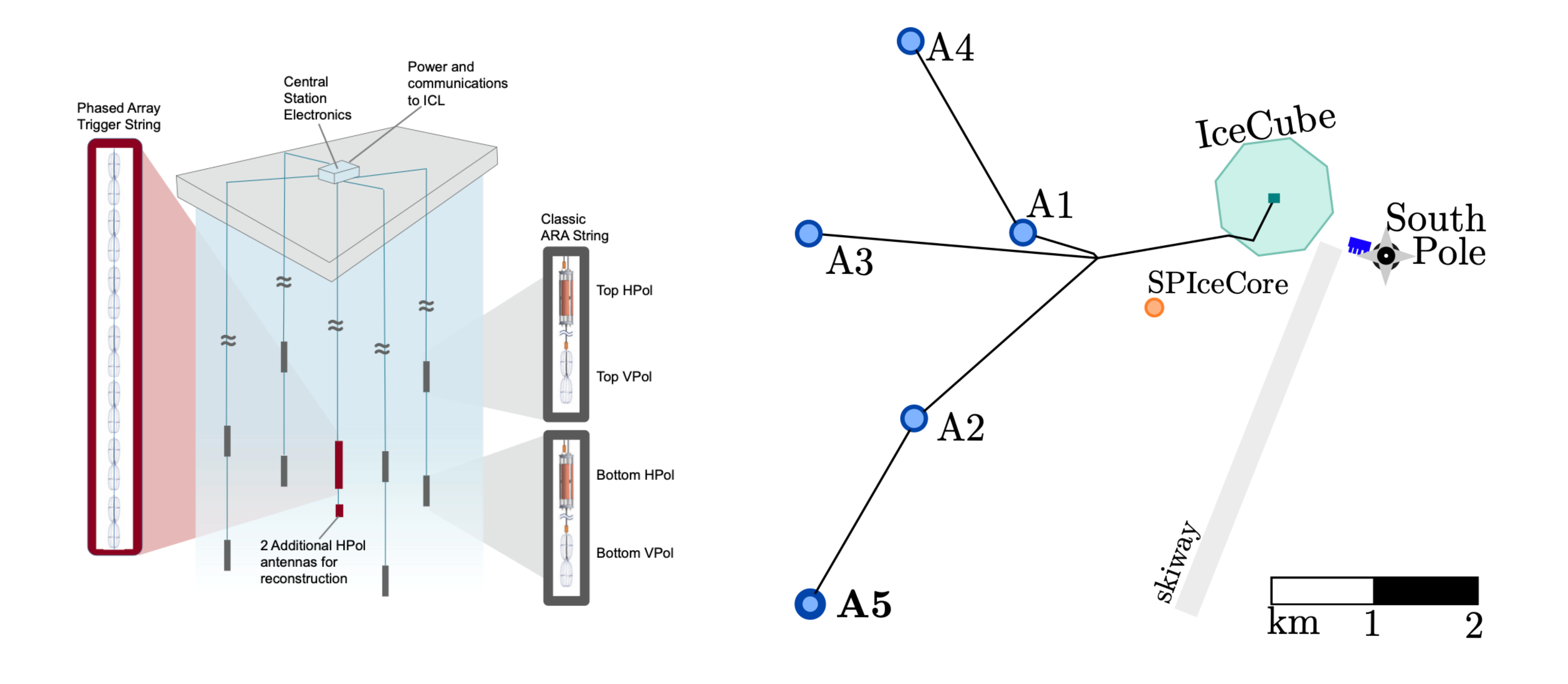}
    \caption{Schematic and map ARA stations at the South Pole. Figure reproduced from Ref.~\cite{Allison:2022lut}}.
    \label{fig:ARA_diagram}
\end{figure}

The Askaryan Radio Array (ARA) targets the UHE neutrino flux using the in-ice radio technique~\cite{Allison:2011wk}. An overview of the experiment is provided in Fig.~\ref{fig:ARA_diagram}. Located near the South Pole in Antarctica, it takes advantage of the deep glacial ice there and the long radio attenuation lengths to search for Askaryan emission from neutrino showers.

In ARA, clusters of radio antennas are deployed 200~m deep below the surface of the ice to access the larger volume of ice visible as the antennas are embedded in denser, glacial ice. 
Each of five ARA stations uses 16 cylindrical antennas to form an interferometer.
The deep antenna deployments at 200\,m give the array a wide field of view, stretching from $\sim \SI{5}{\degree}$ below the horizon to $\sim \SI{45}{\degree}$ above it. Powered by a DC electric grid, ARA continuously monitors almost one fourth of the sky.
One station includes a novel trigger that has been demonstrated to lower the trigger threshold of in-ice Askaryan detectors, thereby lowering the energy threshold and the overall sensitivity of in-ice radio detectors by up to a factor of 2 near 30\,PeV~\cite{Allison:2018ynt, Allison:2022lut}. 
A series of experimental neutrino searches and resultant flux upper limits demonstrate the scalability of the in-ice radio technique~\cite{Allison:2014kha, Allison:2015eky, ARA:2019wcf, Allison:2022lut}, and the ability of the detector to both trigger and analyze with high efficiency ($>50$\%) very low signal to noise ratio data (SNR$\sim 2$), which is essential to achieve the proposed sensitivity of next generation arrays like IceCube-Gen2. ARA anticipates world leading sensitivity to the UHE neutrino flux above 1~EeV by 2023 ~\cite{ARA:2019wcf}.

Besides searching for neutrinos, ARA has also made a series of measurements which demonstrate the feasibility of the deep technology and the suitability of the South Pole experimental site. Both of these are critical for next generation instruments like IceCube-Gen2. By measuring pulsers deployed on the bottom of IceCube, ARA made the longest horizontal-baseline measurement of the South Pole ice attenuation, confirming the exceptional clarity ($>\SI{1}{km}$ attenuation length) of the ice~\cite{Allison:2019rgg}. Using the same pulsers, ARA demonstrated the vertex reconstruction capability of the deep technology, confirming that when a station observes both a direct and reflected pulse, a neutrino vertex can be located to within \SI{1}{\degree} in zenith and azimuth, and within $\sim 30\%$ in range  ~\cite{Allison:2017jpy}. This reconstruction capability is essential to the energy resolution of ARA and future radio instruments. ARA data has also revealed potentially novel properties of the Antarctic ice such as birefringence, which mixes the polarization states of a propagating radio signal~\cite{Allison:2017jpy}. Understanding such properties of the medium are critical to neutrino directional reconstruction~\cite{Connolly:2021cum}.


\subsubsection{RNO-G}

\begin{figure}[t]
    \centering
    \includegraphics[width=0.8\textwidth]{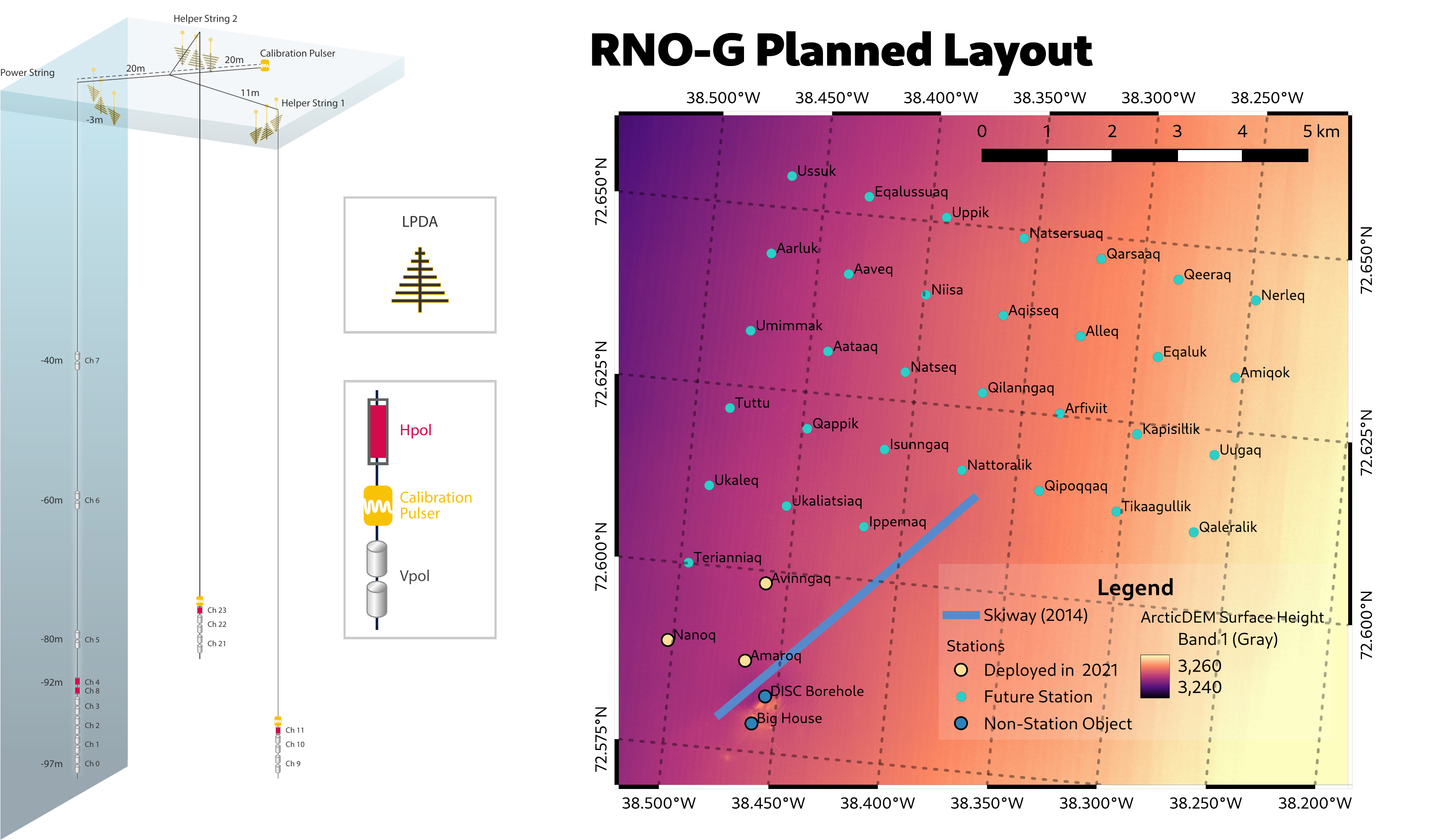}
    \caption{Schematic and map of RNO-G stations being installed at Summit Station in Greenland. Figure reproduced from Ref.~\cite{RNO-G:2021hfx}.}.
    \label{fig:schematic_RNO}
\end{figure}

The Radio Neutrino Observatory in Greenland (RNO-G) is an in-ice detector that measures neutrinos through the Askaryan emission generated by in-ice showers \cite{RNO-G:2020rmc,RNO-G:2021hfx}. The 3~km deep ice sheet above central Greenland with attenuation length of approximately 1~km at the relevant frequencies of 100~MHz to 1~GHz provides a good target material for achieving large effective volumes. The ice will be instrumented with a sparse array of 35 autonomous radio detector stations with a separation of 1.5~km. 
A schematic and map of the RNO-G stations are shown in Fig. \ref{fig:schematic_RNO}.
The stations are solar-powered with additional wind generators under development to power the stations during the dark winter months. The stations are connected through an LTE network to Summit station. The first three stations have been installed in 2021, the remaining stations will be installed over the next three years. 

Each station is equipped with total of 24 antennas. An interferometric phased array provides a low-threshold trigger, consisting of 4 bicone antennas installed in close proximity vertically above each other at a depth of approximately 100~m~\cite{Allison:2018ynt}. Additional bicone (vertical signal polarization) and quad-slot (horizontal signal polarization) antennas above the phased array and horizontally displaced on two additional strings provide additional information to reconstruct the properties of the neutrino~\cite{Aguilar:2021uzt}. LPDA antennas are installed close to the surface providing additional neutrino sensitivity to neutrinos with complementary uncertainties. Each station also comprises three upward facing LPDAs to veto and measure radio emission of air showers which provide in-situ calibration signals~\cite{Arianna:2021lnr}. 

Due to its relatively low latitude of $72^\circ$ North, RNO-G observes the majority of the Northern sky within 24~hours adding UHE neutrino information to multi-messenger observations. Its diffuse flux sensitivity is large enough to start probing the parameter space of GZK neutrino production (see Fig. \ref{fig:diffuse_sensitivity}). Furthermore, RNO-G will be a technical testbed and pave the way for the much larger radio detector array foreseen for IceCube-Gen2.


\subsubsection{\label{sec:icecube_gen2_radio}IceCube-Gen2 radio}

\begin{figure}[t]
    \centering
    \includegraphics[width=\textwidth]{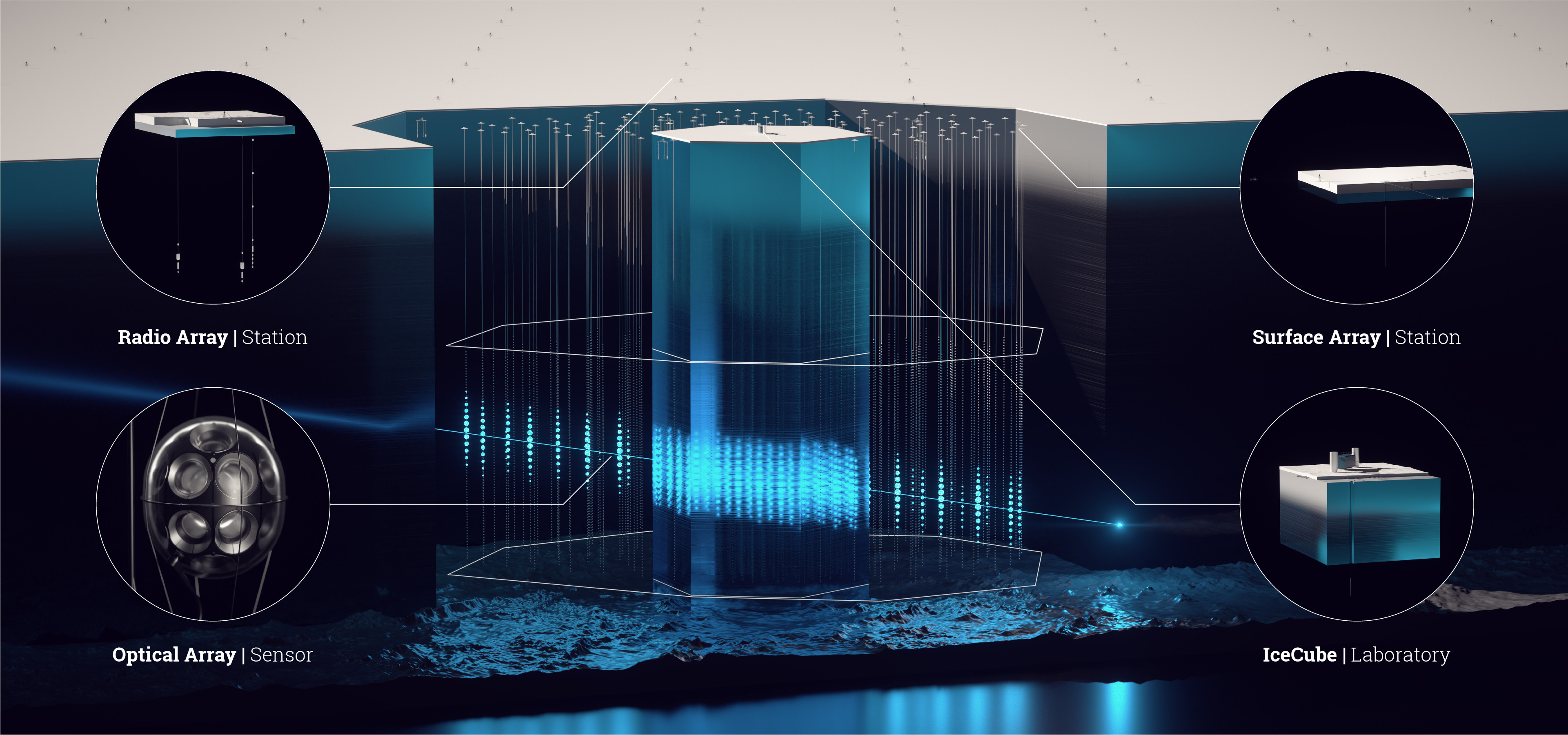}
    \caption{Rendering of the proposed IceCube-Gen2 neutrino observatory combining an optical, surface and radio array.}
    \label{fig:Gen2_overview}
\end{figure}

To extend the energy reach to EeV energies, IceCube-Gen2 will include a sparse array of radio detector stations next to its optical component \cite{IceCube-Gen2:2020qha} 
(see Fig. \ref{fig:Gen2_overview}). 
The radio technique allows for a cost-efficient instrumentation of the large volumes required to measure the low flux of UHE neutrinos. The ice at the South Pole provides an optimal target material for radio detection with attenuation lengths of more than \SI{2}{km} close to the surface where the ice is coldest. 
The radio array will cover an area of approximately \SI{500}{km^2}. The proposed array consists of two types of radio detector stations that measure and reconstruct neutrino properties with complementary uncertainties to maximize the discovery potential by mitigating risks and adding multiple handles for rare background rejection~\cite{IceCube-Gen2:2021rkf}. Hybrid stations combine omni-directional cylindrical antennas (both vertical and horizontal polarization) lowered 150~m below the surface of the ice with high-gain antennas near the surface.  The hybrid stations build on the heritage of the deep stations explored by ARA, the shallow stations being explored by ARIANNA, and the hybrid stations currently being deployed for RNO-G. The array design also uses additional shallow-only stations that use LPDA antennas close to the surface with one additional dipole antenna at \SI{15}{m} to aid event reconstruction. All shallow components are also equipped with upward facing LPDA antennas which provide sensitivity to cosmic rays to veto air-shower~induced background~\cite{Garcia-Fernandez:2020dhb, Glaser:2021hfi}, as well as to monitor the detector performance (see, e.g., Ref.~\cite{Arianna:2021lnr}).

The IceCube-Gen2 radio array will provide sufficient sensitivity to probe GZK neutrino production. 
The Gen2 sensitivity would reach the current best-fit models to  
UHECR data measured by the Pierre Auger Observatory, assuming sources identical in UHECR luminosity, spectrum and composition, 
as well as a rigidity-dependent cut-off and thereby essentially no protons at the
highest-energies~\cite{AlvesBatista:2018zui, Heinze:2019jou}. 
In an only slightly more favorable scenario of 10\% protons, IceCube-Gen2 will detect at least 2 events per year above $\sim$100 PeV~\cite{IceCube-Gen2:2021rkf}. 

For an unbroken astrophysical neutrino spectrum that follows $E^{-2.28}$, as the one shown
in Fig.~\ref{fig:diffuse_sensitivity}, the radio detector of Gen2 will measure close to ten neutrinos per year where most detected neutrinos will have energies between \SI{e17}{eV} and \SI{e18}{eV}~\cite{IceCube-Gen2:2021rkf}.
Due to its location at the South Pole, the instrument continuously observes the same part of the sky with most sensitivity between $\delta \approx \SI{-40}{\degree}$ and $\delta \approx \SI{0}{\degree}$ as the Earth is opaque to neutrinos at ultra-high energies. The instantaneous sensitivity will allow to explore neutrino production in transient events such as neutron-star mergers. 

The large sensitivity to neutrinos arriving from and slightly below the horizon enables the measurement of the neutrino-nucleon cross section at extremely high energies~\cite{Connolly:2011vc, Denton:2020jft, Huang:2021mki, Valera:2022ylt, Esteban:2022uuw}. Also a measurement of the inelasticity seems possible through the detection of high-quality electron neutrino charge-current interaction where the energy of the hadronic shower induced by the breakup of the nucleus can be measured separately from the electromagnetic shower induced by the electron. 
Furthermore, the production of high-energy muons in air showers with energy above a PeV can be probed through a coincident measurement of the air shower via the in-air radio emission and a muon induced particle shower in the ice via the Askaryan emission. 


\subsubsection{BEACON}

\begin{figure}[t]
    \centering
    \includegraphics[height=0.32\textwidth]{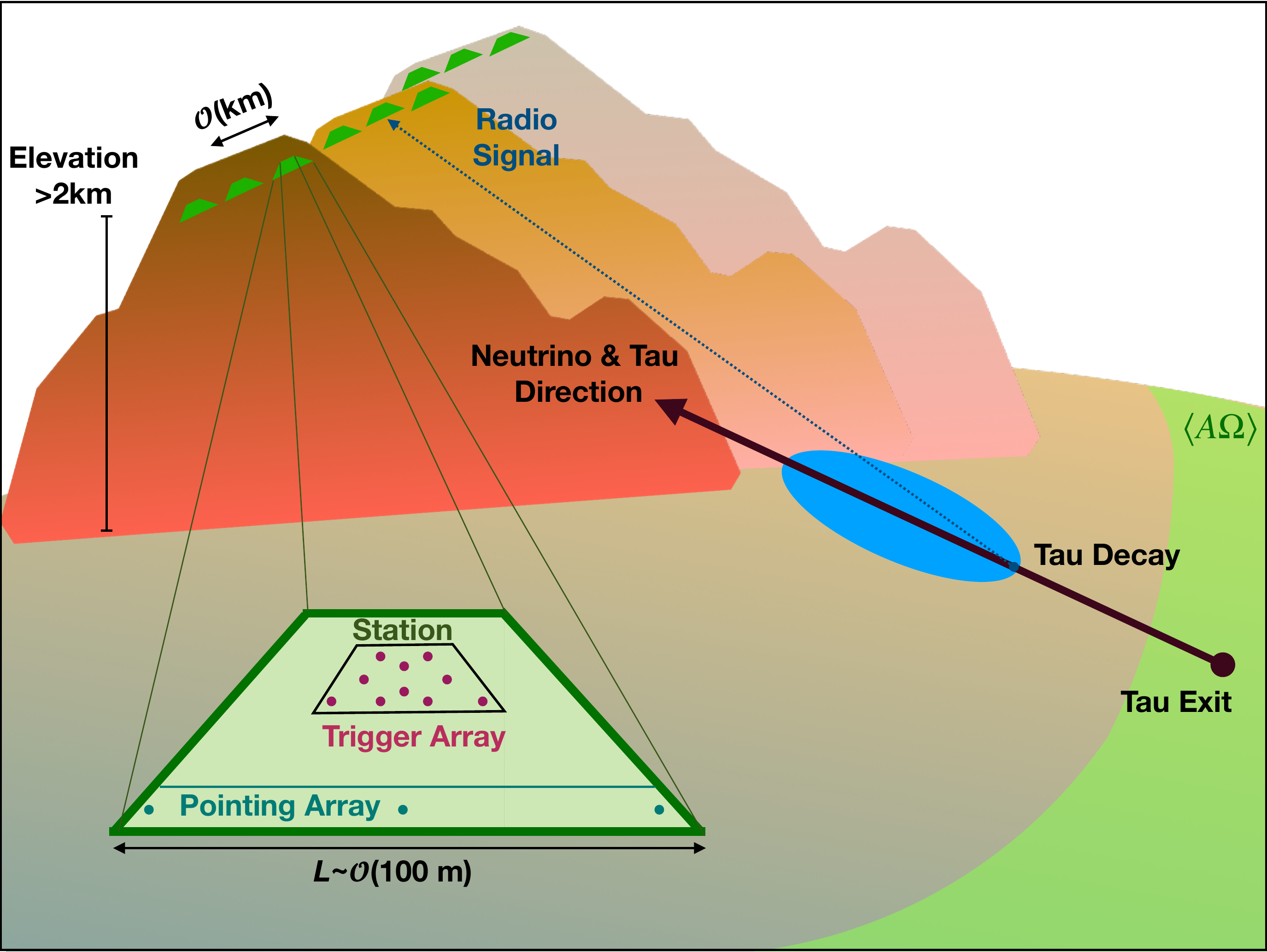}
    \includegraphics[height=0.32\textwidth]{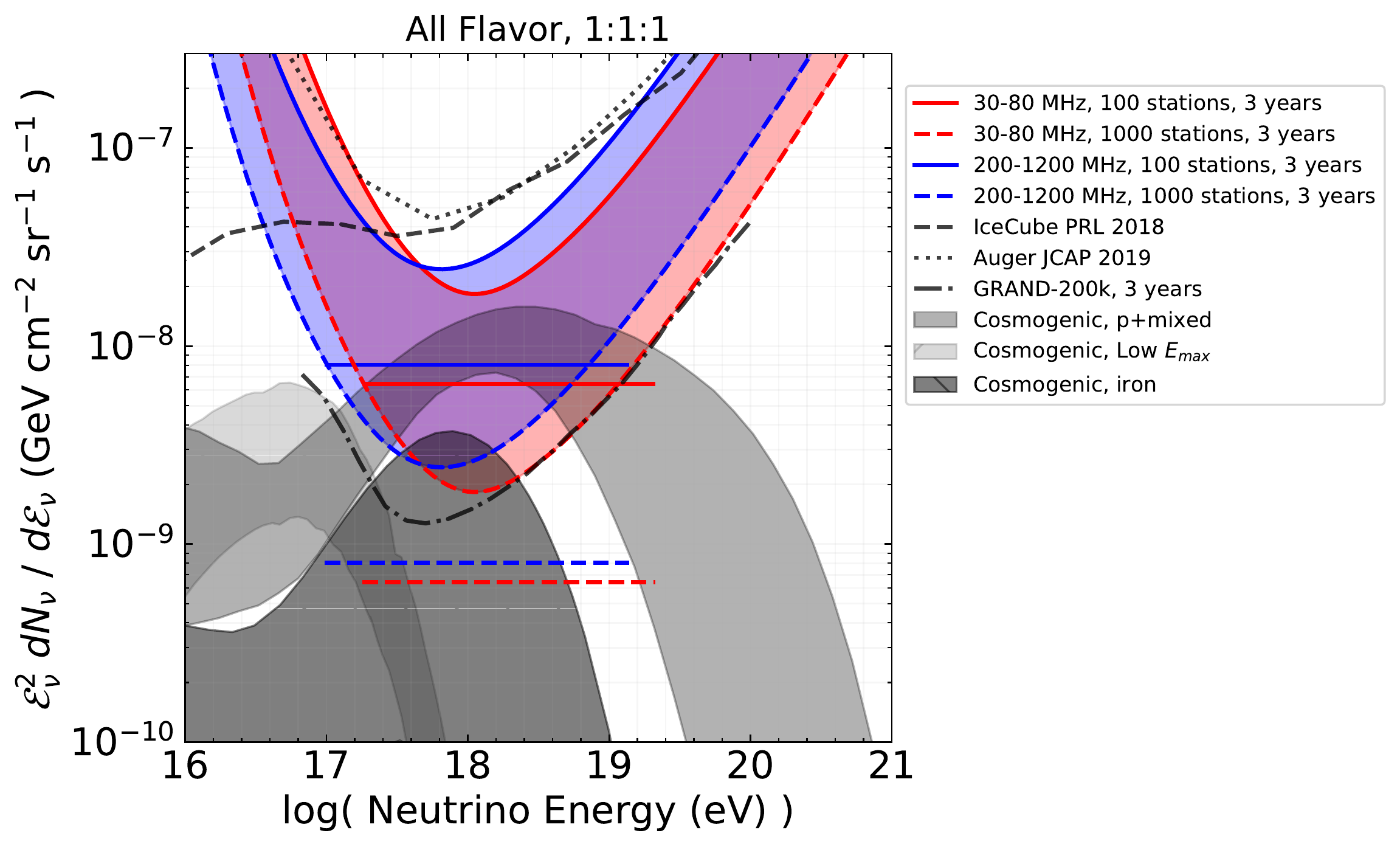}
    \caption{BEACON observes up-going tau neutrinos from a high-elevation mountain. The detection concept is shown on the left while the sensitivity to an all-flavor flux (assuming a 1:1:1 flavor ratio, half-decade energy bins, and a three-year exposure) is shown on the right. The expected 90\% confidence upper limit on the differential flux for 100 stations and 1000 stations of both the low- and high-frequency reference designs are compared with models of the cosmogenic flux~\cite{Kotera:2010yn, Romero-Wolf:2017xqe, Kampert:2012mx} and upper limits from Auger~\cite{PierreAuger:2019ens}, IceCube~\cite{IceCube:2018fhm}, and the proposed GRAND-200k experiment~\cite{GRAND:2018iaj}. Reproduced from Ref.~\cite{Wissel:2020sec}.}
    \label{fig:beacon}
\end{figure}

The Beamforming Elevated Array for COsmic Neutrinos (BEACON) is a detector concept that targets Earth-skimming tau neutrinos. BEACON uses the radio technique to search for air showers induced by tau lepton decay, and will be deployed at several sites around the world for a full sky coverage~\cite{Wissel:2020sec}. Because the BEACON concept relies on low power ($<50$~W), inexpensive, low channel-count instrumentation, the technique is scalable to many stations that can be deployed at one or at many locations around the world. 

BEACON is an efficient design that builds a large neutrino volume by fully exploiting the mountain geometry and by using interferometric phased arrays~\cite{Wissel:2020sec}. The design, shown schematically in Fig.~\ref{fig:beacon}, uses the high-elevation of a mountain to monitor a large area of the ground for up-going tau neutrinos fully exploits the topography at a site to maximize the neutrino interaction volume with sparse instrumentation. Stations consist of a clustered phased array used for triggering and antennas placed at longer baselines for improved pointing resolution (sub-degree). Phased arrays are advantageous in this geometry, because they both improve sensitivity and can be tuned to focus in on the horizon where most of the tau neutrinos are expected. Air showers are expected to be observable up to 100~km away from the detector, making mountains with 2--3~km prominence ideal sites to search for up-going tau neutrinos.

Within 3 years of observations with a full-scale instrument consisting of 1000 stations, BEACON is expected to achieve a sensitivity comparable to pessimistic models of cosmogenic neutrinos, but even with a smaller array of 100 stations, the large effective area will improve upon existing limits within 3 years of observations~\cite{Wissel:2020sec} The sensitivity to a diffuse flux is shown on the right in Fig.~\ref{fig:beacon}. The chosen frequency band is 30--80~MHz, but sensitivity studies have shown that a higher frequency band, for instance $200-1200$\,MHz could also be suitable~\cite{Wissel:2020sec}. This leaves the possibility to deploy hybrid frequency-band arrays in order to combine them to improve the pointing and the reconstructions capabilities.
 
The BEACON experiment is currently in the demonstration phase, with a prototype deployed at the White Mountain Research Station of California. The prototype array consists of 4 dual-polarized electrically-short-dipole antennas and a two stages of amplification. The trigger system is based on a low-power phased array trigger implemented on an FPGA. While the prototype electronics were originally  developed for the ARA experiment~\cite{Allison:2018ynt}, custom modular electronics are under design for future stations. 
 
While the prototype is too small to search for tau neutrinos, the technique can be tested thoroughly through searches for cosmic rays at high zenith angles~\cite{Zeolla:2021cbb}. A cosmic-ray search is underway currently~\cite{BEACON:2021fpe}, and with the observed flux, the projected neutrino sensitivity will be trained on the measured trigger threshold \textit{in situ}. With this validation of instrument sensitivity, the experiment can be readily expanded to the full array making it competitive to detect UHE neutrinos~\cite{AlvesBatista:2018zui}.


\subsubsection{GRAND}

\begin{figure}[t]
    \centering
    \includegraphics[width=0.98\linewidth]{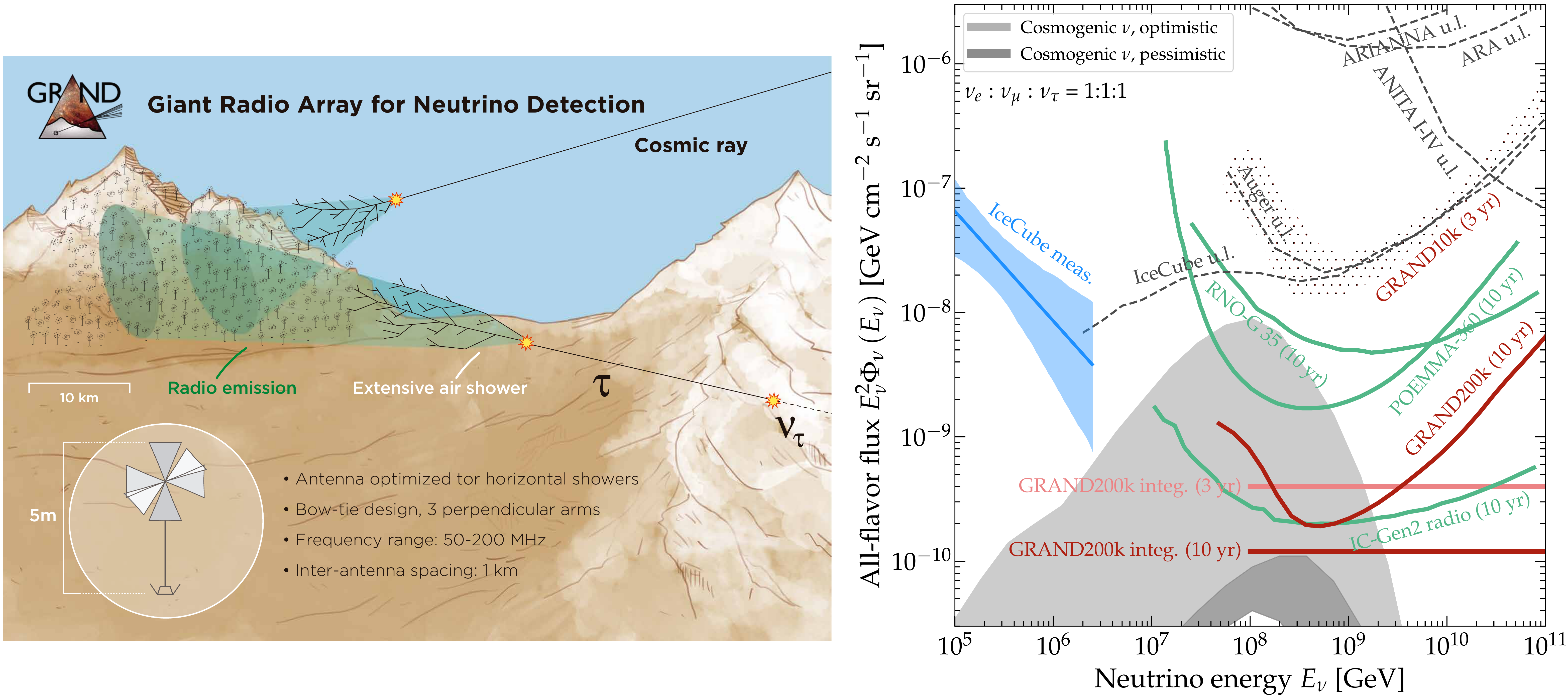}
    \caption{{\it Left:} The GRAND detection principle relies on the observation of EAS induced by both the direct interaction of the cosmic particles in the atmosphere (e.g., cosmic rays and gamma rays), and neutrino underground interaction with subsequent decay of the tau lepton in the atmosphere. The GRAND concept takes advantage of the detector topography in order to increase the detection efficiency~\cite{Decoene:2019izl}, and the reconstruction performances~\cite{Decoene:2021ncf}. {\it Right:} The differential and integrated neutrino sensitivity limits derived from the 10, 000 antennas simulation (GRAND10k, pink area) and its extrapolation to the 20-times larger GRAND array (GRAND200k, maroon line). The gray region represents the expectations for the all-flavor cosmogenic neutrinos flux obtained from the results of the Pierre Auger Observatory~\cite{AlvesBatista:2018zui}. Taken from~\cite{GRAND:2018iaj}.}
    \label{fig:grand}
\end{figure}

The Giant Radio Array for Neutrino Detection (GRAND) is a planned observatory for UHE particles, including neutrinos, cosmic rays, and gamma rays. The instrument is a large-scale radio array sensitive to  the geomagnetic radio emission air showers in the atmosphere. In particular, GRAND targets UHE neutrinos by searching for very inclined showers, i.e., showers coming from directions close to the horizon, expected from Earth-skimming UHE tau neutrinos~\cite{GRAND:2018iaj}.
Figure~\ref{fig:grand} shows the GRAND concept and the differential and integrated neutrino sensitivity limits for several antenna configurations.
The design targets diffuse flux sensitivities predicted by pessimistic models of UHE neutrino flux down to $\sim10^{-10}$\,GeV cm$^{-2}$ s$^{-1}$ sr$^{-1}$.

To reach this target sensitivity, GRAND will instrument a large $200000$\,km$^2$ area with clusters of radio antennas. Each of the 20 clusters will use 10000 radio antennas to monitor the horizon for emerging tau neutrinos. Each antenna operates in the 50--200~MHz band and the local topology will be analyzed to determine the optimal locations of the antennas within the clusters and the array~\cite{Decoene:2019izl} both for improved sensitivity and to enable sub-degree angular resolution~\cite{Decoene:2021ncf,GRAND:2018iaj}. With such precision angular resolution, GRAND will be able to to distinguish between neutrinos and the more prevalent cosmic rays, as the neutrino showers are horizontal or slightly upward-going while cosmic-ray showers are more downward-going. Moreover, GRAND will be an important instrument in transient time-domain radio astronomy, because of its large number of antennas and field of view.

The radio-detection of extensive air showers~\cite{Ardouin:2010gz} has been demonstrated by prior experiments and GRAND builds on this heritage. However, scaling to a large array requires a staged approach to further refine the design. The first prototype array, GRANDProto300, will use 300 antennas to validate autonomous radio detection and reconstruction of extensive air showers. This large array will also enable studies of other UHE particles including cosmic rays and photons as well as radio astronomy~\cite{Decoene:2019sgx}. The second prototype stage will scale up to 1000 antennas with GRAND10k, building on the design from the 300-antenna array and will explore the feasibility of the radio detection with large-scale and sparse arrays. GRAND10k is expected to reach a diffuse flux sensitivity of $8 \times 10^{-9}$\,GeV cm$^{-2}$ s$^{-1}$ sr$^{-1}$ after 3 years~\cite{GRAND:2018iaj}, constraining or discovering UHE neutrinos expected from optimistic models~\cite{AlvesBatista:2018zui, Rodrigues:2020pli}. With this staged approach, the design can be refined as the instrument scales up to the full 200000 antennas while simultaneously addressing key science questions in physics and astrophysics.


\subsubsection{POEMMA} 

The Probe Of Extreme Multi-Messenger Astrophysics (POEMMA) is a NASA Astrophysics Probe-class, space-based mission designed to perform definitive measurements of Ultra-High Energy Cosmic Rays (UHECRs) and searches for cosmic neutrinos. POEMMA was developed as one of the examples of NASA-funded missions to support the definition of the NASA Probe-class, and it is a potential candidate for a future NASA Probe Announcement of Opportunity. 
POEMMA utilizes the vast amount of atmosphere and Earth viewed using wide Field-of-View (FoV) as neutrino targets and measures the optical signals from the extensive air showers (EAS) from the products of the neutrino interactions \cite{POEMMA:2020ykm}. Over a five-year period, POEMMA is anticipated to measure $\sim 1,400$ UHECRs above 40~EeV and either detect the first neutrino with $E_{\nu}>20$~PeV for tau neutrinos and $E_{\nu}>20$~EeV for any flavor of neutrinos or set limits on the diffuse flux and stringent limits on transient source models with significant neutrino flux above 20~PeV.

\begin{figure}[t]
\begin{center}
   \includegraphics[width=0.65\textwidth]{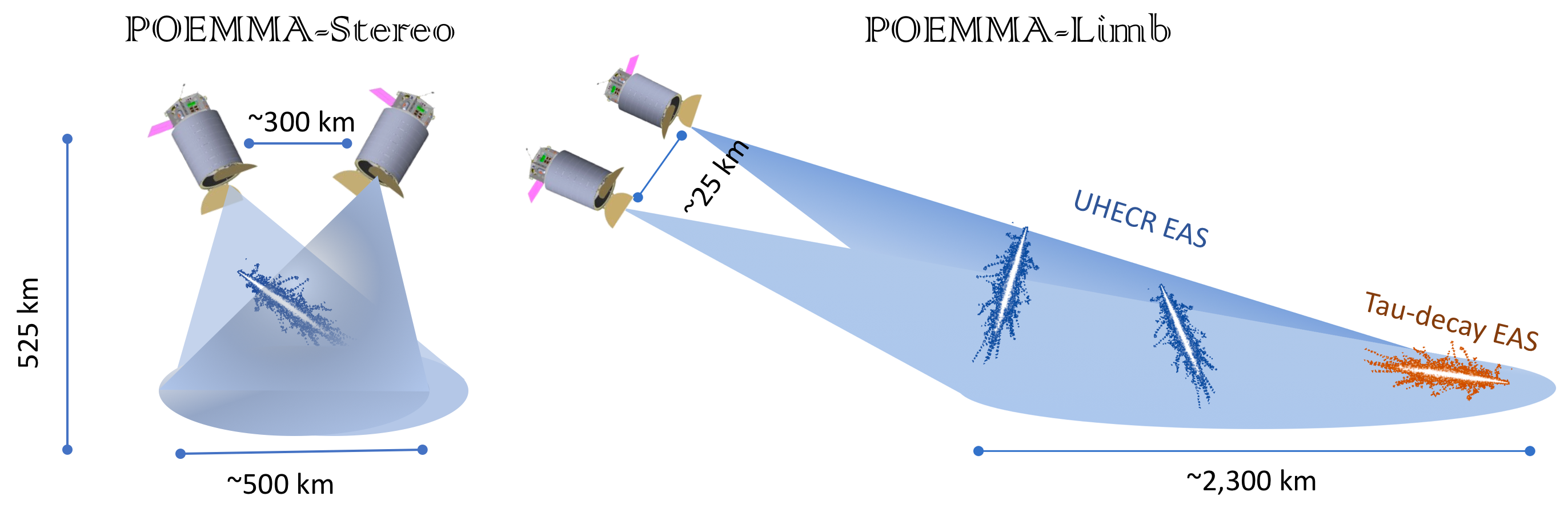}\\
        \includegraphics[width=0.34\textwidth]{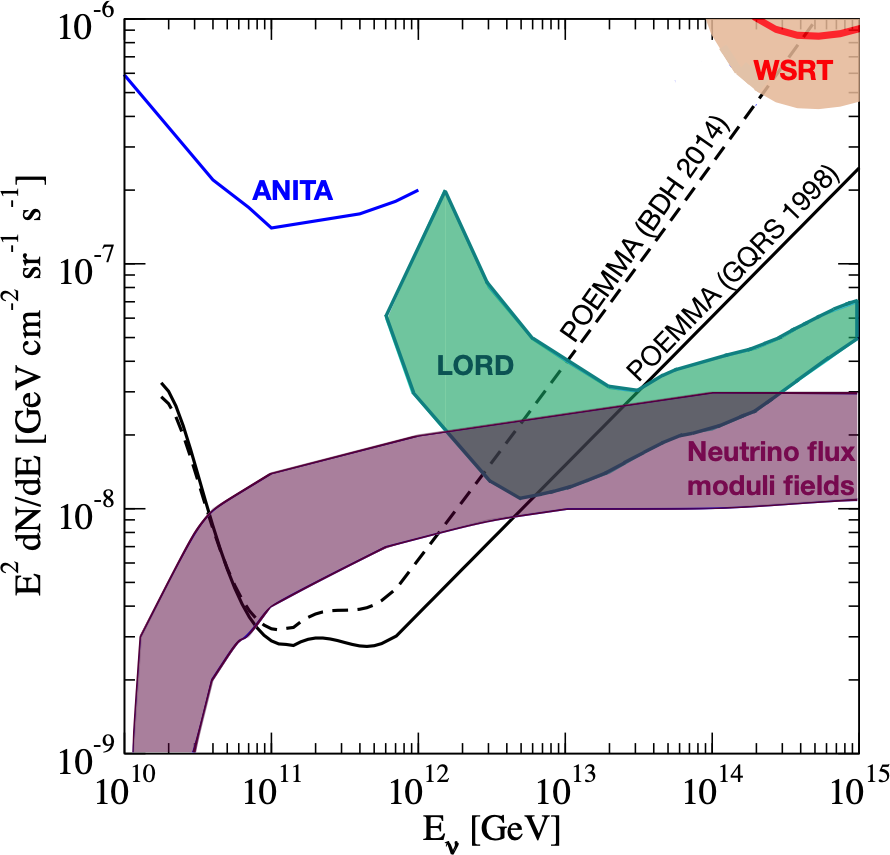}   
    \includegraphics[width=0.45\textwidth]{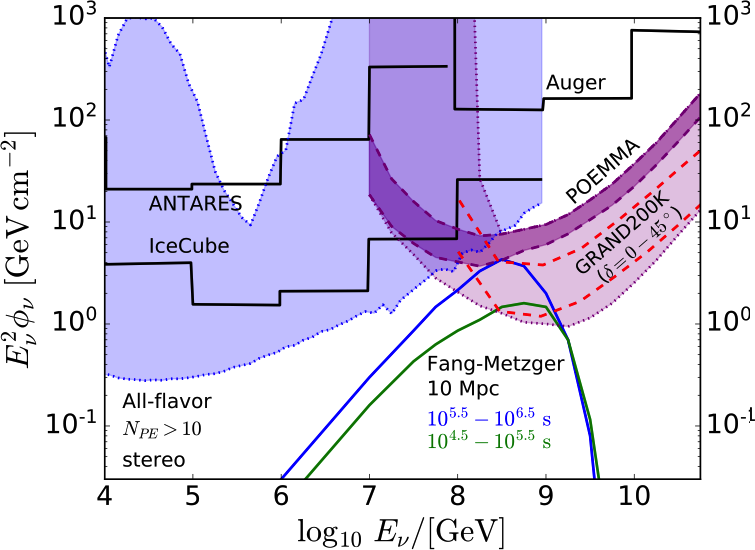}
\end{center}
\caption{{\it Top:} POEMMA observing modes: POEMMA-Stereo for UHECRs and UHE $\nu$-induced extensive air showers, and POEMMA-Limb for $\nu_\tau\to\tau$-sourced up-going extensive air showers. Figure reproduced from Ref.~\cite{POEMMA:2020ykm}.
{\it Bottom left:} POEMMA 90\% CL per decade, 5-year sensitivity to EAS showers arising from neutrino
interactions in the atmosphere and detected via the stereo fluorescence measurements from charged-current and neutral-current
interactions from all three neutrino flavors. The solid (dashed) curve is calculated using cross sections from Ref.~\citenum{Gandhi:1998ri} (Ref.~\citenum{Block:2014kza}). Predictions for strongly coupled string moduli (maroon
band) \cite{Berezinsky:2011cp} are also shown along with upper limits from ANITA I-IV (blue line)~\cite{Gorham:2019guw} and Westerbork
Synthesis Radio Telescope (WSRT; red line with tan band) \cite{Scholten:2009ad}, and the projected sensitivity for LORD
(green band) \cite{Ryabov:2016fac}. 
{\it Bottom right:} POEMMA 90\% CL per-decade tau neutrino sensitivity assuming the binary neutron star merger (BNS) model of Fang and Metzger~\cite{Fang:2017tla} for a source at a distance of 10~Mpc.  The sky-location sensitivity range for IceCube is shown in the blue band while that for GRAND is in the area defined by the red dash. The black lines show the measured limits to neutrinos from GW170817 set by IceCube and Auger \cite{ANTARES:2017bia}. Figure reproduced from Ref.~\cite{POEMMA:2020ykm}. 
  }
   \label{fig:POEMMAlimits}
 \end{figure}
 
The POEMMA mission consists of two identical spacecraft that fly in a loose formation at 525~km altitude, 28.5$^\circ$ inclination orbits, separated by a nominal lateral distance of 300~km. Each POEMMA spacecraft contains a Schmidt telescope with $6 \, \mathrm{m}^{2}$ optical collecting area and a $45^{\circ}$ full FoV. The focal plane of each telescope is segmented into two sections, each optimized for different EAS optical measurements, and with each section comprised of pixels with 0.086$^\circ$ FoV. The POEMMA Fluorescence Camera occupies 80\% of each focal surface and is optimized to record the fluorescence light from EAS initiated by UHECR in the atmosphere. The POEMMA Cherenkov Camera  occupies the remaining 20\% of the focal surface and is oriented to observe near the Earth's limb, when the POEMMA telescopes are appropriately tilted, to record the beamed, Cherenkov emission produced by EAS sourced from Earth-skimming neutrino interactions above 20~PeV. The POEMMA telescopes can slew in both azimuth ($90^{\circ}$ in $\sim 8$ minutes) and zenith, allowing for unprecedented follow-up on transient astrophysical events by tracking sources as they move across the sky as viewed below the limb of the Earth \cite{Venters:2019xwi}. The separation of the POEMMA spacecraft can also be decreased to $\sim$25~km to put both telescopes in the upward-moving EAS light pool for each event, thus reducing the neutrino detection energy threshold. The period of the POEMMA spacecraft orbit is 95 minutes. Due to this and the POEMMA telescopes viewing orientation, one of the main advantages of the POEMMA mission is being able to achieve full-sky coverage for both UHECR and UHE neutrino sources.
A schematic of the fluorescence telescope configuration viewing down and the Cherenkov telescope viewing towards the limb in the upper panel of Fig. \ref{fig:POEMMAlimits} shows POEMMA's two distinct science modes.  

The first science operating mode is a precision UHECR and UHE neutrino stereo mode where the telescopes are oriented to co-measure the EAS air fluorescence signal in a common volume corresponding to nearly $10^{13}$ tons of atmosphere. Due to the high accuracy of the EAS reconstruction from the stereo fluorescence technique when viewing the entire EAS development using the large FoV from low-Earth orbit, POEMMA can accurately reconstruct the development of the EAS with $\lesssim 20^\circ$ angular resolution, $\lesssim 20\%$ energy resolution, and $\lesssim 30$ g/cm$^2$ $X_{\rm max}$ resolution \cite{Anchordoqui:2019omw}. This performance yields excellent sensitivity for all neutrino flavors for UHE EAS that begin deeper in the atmosphere, well-separated from the dominant UHECR flux. 

At neutrino energies above EeV, in both charged-current and neutral-current interactions, $\sim$80\% of the neutrino energy is carried by the emergent lepton and $\sim 20$\% goes into a hadronic cascade \cite{Gandhi:1998ri}. Thus, the properties of the composite EAS are determined by the emergent UHE lepton for both charged- and neutral-current interactions in the atmosphere. The wide FoV for POEMMA allows the measurement of long, more horizontal EAS generated by neutrino interactions deep in the atmosphere, providing strong rejection of UHECR-induced EAS~\cite{Anchordoqui:2019omw}. Fig.~\ref{fig:POEMMAlimits} shows the 90\% CL per-decade all-flavor limits for 5-years of POEMMA fluorescence UHE measurements assuming two different neutrino cross sections~\cite{Gandhi:1998ri,Block:2014kza} at the highest energies. Comparison to the published limits from ANITA~\cite{Gorham:2019guw} and WSRT~\cite{Scholten:2009ad} observations are shown as well as theoretical predictions of the neutrino flux from strongly coupled string moduli generated early in the formation of the Universe~\cite{Berezinsky:2011cp}.
     
The second POEMMA science mode has the telescopes pointed to view slightly below the limb of the Earth to be sensitive to the beamed, optical Cherenkov signal sourced by tau-neutrino interactions in the Earth.
Tau leptons produced by neutrinos in the Earth can emerge and decay to produce up-moving EAS. 
This $\nu_{\tau} \rightarrow \tau$ detection channel allows POEMMA to have a sensitivity to energies $E_{\nu} \gtrsim 20$~PeV. POEMMA's FoV constrains viewing to an azimuth band of $\sim 30^\circ$; however, for an idealized $2 \pi$ azimuth telescope configuration with POEMMA's optical Cherenkov performance, the limits on the diffuse neutrino flux set by IceCube could be improved by roughly an order of magnitude for $E_{\nu}>$100~PeV~\cite{Reno:2019jtr, Cummings:2020ycz}. In principle, POEMMA is also sensitive to Earth-emergent neutrinos through the $\nu_{\mu} \rightarrow \mu$ and $\nu_{\tau} \rightarrow \tau \rightarrow \mu$ interaction channels. These channels improve POEMMA's sensitivity for $E_{\nu}<10$~PeV due to the relatively long interaction lengths of the muon around 1~PeV, and in the case of the primary $\nu_{\mu}$, increased Earth-emergence probabilities.

The capability of the POEMMA satellites to quickly slew to the direction of an astrophysical transient neutrino source and follow the source give POEMMA unique observational capabilities using the Cherenkov telescopes. As shown in Fig.~\ref{fig:POEMMAlimits}, it is expected that, compared to ground-based experiments, POEMMA will improve upon the sensitivity to long-burst transient events (duration of $10^{5-6}$~s from, e.g., binary neutron star mergers and tidal disruption events) for $E_{\nu}>100$~PeV by nearly an order of magnitude. Additionally, POEMMA will improve on the sensitivity to short-burst (duration of $\sim 10^{3}$~s from, e.g, short gamma ray bursts with extended emission) by at least an order of magnitude~\cite{Venters:2019xwi}, as illustrated in Fig.~\ref{fig:short_burst_nu}.


\subsubsection{EUSO-SPB2} 

The Extreme Universe Space Observatory aboard a Super Pressure Balloon 2 (EUSO-SPB2) is the follow-up mission to the EUSO-Balloon  and EUSO-SPB1 missions, flown in 2014, and 2017, respectively. EUSO-SPB2 will implement the technologies utilized during past EUSO missions in a near-space environment to validate the detection strategy of POEMMA and future space-based observatories by measuring cosmic rays via both fluorescence and optical Cherenkov emission~\cite{Scotti:2020, Eser:2021}.

The EUSO-SPB2 instrument contains two telescopes, each with a different science target: the Fluorescence Telescope (FT) points downward and will measure the microsecond-scale fluorescence tracks of EAS induced by UHECR interactions in the atmosphere, while the Cherenkov Telescope (CT) points towards the limb to measure the nanosecond-scale Cherenkov emission produced by EAS induced by above-the-limb cosmic rays and Earth-skimming neutrinos. The Field of View (FoV) of the EUSO-SPB2 FT is $11^{\circ} \times 35^{\circ}$, while that of the CT is $4^{\circ} \times 12.8^{\circ}$. The EUSO-SPB2 flight train will include an azimuth rotator configured for day and night pointing and an elevation angle tilting mechanism on the CT. EUSO-SPB2 will launch from Wanaka, NZ in Spring of 2023, and aims for a 100-day flight at an altitude of 33~km.

In addition to quantifying the night-sky airglow background near the Earth limb for the optimization of future space and near-space missions, the EUSO-SPB2 CT is expected to observe hundreds of direct cosmic rays per hour of operational live time during its flight \cite{Cummings:2021}. These events share many characteristics with the signals generated from neutrino sourced events and are a prime candidate for event reconstruction and evaluation of detector optics, electronics, and triggering algorithms. A larger data set of these above-the-limb events will help to quantify refraction near the Earth limb, and may help evaluate similar events that have been observed by ANITA, and are likely to be observed by PUEO. Due to its relatively short flight duration and small azimuth FoV, EUSO-SPB2 is not competitive with existing ground-based experiments in regards to setting limits on the diffuse neutrino flux~\cite{Cummings:2020ycz}. However, by having the ability to slew in azimuth and zenith, EUSO-SPB2 can perform ``target of opportunity" observations, where it will search for neutrino sources following alerts of astrophysical transients \cite{Venters:2021yex,Reno:2021veo}.


\subsubsection{ANITA}

The Antarctic Impulsive Transient Antenna (ANITA) is a NASA mission that searched for ultra-high energy particles from a long-duration balloon in Antarctica~\cite{ANITA:2008wdk}. Originally designed to search for neutrino-induced Askaryan radiation that refracts out of the ice, successive flights and studies demonstrated its capability of detecting geomagnetic radiation from cosmic rays~\cite{ANITA:2010ect, Schoorlemmer:2015afa, ANITA:2016vrp, ANITA:2018sgj, Gorham:2020zne} and its sensitivity to Earth-skimming tau neutrinos~\cite{Romero-Wolf:2018zxt, ANITA:2021xxh}. The upper limits from ANITA's Askaryan channel remain the strongest in the energy range above 30~EeV~\cite{Gorham:2008dv, ANITA:2010hzc, ANITA:2018vwl, Gorham:2019guw}. In three of the flights, its trigger was sensitive to the geomagnetic radiation from air showers and in each of these flights it reported anomalies that remain outstanding questions in the field~\cite{ANITA:2016vrp, ANITA:2018sgj, Gorham:2020zne}. It has additionally set constraints on the search for magnetic monopoles~\cite{ANITA-II:2010jck} and Lorentz-invariance violation~\cite{Gorham:2012qs}.

The great advantage of the sub-orbital platform is that the high float altitude enables the largest instantaneous effective areas in the 10~EeV--ZeV energy range and a unique observing geometry. The ANITA payloads consisted of broadband quad-ridged horn antennas with a maximum bandwidth of 180--1200 MHz with improved trigger sensitivities, lower noise temperatures, and extensions with each subsequent flight. While the trigger used a combinatoric trigger, interferometric analysis of the signal enabled ANITA to reconstruct the radio-signal to within $0.1^{\circ}$-$0.2^{\circ}$~\cite{Romero-Wolf:2014pua}, which translates to an elongated ellipse in the Askaryan channel with a maximum extent of a few degrees and to ${\sim}1^{\circ}$ for the in-air tau neutrino channel. This beam-forming technique is carried forward and implemented at the trigger level for PUEO, dramatically lowering its trigger and energy threshold~\cite{PUEO:2020bnn}.


\subsubsection{PUEO}

\begin{figure}[t]
    \centering
    \includegraphics[height=0.32\textwidth]{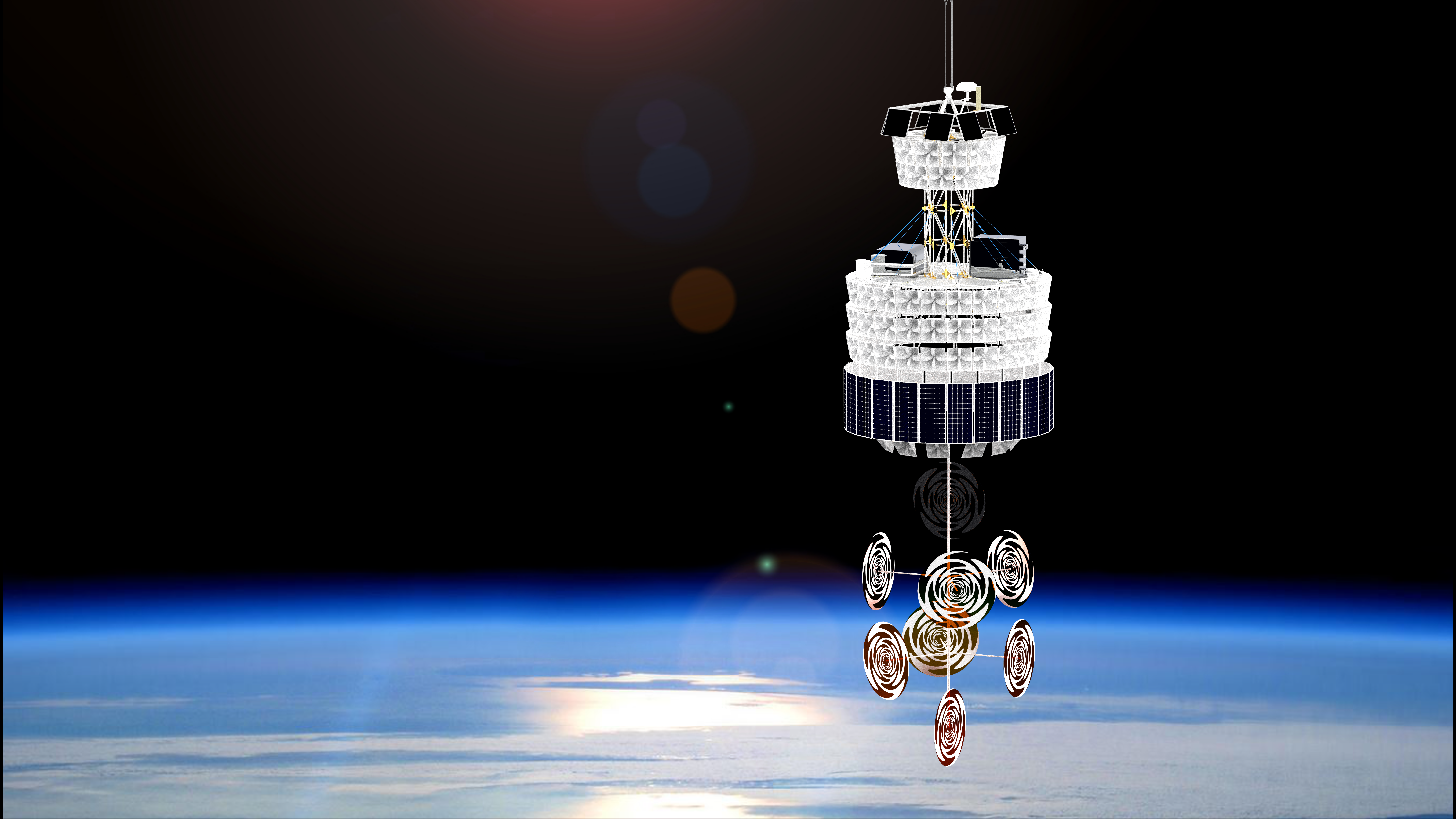}
    \includegraphics[height=0.33\textwidth]{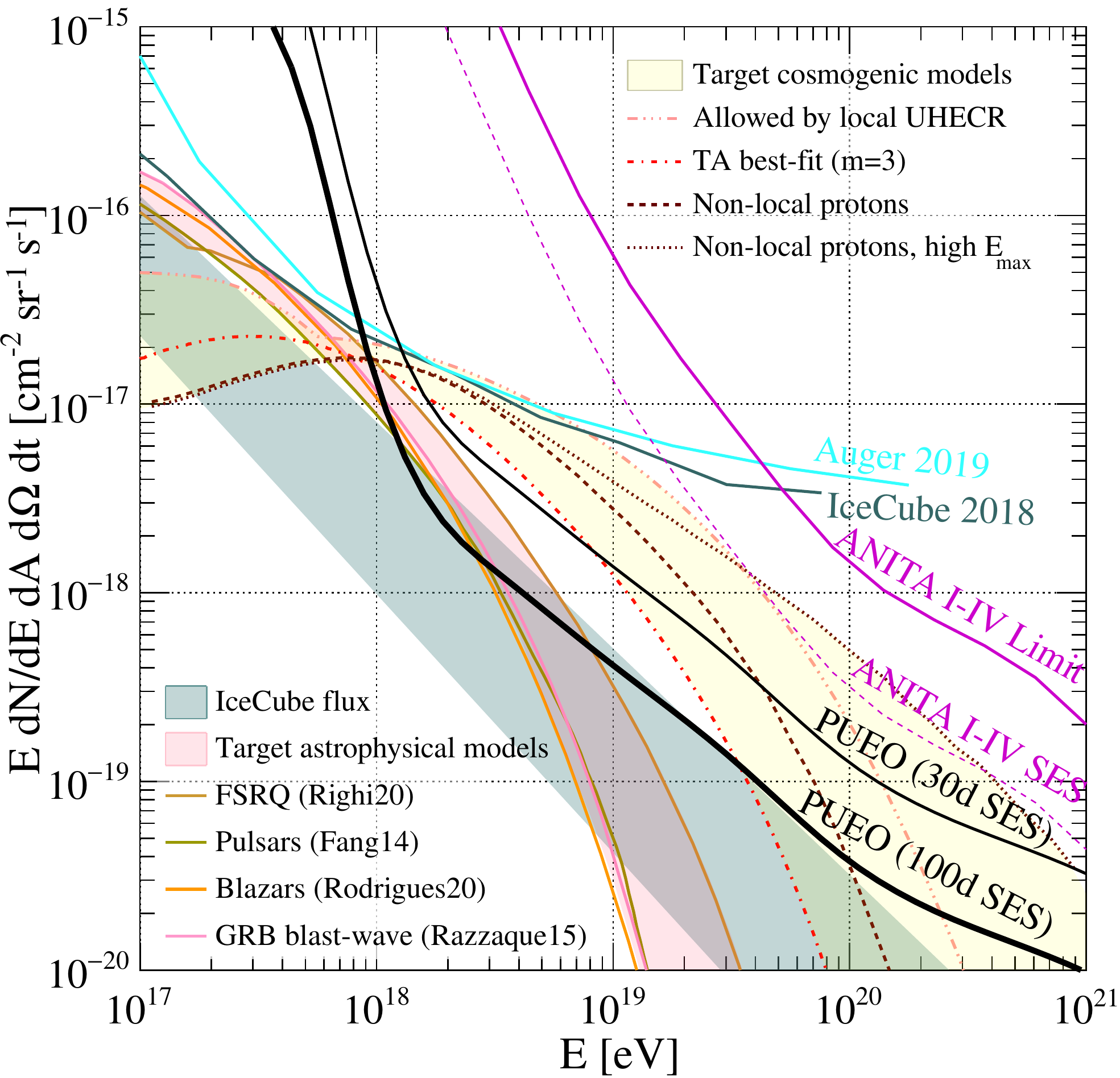}
    \caption{PUEO, the successor to the ANITA experiment. \textit{left} PUEO has two different antenna arrays: the main instrument, comprised of over 100 quad-ridged horn antennas; and a low-frequency instrument, consisting of several channels of sinuous antennas, that drops down below the payload (seen here) and is designed to increase PUEO's sensitivity to extensive air shower-like signals from $\tau$-lepton-decays in the air, and UHECRs. From Ref.~\cite{PUEO:2020bnn}. \textit{right} PUEO is designed to search for UHE neutrino interactions in the Antarctic ice via the Askaryan effect, and will have world-leading sensitivity to UHE neutrinos with energies above 1 EeV \cite{PUEO:2020bnn, Deaconu:2019rdx}.
    \label{fig:pueo}}
\end{figure}

The Payload for Ultrahigh Energy Observations (PUEO) (Fig.~\ref{fig:pueo}) 
is a long-duration Antarctic
balloon experiment designed to have world-leading sensitivity to
UHE neutrinos at energies above \SI{1}{EeV}. The direct successor to the ANITA experiment, it expects more
than an order of magnitude more sensitivity than its predecessor below
\SI{30}{EeV}~\cite{PUEO:2020bnn, Deaconu:2019rdx}. A multi-purpose instrument for detecting ultra-high-energy particles, PUEO can search for radio emission from neutrinos both from when they interact in the ice and from Earth-skimming tau neutrinos. PUEO is also sensitive to geomagnetic radio emission from UHECRs, including showers that evolve in the stratosphere. The sensitivity of
PUEO allows it to probe a host of diffuse cosmogenic and astrophysical
neutrino flux models, with multiple neutrinos expected during the
nominal \SI{30}{day} flight under the most optimistic flux models
allowed by current limits~\cite{PUEO:2020bnn}. PUEO’s very large
instantaneous aperture makes it well-suited to measuring UHE neutrino
fluences from transient astrophysical sources, if they occur in PUEO's
field-of-view during the flight.

The main PUEO instrument, seen in Fig.~\ref{fig:pueo}, follows the same overall design of ANITA but
with significant changes to the internal electronics and improvements
to the signal chain to significantly increase the sensitivity
in all of its four detection channels (in-ice Askaryan,
above-horizon stratospheric UHECR, reflected UHECR, and Earth-skimming
$\tau$-induced EAS). Like the ANITA experiment, PUEO will be sensitive
to tau neutrinos through the radio emission from both in-ice
(Askaryan) events and in-air $\tau$-decay-induced showers.

Unlike ANITA, PUEO is composed of two instruments, each
dedicated to different detection channels.  The ``main instrument''
consists of 96~triggered quad-ridged horn antennas (compared to the~48 used in
the last flight of ANITA), with an additional ring of downward-canted antennas that increase
PUEO's sensitivity to the
anomalous steeply upcoming cosmic-ray-like events observed in past
ANITA flights~\cite{ANITA:2018sgj,ANITA:2017qmn}. The main instrument targets the \SI{300}{MHz} to
\SI{1200}{MHz} frequency band, similar to the ANITA instrument, but with a slightly higher low-frequency limit. The multi-channel ``low-frequency
instrument'' targets the 50--300~MHz band designed
to detect EAS signals, from UHECRs or $\tau$-lepton decays, extending the frequency range and allowing for an independent measure of polarity for events due to air showers. PUEO will also
deploy an improved radio-frequency
signal chain and more accurate attitude determination systems, when
compared to ANITA-IV.

Despite deploying more than twice the number of antennas as the last
flight of ANITA, the biggest improvement in PUEO's sensitivity comes
from the real-time interferometric beamforming used in the
trigger subsystem. This beamforming trigger computes highly
directional beams on the sky, in real-time during the flight, by
coherently summing waveforms with different time delays, allowing PUEO
to have a 50\% trigger efficiency level at a signal-to-noise ratio
(SNR) of \({\sim}1\)~\cite{PUEO:2020bnn}. PUEO's beamforming trigger
system, with extensive heritage from the phased-array trigger of the
Askaryan Radio Array~\cite{Allison:2018ynt}, is built on the Xilinx
RFSoC (Radio-Frequency System-on-Chip) platform which combines
high-bandwidth digitizers, large field-programmable gate-arrays
(FPGAs), and digital signal processing (DSP) cores onto a single die.


\subsubsection{AugerPrime}

\begin{figure}[t]
    \centering
    \includegraphics[width=.39\textwidth, trim={1cm 1cm 0 0},clip]{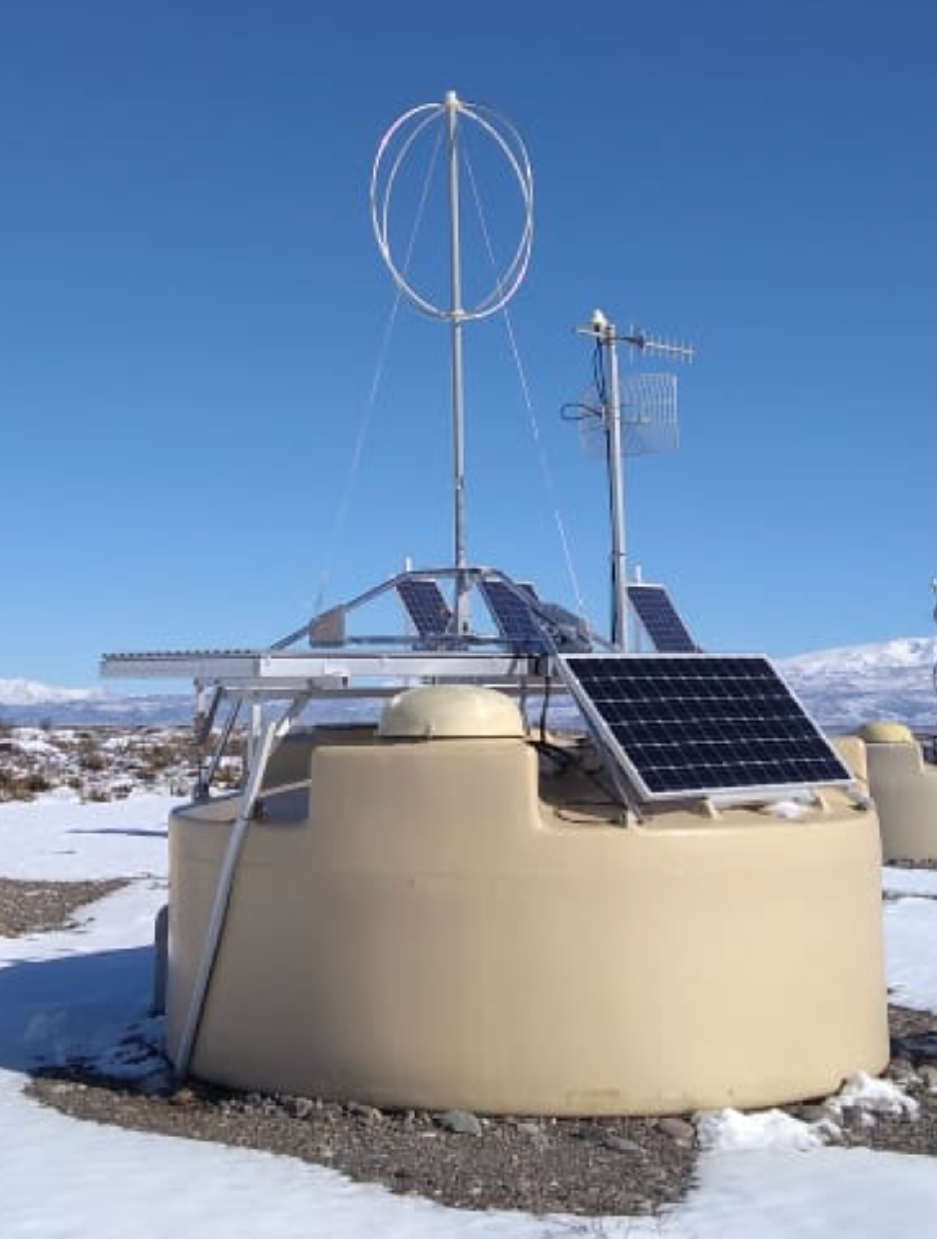}
     \includegraphics[width=.57\textwidth, trim={0.5cm 0.5cm 0 0},clip]{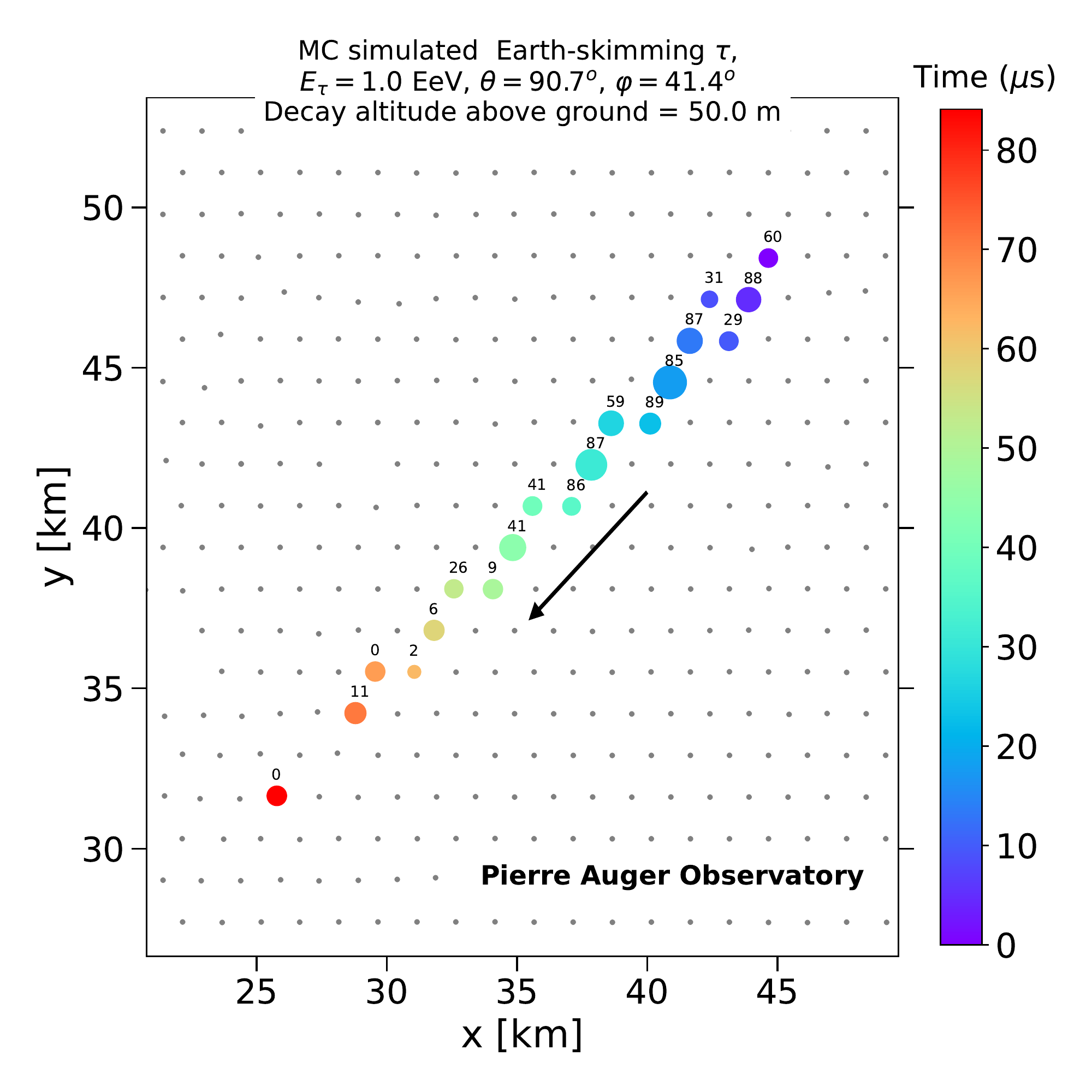}
    \caption{{\it Left:} Picture of an upgraded water-Cherenkov station of AugerPrime. The flat scintillator detector and the SALLA antenna are visible on top of the station, along with the solar panel and the communications Yagi antenna. Source: Pierre Auger Observatory. {\it Right:} Footprint of a simulated upward-going tau-induced shower produced by a tau neutrino skimming the Earth. The colors indicate the time of arrival of the shower particles at the water-Cherenkov stations from earliest (violet) to latest (red). The black arrow is parallel to the projection of the shower axis on the ground. The radii of the colored stations are proportional to the logarithm of the signal deposited, and the numbers on top are the fraction of the signal due to the electromagnetic component of the shower. The gray dots represent stations of the array without signal. A large fraction of electromagnetic signal in inclined events such as this one, concentrated in the earliest part of the shower, constitute a signature of neutrino-induced showers.}
    \label{fig:AugerPrime}
\end{figure}

With the present Surface Detector (SD) of the Pierre Auger Observatory---an array of $\sim 1600$ Water-Cherenkov Detectors (WCD) spread over a surface of $\sim\,3000~\mathrm{km^2}$ in an hexagonal grid with 1.5 km separation---UHE neutrinos can be efficiently discriminated in the much larger background of cosmic ray-induced EAS. The neutrino search capabilities are enhanced at EeV energies for EAS developing in the inclined directions with respect to the vertical to ground, at zenith angles $\theta\in(60^\circ,95^\circ)$, by looking for penetrating showers with a significant amount of electromagnetic component, characteristic of neutrino-induced EAS close to the ground~\cite{PierreAuger:2019ens}.

AugerPrime is the upgrade of the Pierre Auger Observatory~\cite{Aab:2016vlz}. It is fully funded, under construction, and foreseen to take data until 2030. A complementary measurement of the shower particles will be provided by an array of plane plastic scintillators (SSD)~\cite{PierreAuger:2021ccl} above the existing WCD (Fig.~\ref{fig:AugerPrime}). This allows sampling the shower particles simultaneously with two detectors that have different responses to the muonic and electromagnetic components of the shower, potentially improving the identification of neutrino-induced EAS. The geometrical acceptance of the flat scintillators is reduced for very inclined showers $\theta>75^\circ$ where neutrino identification is currently performed with the highest efficiency~\cite{PierreAuger:2019ens}. However, for moderate inclination angles $\theta\in(\sim60^\circ,\,\sim75^\circ)$ the expectation is that the SSD can improve neutrino selection in an angular range where the current search algorithms reach efficiencies typically below $75\%$, and even extend it to $\theta < 60^\circ$, where searches for neutrinos are not performed due to the large cosmic-ray background not reducible in an efficient manner with current algorithms.

Besides more than doubling the current statistics, AugerPrime will have superior cosmic-ray composition measurement capabilities compared to the WCD array~\cite{PierreAuger:2021ccl}. Two of the major objectives of AugerPrime are to elucidate the mass composition and the origin of the flux suppression at the highest energies as well as search for a flux contribution of protons up to the highest energies as low as $10\%$. The proton fraction is known to have a large effect on the intensity of the cosmogenic neutrino flux~\cite{vanVliet:2019nse}, constituting a decisive ingredient for estimating the physics potential and design of future detectors searching for UHE neutrinos. 

An integral part of the AugerPrime detector is the Radio Detector (RD)~\cite{Pont:2021pwd}. Built on the experience gained with the construction and operation of the Auger Engineering Radio Array~\cite{PierreAuger:2012ker}, the goal is to install an antenna on top of each WCD to detect the radio signal induced in EAS by geomagnetic charge separation in the shower~\cite{Kahn-Lerche_not_in_inspires} and the Askaryan effect~\cite{Askaryan:1961pfb}. The RD is most efficient for the detection of inclined showers~\cite{PierreAuger:2018pmw}, and the radio footprint at ground is known to contain information on $X_\mathrm{max}$~\cite{PierreAuger:2021rio}, the depth at which the shower reaches the maximum in its development. The RD will provide a clean measurement of the electromagnetic shower component, ideal to separate neutrino-induced cascades from hadronic showers. These two facts open up the possibility of using the radio detector in the search for penetrating inclined showers induced by UHE neutrinos, complementing and aiding in current searches.

The surface detector stations will be upgraded with faster, better timing accuracy and increased dynamic range electronics~\cite{Nitz:2021kdx}. The new electronics has the resources to implement further trigger algorithms on top of the existing ones, targeted to physics tasks such as searches for photons and neutrinos at sub-EeV energies, including the possibility of incorporating the WCD, SSD, and RD to the trigger providing a wealth of information on a single event. This is only starting now to be exploited in Auger~\cite{PierreAuger:2021ywt}.


\subsubsection{GCOS}

\begin{figure}[t]
 \includegraphics[width=0.7\textwidth]{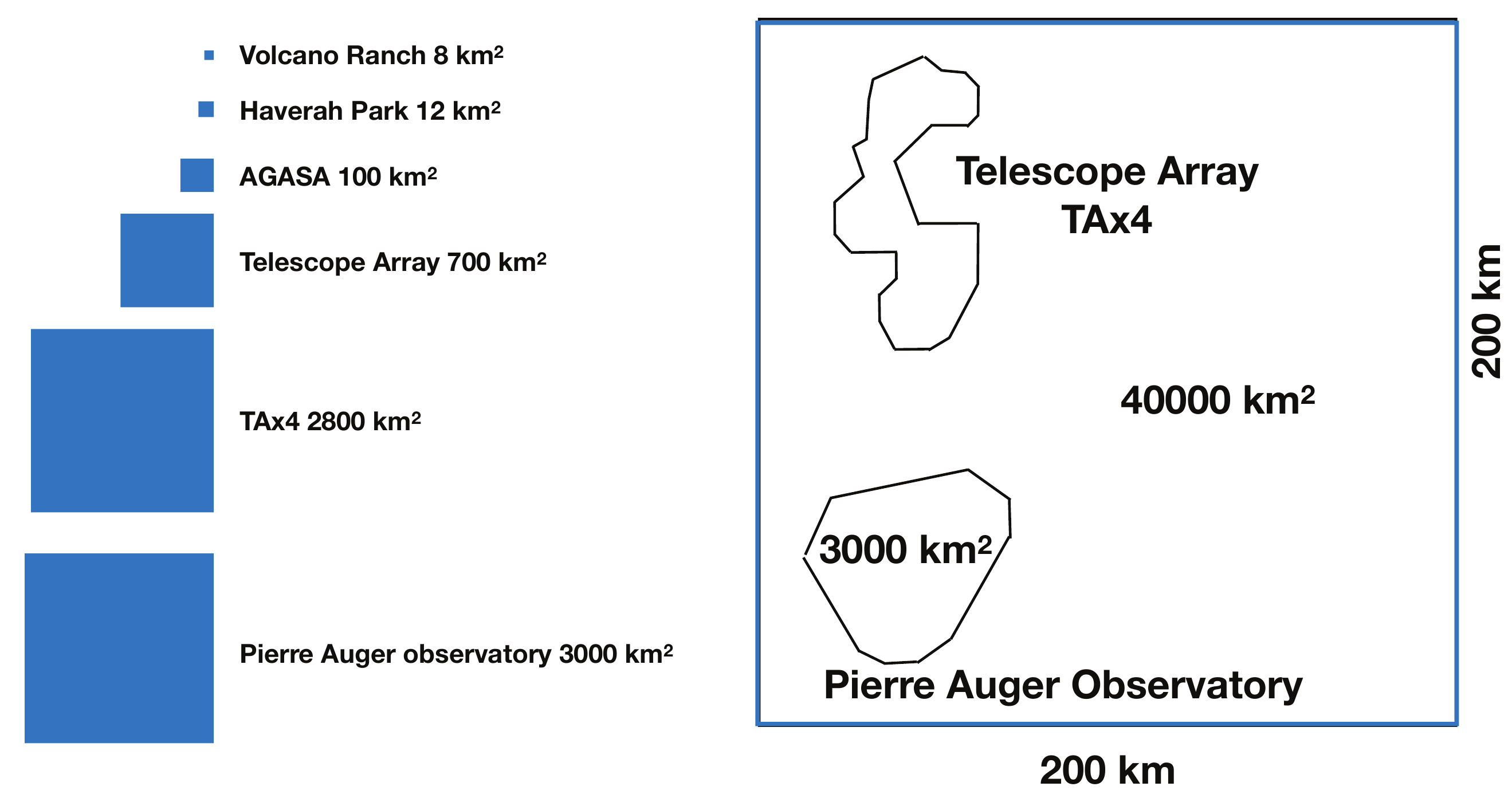}
 \caption{Illustration of total collection area for existing installations and GCOS. From Ref.~\cite{Horandel:2021prj}.}
 \label{fig:gcos-area}
\end{figure}

The Global Cosmic Ray Observatory (GCOS) is presently in the design phase, with the final detection concept and setup being worked out.  More detailed considerations can be found in Ref.~\cite{Horandel:2021prj} and references therein. With the goal of reaching  an exposure of  at least $2\times 10^5$~km$^2$ yr in a period of 10 years and full-sky coverage a set of surface arrays with a total area of about 40000~km$^2$  is anticipated as shown in Fig.\,\ref{fig:gcos-area}. 

GCOS will be designed as a multi-messenger observatory, aiming to register charged cosmic rays, gamma rays, and neutrinos with energies above $10^{19}$~eV.
Different detection concepts are at hand. They need to be optimized to reach the targeted physics case. The design will include a surface detector array with segmented/nested water Cherenkov detectors and radio antennas.
Similarly to AugerPrime, as described above, but with a much bigger area/acceptance, the surface detector array of GCOS will be able to search for neutrinos by identifying electron-rich horizontal air showers and distinguish them from hadronic cascades.
The radio detector will provide a clean measurement of the electromagnetic component, ideal to identify neutrino-induced cascades.
Nested water-Cherenkov detectors will also provide enhanced mass and neutrino sensitivity for horizontal air showers.
Identification of the sources of UHE particles will require a good angular resolution.
Assuming a detector spacing of the order of $1.6-2$~km, an angular resolution $<0.5^\circ$ is realistic.
 
GCOS is at present in the phase of developing its science case and studies are being conducted to optimize a suitable detector design. 
The performance of existing detector systems, in particular AugerPrime, gives concrete and proven design examples to achieve the needed sensitivity for UHE neutrinos.
The final design of GCOS will depend on the results that will be obtained with the Auger Observatory and the Telescope Array in the coming decade. New findings will influence the  science case and, thus, the design of the observatory. A promising approach towards a full-scale GCOS could be to gradually increase the aperture of the existing arrays. For example, discoveries brought about by increasing the aperture of present-day Auger a few times will improve the understanding about the highest-energy particles and will clarify the neutrino-related design goals for an even larger observatory.
Prototype detectors are expected to be built after 2025. The construction of GCOS at multiple sites is expected to start after 2030, with an anticipated operation time of at least twenty years.


\section*{Acknowledgements}

MA is supported by the Initiative and Networking Fund of the Helmholtz Association, the Deutsches Elektronen Synchrotron (DESY).
MB is supported by the {\sc Villum Fonden} under project no.~29388. 
LL is supported by the US National Science Foundation (US NSF)-Office of Polar Programs, the US NSF Physics Division through Awards PHY-1913607 and the Wisconsin Alumni Research Foundation.
NO is supported by the US NSF through Award 2112769.
MHR is supported in part by the US Department of Energy (US DOE) grant DE-SC-0010113. 
SAW is supported by the US NSF through Awards 2111232 and CAREER 2033500 and NASA through Awards 80NSSC20K1288, 80NSSC20K0925, and 80NSSC20K0925.   
SKA and AN acknowledge the financial support from the Swarnajayanti Fellowship Research Grant (File no.\~SB/SJF/2020-21/21) provided 
by the Science and Engineering Research Board (SERB) and Department of Science and Technology (DST), Govt.~of India. SKA would like 
to thank the United States-India Educational Foundation for providing the financial support through the Fulbright-Nehru Academic and 
Professional Excellence Fellowship (Award No.~2710/F-N APE/2021).
JAM and EZ have received financial support from Xunta de Galicia (Centro singular de investigación de Galicia accreditation 2019-2022), by European Union ERDF, and by the ``Mar\'ia de Maeztu” Units of Excellence program MDM-2016-0692 and the Spanish Research State Agency, and from Ministerio de Ciencia e Innovaci\'on PID2019-105544GB-I00
and RED2018-102661-T (RENATA).
RAB is funded by the ``la Caixa'' Foundation (ID 100010434) and the European Union's Horizon~2020 research and innovation program under the Marie Skłodowska-Curie grant agreement No 847648, fellowship code LCF/BQ/PI21/11830030. 
BAC thanks the National Science Foundation for support through the Astronomy and Astrophysics Postdoctoral Fellowship under Award AST-1903885.
PBD acknowledges support from the US DOE under Grant Contract DE-SC0012704. 
YF would like to acknowledge support from the ICTP through the Associates Programme and from the Simons Foundation through grant number
284558FY19.
YSJ is supported by the National Research Foundation of Korea (NRF) grant funded by the Korea government through Ministry of Science and ICT Grant 2021R1A2C1009296.
The work of SRK was supported in part by the US NSF under grant number PHY-1307472 and the US DOE under contract number DE-AC-76SF00098.
JFK acknowledges support by NASA grant 80NSSC19K0626 at the University of Maryland, Baltimore County and under proposal 17-APRA17-0066 at NASA/GSFC.
ER is funded by the Deutsche Forschungsgemeinschaft (DFG, German Research Foundation) SFB 1258, Cluster of Excellence 2094 ORIGINS, and the European Research Council (ERC) (Grant agreement No.~101053168).
ZhD and OVS are supported by the Ministry of Education and Science of the Russian Federation within the financing program of large scientific projects of the ``Science'' National Project (grant no.~075-15-2020-778).
The work of SY is supported by JSPS KAKENHI Grant No.~18H05206.
BZ is supported by the Simons Foundation.


\bibliographystyle{JHEP}
\input{main-v2.bbl}
\end{document}

%% file: authors.tex

\author[1]{Markus Ackermann,}
\author[2,3,4]{Sanjib K. Agarwalla,}
\author[5]{Jaime Alvarez-Mu\~niz,}
\author[6]{Rafael {Alves Batista},}
\author[7]{Carlos A. Arg\"uelles,}
\author[8]{Mauricio Bustamante,}
\author[9]{Brian A.~Clark,}
\author[10]{Austin Cummings,}
\author[2,3]{Sudipta Das,}
\author[10]{Valentin Decoene,}
\author[11]{Peter B.~Denton,}
\author[12]{Damien Dornic,}
\author[13]{Zhan-Arys Dzhilkibaev,}
\author[14]{Yasaman Farzan,}
\author[6,15]{Alfonso~Garcia,} 
\author[16]{Maria Vittoria Garzelli,}
\author[17]{Christian Glaser,}
\author[18,19]{Aart Heijboer,}
\author[18,20]{J\"org R.\ H\"orandel,}
\author[21,22]{Giulia Illuminati,}
\author[23]{Yu Seon Jeong,}
\author[24]{John L.~Kelley,}
\author[25]{Kevin J. Kelly,}
\author[10]{Ali Kheirandish,}
\author[26,27]{Spencer R. Klein,}
\author[28]{John F. Krizmanic,}
\author[29]{Michael J. Larson,}
\author[24]{Lu Lu,}
\author[10]{Kohta Murase,}
\author[2]{Ashish Narang,}
\author[30]{Nepomuk Otte,}
\author[31]{Remy L. Prechelt,}
\author[32,33]{Steven Prohira,}
\author[34]{Mary Hall Reno,}
\author[35]{Elisa Resconi,}
\author[36]{Marcos Santander,}
\author[8]{Victor B. Valera,}
\author[24]{Justin Vandenbroucke,}
\author[13]{Olga Vasil'evna Suvorova,}
\author[37]{Lawrence Wiencke,}
\author[10]{Stephanie Wissel,}
\author[38]{Shigeru Yoshida,}
\author[24]{Tianlu Yuan,}
\author[5]{Enrique Zas,}
\author[7]{Pavel Zhelnin,}
\author[39]{Bei Zhou}

%% file: affiliations.tex

\affiliation[1]{DESY, D-15738 Zeuthen, Germany}
\affiliation[2]{Institute of Physics, Sachivalaya Marg, Sainik School Post, Bhubaneswar 751005, India}
\affiliation[3]{Homi Bhabha National Institute, Anushakti Nagar, Mumbai 400094, India}
\affiliation[4]{International Centre for Theoretical Physics, Strada Costiera 11, 34151 Trieste, Italy}
\affiliation[5]{Instituto Galego de Fisica de Altas Enerxias, Universidade de Santiago de Compostela, Spain}
\affiliation[6]{
Instituto de F\'isica Te\'orica UAM/CSIC, Universidad Aut\'onoma de Madrid, 28049 Madrid, Spain}
\affiliation[7]{Department of Physics \& Laboratory for Particle Physics and Cosmology, Harvard University, Cambridge, MA 02138, USA}
\affiliation[8]{Niels Bohr International Academy, Niels Bohr Institute, University of Copenhagen, Copenhagen 2100, Denmark}
\affiliation[9]{Dept. of Physics and Astronomy, Michigan State University, East Lansing, MI 48824, USA}
\affiliation[10]{Departments of Physics and Astronomy \& Astrophysics, Institute for Gravitation and the Cosmos, Pennsylvania State University, State College, Pennsylvania 16802 USA}
\affiliation[11]{High Energy Theory Group, Physics Department, Brookhaven National Laboratory, Upton, NY 11973, USA}
\affiliation[12]{Aix Marseille Univ, CNRS/IN2P3, CPPM, Marseille, France}
\affiliation[13]{Institute for Nuclear Research, Russian Academy of Sciences, Moscow, Russia}
\affiliation[14]{School of Physics, Institute for Research in Fundamental Sciences (IPM), Tehran, Iran}
\affiliation[15]{Instituto de Física Corpuscular (IFIC), Universitat de València (UV), 46980 Paterna, València, Spain}
\affiliation[16]{II. Institut f\"{u}r Theoretische Physik, Universit\"{a}t Hamburg, D–22761 Hamburg, Germany}
\affiliation[17]{Uppsala University Department of Physics and Astronomy, Uppsala SE-752 37, Sweden}
\affiliation[18]{Nikhef,
Science Park 105, 1098 XG Amsterdam, The Netherlands}
\affiliation[19]{University of Amsterdam,
Science Park 904, 1098 XH Amsterdam, The Netherlands}
\affiliation[20]{Radboud University, Department of Astrophysics/IMAPP, 6500GL Nijmegen, The Netherlands}
\affiliation[21]{INFN - Sezione di Bologna, Viale Berti-Pichat 6/2, 40127 Bologna, Italy}
\affiliation[22]{Dipartimento di Fisica e Astronomia dell’Università, Viale Berti Pichat 6/2, 40127 Bologna, Italy}
\affiliation[23]{High Energy Physics Center, Chung-Ang University, Seoul 06974, Korea}
\affiliation[24]{Department of Physics \& Wisconsin IceCube Particle Astrophysics Center,
University of Wisconsin–Madison, Madison, WI 53706, USA}
\affiliation[25]{Theoretical Physics Department, CERN, Esplande des Particules, 1211 Geneva 23, Switzerland}
\affiliation[26]{Department of Physics, University of California, Berkeley, CA 94720, USA}
\affiliation[27]{Lawrence Berkeley National Laboratory, Berkeley, CA 94720, USA}
\affiliation[28]{Laboratory for Astroparticle Physics, NASA/GSFC, Greenbelt, MD 20771,USA}
\affiliation[29]{Department of Physics, University of Maryland, College Park, MD 20742, USA}
\affiliation[30]{Georgia Institute of Technology, School of Physics \& Center for Relativistic Astrophysics, Atlanta, Georgia 30332-0430, USA}
\affiliation[31]{University of Hawai'i M\=anoa, Honolulu, HI 96822, USA}
\affiliation[32]{Department of Physics, Center for Cosmology and AstroParticle Physics, Ohio State University, Columbus, OH 43210}
\affiliation[33]{Department of Physics and Astronomy, University of Kansas, Lawrence, KS 66045}
\affiliation[34]{Department
of Physics and Astronomy, University of Iowa, Iowa City, Iowa 52242 USA}
\affiliation[35]{Department of Physics, Technical University of Munich, D-86748 Garching, Germany}
\affiliation[36]{Department of Physics and Astronomy, University of Alabama, Tuscaloosa, AL 35487, USA}
\affiliation[37]{Colorado School of Mines, Golden, CO 80401, USA}
\affiliation[38]{International Center for Hadron Astrophysics, Chiba University, Chiba 268-8522, Japan}
\affiliation[39]{Department of Physics and Astronomy, Johns Hopkins University, Baltimore, Maryland 21218, USA}

%% file: endorsers.tex
\eauthor[40]{Luis A. Anchordoqui,} 
\eauthor[24]{Yosuke Ashida,}
\eauthor[30]{Mahdi Bagheri,} 
\eauthor[24]{Aswathi Balagopal V.,}
\eauthor[24]{Vedant Basu,}
\eauthor[32]{James Beatty,} 
\eauthor[41]{Nicole Bell,} 
\eauthor[24]{Abigail Bishop,} 
\eauthor[7]{Julia Book,}
\eauthor[42]{Anthony Brown,} 
\eauthor[76]{Michael Campana,}
\eauthor[67]{Nhan Chau,}
\eauthor[43]{Alan Coleman,} 
\eauthor[32]{Amy Connolly,} 
\eauthor[68]{Pablo Correa,}
\eauthor[39]{Cyril Creque-Sarbinowski,} 
\eauthor[10]{Austin Cummings,} 
\eauthor[45]{Zachary Curtis-Ginsberg,} 
\eauthor[67]{Paramita Dasgupta,} 
\eauthor[68]{Simon De Kockere,} 
\eauthor[68]{Krijn de Vries,} 
\eauthor[9]{Tyce DeYoung,}
\eauthor[45]{Cosmin Deaconu,} 
\eauthor[24]{Abhishek Desai,}
\eauthor[52]{Armando di Matteo,} 
\eauthor[80]{Dominik Elsaesser,}
\eauthor[29]{Kwok Lung Fan,}
\eauthor[78]{Anatoli Fedynitch,}
\eauthor[10]{Derek Fox,}
\eauthor[33]{Mohammad Ful Hossain Seikh,} 
\eauthor[77]{Philipp F\"urst,}
\eauthor[77]{Erik Ganster,}
\eauthor[35]{Christian Haack,} 
\eauthor[1]{Steffen Hallman,} 
\eauthor[24]{Francis Halzen,}
\eauthor[80]{Andreas Haungs,} 
\eauthor[38]{Aya Ishihara,}
\eauthor[53]{Eleanor Judd,} 
\eauthor[1]{Timo Karg,} 
\eauthor[24]{Albrecht Karle,}
\eauthor[54]{Teppei Katori,} 
\eauthor[9]{Alina Kochocki,}
\eauthor[9]{Claudio Kopper,} 
\eauthor[1]{Marek Kowalski,} 
\eauthor[66]{Ilya Kravchenko,} 
\eauthor[76]{Naoko Kurahashi,}
\eauthor[65]{Mathieu Lamoureux,} 
\eauthor[73]{Hermes Le\'on Vargas,} 
\eauthor[81]{Massimiliano Lincetto,}
\eauthor[75]{Qinrui Liu,}
\eauthor[24]{Jim Madsen,}
\eauthor[24]{Yuya Makino,}
\eauthor[66]{Joseph Mammo,} 
\eauthor[24]{Kevin Meagher,}
\eauthor[38]{Maximilian Meier,}
\eauthor[35]{Martin Ha Minh,} 
\eauthor[56,57]{Lino Miramonti,} 
\eauthor[24]{Marjon Moulai,}
\eauthor[20]{Katharine Mulrey,} 
\eauthor[10]{Marco Muzio,} 
\eauthor[43]{Ek Narayan Paudel,}
\eauthor[60]{Ek Narayan Paudel,} 
\eauthor[1]{Anna Nelles,} 
\eauthor[24]{William Nichols,} 
\eauthor[33]{Alisa Nozdrina,} 
\eauthor[17]{Erin O'Sullivan,} 
\eauthor[24]{Vivian O\'Dell,}
\eauthor[24]{Jesse Osborne,}
\eauthor[58]{Vishvas Pandey,} 
\eauthor[9]{Jean Pierre Twagirayezu,}
\eauthor[24]{Alex Pizzuto,} 
\eauthor[59]{Mattias Plum,} 
\eauthor[79]{Carlos Pobes Aranda,}
\eauthor[1]{Lilly Pyras,} 
\eauthor[74]{Christoph Raab,}
\eauthor[24]{Zoe Rechav,}
\eauthor[30]{Oscar Romero Matamala,} 
\eauthor[36]{Marcos Santander,} 
\eauthor[24]{Pierpaolo Savina,}
\eauthor[60]{Frank Schroeder,} 
\eauthor[35]{Lisa Schumacher,} 
\eauthor[61]{Sergio Sciutto,} 
\eauthor[24]{Manuel Silva,}
\eauthor[72]{Rajeev Singh,} 
\eauthor[45]{Daniel Smith,} 
\eauthor[44]{Juliana Stachurska,}
\eauthor[63]{Olga Suvorova,} 
\eauthor[30]{Ignacio Taboada,} 
\eauthor[70]{Samuel Timothy Spencer,} 
\eauthor[67]{Simona Toscano,} 
\eauthor[61]{Matias Tueros,} 
\eauthor[68]{Nick van Eijndhoven,} 
\eauthor[69]{Peter Veres,}
\eauthor[69]{P\'eter Veres,} 
\eauthor[45]{Abigail Vieregg,} 
\eauthor[24]{Winnie Wang,}
\eauthor[62]{Robert Wayne Springer,} 
\eauthor[9]{Nathan Whitehorn,}
\eauthor[1]{Walter Winter,}
\eauthor[24]{Emre Yildizci,} 
\eauthor[9]{Shiqi Yu,}
\eauthor[64]{Marka Zsuzsa} 

%% file: affiliations-more.tex
\affiliation[40]{Department of Physics and Astronomy, Lehman College, City University of New York, Bronx, NY 10468, USA}
\affiliation[41]{School of Physics, The University of Melbourne, Victoria 3010, Australia}
\affiliation[42]{Centre for Advanced Instrumentation, Department of Physics, University of Durham, South Road, Durham, DH1 3LE, UK}
\affiliation[43]{Bartol Research Institute, Department of Physics and Astronomy, University of Delaware, Newark DE, USA}
\affiliation[44]{Dept. of Physics, Massachusetts Institute of Technology, Cambridge, MA 02139, USA}
\affiliation[45]{Dept. of Physics, Enrico Fermi Institue, Kavli Institute for Cosmological Physics, University of Chicago, Chicago, IL 60637, USA}
\affiliation[52]{INFN, Sezione di Torino, Torino, Italy}
\affiliation[53]{Space Sciences Laboratory, University of California, Berkeley, CA, USA}
\affiliation[54]{Department of Physics, King’s College London, WC2R 2LS London, UK}
\affiliation[55]{Columbia Astrophysics Laboratory, Columbia University in the City of New York, New York, NY 10027, USA}
\affiliation[56]{INFN, Sezione di Milano, Milano, Italy}
\affiliation[57]{Universit\`{a} di Milano, Dipartimento di Fisica, Milano, Italy}
\affiliation[58]{Department  of  Physics,  University  of  Florida,  Gainesville,  FL  32611,  USA}
\affiliation[59]{Physics Department, South Dakota School of Mines and Technology, Rapid City, SD 57701, USA}
\affiliation[60]{University of Delaware, Department of Physics and Astronomy, Bartol Research Institute, Newark, DE, USA}
\affiliation[61]{Departamento de F\'{i}sica, Universidad Nacional de La Plata, C. C.67 - 1900 La Plata, Argentina}
\affiliation[62]{Department of Physics and Astronomy, University of Utah, Salt Lake City, UT 84112, USA}
\affiliation[63]{Institute for Nuclear Research of Russian Academy of Sciences, Moscow, Russia}
\affiliation[64]{Columbia Astrophysics Laboratory, Columbia University in the City of New York, 550 W 120th St., New York, NY 10027, U.S.A.}
\affiliation[65]{Centre for Cosmology, Particle Physics and Phenomenology - CP3, Université catholique de Louvain, Louvain-la-Neuve, Belgium}
\affiliation[66]{Dept. of Physics and Astronomy, University of Nebraska, Lincoln, Nebraska 68588}
\affiliation[67]{Universit ́e Libre de Bruxelles, Science Faculty CP230, B-1050 Brussels, Belgium}
\affiliation[68]{Vrije Universiteit Brussel (VUB), Dienst ELEM, B-1050 Brussels, Belgium}
\affiliation[69]{Center for Space Plasma and Aeronomic Research, University of Alabama in Huntsville, 320 Sparkman Drive, Huntsville, AL 35899, USA}
\affiliation[70]{Erlangen Centre for Astroparticle Physics, Friedrich-Alexander-Universit{\"a}t Erlangen-N{\"u}rnberg, D-91058 Erlangen, Germany}
\affiliation[71]{Institute of Nanoscience and Materials of Aragon (INMA)}
\affiliation[72]{Institute of Nuclear Physics Polish Academy of Sciences, PL 31-342 Krak\'ow, Poland}
\affiliation[73]{Institute of Nuclear Physics Polish Academy of Sciences, PL 31-342 Krak\'ow, Poland}
\affiliation[74]{Institute of Physics of the Czech Academy of Sciences, 182 21 Prague 8, Czechia}
\affiliation[75]{Department of Physics, Engineering Physics and Astronomy, Queen’s University, Kingston, Ontario, Canada}
\affiliation[76]{Department of Physics, Drexel University, Philadelphia, Pennsylvania 19104}
\affiliation[77]{III. Physikalisches Institut B, RWTH Aachen University, 52056 Aachen, Germany}
\affiliation[78]{Institute of Physics, Academia Sinica, Taipei, Taiwan}
\affiliation[79]{INMA, CSIC-Universidad de Zaragoza, Pedro Cerbuna 12, 50009 Zaragoza, Spain}
\affiliation[80]{Karlsruhe Institute of Technology, Institute for Astroparticle Physics, D-76021 Karlsruhe, Germany}

%% file: main-v2.bbl
\hyphenation{Post-Script Sprin-ger}

\providecommand{\href}[2]{#2}\begingroup\raggedright\endgroup